\documentclass[preprint2]{aastex}
\usepackage{fixltx2e}
 
\shorttitle{Circumstellar medium around massive stars}
\shortauthors{Toal\'a \& Arthur}
\begin{document}
\title{Radiation-Hydrodynamic Models of the evolving Circumstellar Medium around Massive Stars}
\author{J.~A. Toal\'{a} \& S.~J. Arthur}
\affil{Centro de Radioastronom\'{i}a y Astrof\'{i}sica, Universidad Nacional Aut\'{o}noma de M\'{e}xico, Campus Morelia}
\affil{Apartado Postal 3-72, 58090, Morelia, Michoac\'{a}n, M\'{e}xico}
\email{j.toala@crya.unam.mx, j.arthur@crya.unam.mx}

\begin{abstract}
We study the evolution of the interstellar and circumstellar media
around massive stars ($M\geqslant 40 M_{\odot}$) from the main
sequence through to the Wolf-Rayet stage by means of
radiation-hydrodynamic simulations. We use publicly available stellar
evolution models to investigate the different possible structures that
can form in the stellar wind bubbles around Wolf-Rayet stars. We find
significant differences between models with and without stellar
rotation, and between models from different authors. More
specifically, we find that the main ingredients in the formation of
structures in the Wolf-Rayet wind bubbles are the duration of the Red
Supergiant (or Luminous Blue Variable) phase, the amount of mass lost,
and the wind velocity during this phase, in agreement with previous
authors. Thermal conduction is also included in our models. We find
that main-sequence bubbles with thermal conduction are slightly
smaller, due to extra cooling which reduces the pressure in the hot,
shocked bubble, but that thermal conduction does not appear to
significantly influence the formation of structures in
post-main-sequence bubbles.  Finally, we study the predicted X-ray
emission from the models and compare our results with observations of
the Wolf-Rayet bubbles S\,308, NGC\,6888, and RCW\,58. We find that
bubbles composed primarily of clumps have reduced X-ray luminosity and
very soft spectra, while bubbles with shells correspond more closely
to observations.

\end{abstract}

\section{Introduction}

Massive stars ($M \geqslant 30 M_{\odot}$) affect the interstellar
(ISM) and their circumstellar (CSM) media due to the combined result
of the action of their strong stellar winds and ionizing photons. At
early times, massive stars photoionize the surrounding interstellar
medium, forming an \ion{H}{2} region of $10^4$~K gas around
themselves, which then begins to expand. The stellar wind from the
main-sequence (MS) phase interacts with this photoionized region,
producing a hot, tenuous bubble of shocked stellar wind material and a
shell of swept-up photoionized gas \citep{1997pism.book.....D}. By the
end of the main-sequence phase, the massive star will be surrounded by
a low density, hot bubble of some $> 20$~pc radius for typical
densities in star-forming regions, bordered by a thick shell of
swept-up neutral material expanding at a few km~s$^{-1}$
\citep{2005A&A...436..155C,2007dmsf.book..183A}. Of course, if the
underlying ISM was not homogeneous, or if the star has a substantial
velocity relative to the ISM ($>10$~km~s$^{-1}$), the details will
vary but the general picture remains.

When the star evolves off the main sequence, the mass-loss rate
increases substantially as the star evolves first towards the red side
of the HR diagram and then back to the blue. During this short-lived
phase, which can lead to the star becoming a red or yellow supergiant
or a luminous blue variable depending on its initial mass, the star
can deposit half of its mass as a slow, dense, most-likely neutral
wind into its immediate environment. The most intensive phases of
stellar mass loss during this period will be episodic and
non-spherical ejections of material \citep{2010ASPC..425..247H}. In
the final Wolf-Rayet (WR) phase of evolution, this dense circumstellar
medium is photoionized by the hot WR star and swept up by the fast WR
wind. The circumstellar medium of Wolf-Rayet stars, therefore,
contains the history of the most recent, and intense, phases of
massive star mass loss.

The slow wind-fast wind interaction has been studied numerically by
many authors, who all show that the interaction leads to instabilities
(Rayleigh-Taylor and thin shell), which result in the corrugation and
eventual break up of the dense shell of swept-up circumstellar
material into filaments and clumps. Optical images of the nebulae
S\,308, NGC\,6888 and RCW\,58 around the WR stars WR\,6, WR\,136 and
WR\,40, respectively, show clear evidence of such a process
\citep{2000AJ....120.2670G}.

In particular, \citet{1996A&A...305..229G,1996A&A...316..133G} were
the first to study in detail the purely hydrodynamical evolution of WR
bubbles and the instabilities which lead to the formation of clumps
and filaments. They took into account the evolution of the stellar
wind properties from the main sequence through the WR stage for a $35
M_\odot$ and a $60 M_\odot$ star as input to their models. Because the
winds in the main-sequence and red supergiant stages are comparatively
steady, these stages were studied in one-dimensional spherical
symmetry, but the formation of instabilities in the slow wind-fast
wind interaction at the onset of the WR stage was studied in 2D
axisymmetric spherical coordinates.
\citet{2003ApJ...594..888F,2006ApJ...638..262F} studied the
interaction of massive stars ($35 M_\odot$ and $60 M_\odot$) with
their environment with particular regard to their effect on the energy
balance of the interstellar medium. These simulations were performed
in two-dimensional cylindrical symmetry from the beginning of the main
sequence stage and included the radiative transfer of
hydrogen-ionizing radiation. In this way,
\citet{2003ApJ...594..888F,2006ApJ...638..262F} were able to study the
evolution of the \ion{H}{2} region around the stellar wind bubble as
well as the wind-wind interactions. The stellar wind parameters in
this work were the same as those used by
\citet{1996A&A...305..229G,1996A&A...316..133G}.

More recent work has related features in supernova and gamma-ray burst
(GRB) afterglow absorption spectra to structures in the circumstellar
medium around the progenitor stars, where the progenitor of the GRB
event is a WR star and the structures are formed during the
interaction of the WR wind with the dense, slow wind of the previous
evolutionary stage
\citep{2005ApJ...631..435R,2005A&A...444..837V,2006MNRAS.367..186E,2007A&A...469..941V}.
\citet{2007ApJ...667..226D} and \citet{2010MNRAS.407..812D} explored
the interaction of the shock wave from the final supernova explosion
of the star with the evolved circumstellar medium and compared the
simulated X-ray emission with observations of very young supernova
remnants.

The evolving circumstellar media around B stars, which do not undergo
a WR stage, was modeled by
\citet{2007RMxAC..30...80C,2008A&A...488L..37C} where the anisotropy
of the stellar winds due to fast stellar rotation was taken into
account, producing distinctive hourglass-shaped structures.
\citet{2008A&A...478..769V} also investigated the effect of rapid
rotation on the evolution of a low metallicity $20 M_\odot$ star, the
resultant non-spherical density profiles in the CSM, and the
consequences for the GRB afterglow spectra.

In this paper, we are particularly interested in the circumstellar
media around Wolf-Rayet stars. Many of the $> 200$ known Galactic
Wolf-Rayet stars are surrounded by nebulae, visible in optical
images. Of these, spectroscopy identifies only 10 as wind-blown
bubbles \citep{1982ApJ...254..578C,1992A&A...259..629E}\footnote{The
  remainder are classified as photoionized regions
  \citet{1982ApJ...254..578C}.}, of which only 2 have been detected in
X-rays. We expect X-rays to arise from the hot, shocked WR wind during
the interaction with the ejected slow wind material. The nebula S\,308
around WR\,6 was detected in soft X-rays by \textit{ROSAT}
\citep{1999A&A...343..599W} and \textit{XMM Newton}
\citep{2003ApJ...599.1189C}, while NGC\,6888 around WR\,136 was
observed by \textit{ROSAT}, \textit{ASCA}, and \textit{SUZAKU}
\citep{1988Natur.332..518B,1994A&A...286..219W,1998LNP...506..425W,2002A&A...391..287W,2005ApJ...633..248W,
  2011ApJ...728..135Z}.  Both WR bubbles show limb-brightened X-ray
morphologies and spectra dominated by gas at $T \sim$ 10$^{6}$~K, with
the X-ray emission interior to the optical [\ion{O}{3}] emision. This
X-ray emission is very soft and will be strongly affected by
absorption due to neutral hydrogen along the line of sight. S\,308 and
NGC\,6888 are two of the closer WR bubbles (about 1.5~kpc distant) and
the greater distance to RCW\,58 (about 3~kpc) could explain why it has
not been detected in X-rays.

The observed X-ray emission from wind-blown bubbles has proved
difficult to model successfully. There are two main schools of
thought: on the one hand, if we take a simple hot, wind-blown bubble,
such as that described in \citet{1997pism.book.....D}, the temperature
of the hot gas is determined by the stellar wind velocity, $kT = 3 \mu
m_\mathrm{H} V_\mathrm{w}^2/16$, where $\mu \sim 0.5$ is the mean
particle mass for fully ionized gas. A typical Wolf-Rayet wind
velocity is $\sim 1500$~km~s$^{-1}$, which implies temperatures of $T
> 3 \times 10^7$~K in the hot gas. The densities in the hot bubble are
very low $n < 0.01$~cm$^{-3}$, and so this model would predict low
luminosity, hard X-rays. On the other hand, there is the oft-used
\citet{1977ApJ...218..377W} model, which assumes thermal conduction
between the hot bubble and the ``cold'' ($10^4$~K), swept-up
shell. Thermal evaporation of dense, cold material into the hot bubble
raises the density and lowers the temperature here. This model
predicts high luminosity, soft X-ray emission. Both of these models
are one-dimensional (i.e., radial) in concept and do not take into
account the break-up of the shell due to instabilities nor the
shock-clump interactions which result. The main argument against
thermal conduction is that even a weak magnetic field would restrict
the movement of the thermal electrons. Diffuse X-ray observations of
young star clusters
\citep{2003ApJ...590..306D,2003ApJ...593..874T,2008Sci...319..309G},
planetary nebulae
\citep{2003IAUS..209..415C,2006IAUS..234..153G,2007apn4.confE...5K},
as well as the wind-blown bubbles S\,308 and NGC\,6888 suggest that
neither model is entirely applicable, since the observations indicate
that the hot gas has a temperature $10^6 < T < 10^7$~K but the
luminosities are far lower than those predicted by the
\citet{1977ApJ...218..377W} model.

In this paper, we investigate the possible structures that can form
around Wolf-Rayet stars by means of radiation-hydrodynamical
simulations taking into account the full time evolution of the stellar
wind mass-loss rate, terminal velocity and stellar ionizing photon
rate, as studied by previous authors. In addition, we include a
time-dependent treatment of thermal conduction in our models, which
has not been done before. We compare results obtained from three sets
of stellar evolution models
\citep{2003A&A...404..975M,2006MNRAS.367..186E} for stars with masses
between $40 M_\odot$ and $60 M_\odot$, which include a set of models
that take into account stellar rotation. By calculating the chemical
abundances expected in the circumstellar nebulae in the
post-main-sequence stages of these stellar evolution models, we can
make comparisons with optical observations. We also calculate the
expected X-ray emission from our numerical simulations under the
assumption of collisional ionization equilibrium and compare our
results to the wind-blown bubbles S\,308, NGC\,6888 and RCW\,58. By
analyzing the structures formed in the circumstellar media of our
simulations, the chemical abundance history of the CSM and the time
evolution of the X-ray luminosites, we can connect observations to
possible mass-loss histories, and hence discriminate between different
stellar evolution models.

In \S~\ref{sec:num} we describe the numerical method including the
treatment of thermal conduction, and in \S~\ref{sec:evol} we show the
main characteristics of the stellar evolution models. Section
\S~\ref{sec:results} presents our one-dimensional and two-dimensional
results and in \S~\ref{sec:comparison} we describe our numerical
treatment for the simulated luminosity and synthetic spectra, and we
make comparisons with both optical and X-ray observations. Section
\S~\ref{sec:discuss} discusses our results and the effect that
differences in the stellar evolution models make to the structures and
observational properties of these objects. Finally,
\S~\ref{sec:conclude} summarizes and concludes our findings.


\section{Numerical Method}
\label{sec:num}

In this section we provide a general description of the numerical
scheme, highlighting new features such as thermal conduction. Full
details of the radiation-hydrodynamics scheme are published in the
cited references and we describe the main features in
Appendix~\ref{app:A}.

\subsection{General description of the numerical scheme}
\label{sec:general}
Our simulations are carried out in one and two dimensions. First, the
time-dependent, spherically symmetric system of radiation-hydrodynamic
equations is solved for the main-sequence and RSG phases, during
which the stellar winds are reasonably steady and instabilities are
relatively unimportant. Then, the 1D results are remapped to a
two-dimensional, cylindrically symmetric $(r,z)$ grid, to continue the
evolution of the circumstellar medium through the WR stage, where the
density gradients and wind-wind interactions lead to instabilities,
which severely disrupt the circumstellar medium.

The 1D spherically symmetric Eulerian hydrodynamic conservation
equations are solved using a second order, finite volume Godunov-type
scheme with outflow-only boundary conditions at the outer boundary. In
2D, the cylindrically symmetric hydrodynamic conservation equations
are solved using a hybrid scheme, in which alternate time steps are
calculated by a Godunov method and the van Leer flux-vector splitting
method \citep{1982LNP...170..507V}. This reduces numerical artefacts,
which can be produced in Godunov-type schemes when slow shocks are
propagating parallel to one of the grid directions. The boundary
conditions in the 2D scheme are those of no flux through the axis of
symmetry and outflow-only conditions at the outer radial and $z$
boundaries.

The 1D simulations start on a grid of 1000 cells with a physical size
of 5 pc, which is large enough to contain the initial Str\"omgren
sphere. We use an expanding grid to avoid large computational domains
at the beginning of the simulation: when the outer shock nears the
edge of the grid, more (uniform ambient medium) cells are added. Each
1D calculation has the same physical resolution (grid cell size $dr =
0.002$~pc). By the end of the RSG stage, the stellar wind bubble,
\ion{H}{2} region and swept-up neutral shell extend up to 30~pc,
meaning that the corresponding grids have of order 6000 cells. The full
evolution of the CSM around the massive stars is followed in 1D from
the beginning of the main-sequence stage through to the WR
stage. However, once the star enters the WR stage, 1D is not adequate
to capture the full richness of the structures formed during the
wind-wind interactions and so we remap the results onto a 2D grid to
continue the calculation in cilindrical $r-z$ coordinates.

The majority of the 2D simulations are carried out on a fixed grid of
500 radial by 1000 $z$-direction cells with a corresponding physical
size of $10\times20$~pc$^{2}$ (i.e, cell size $dr\times dz=
0.02$~pc$\times 0.02$~pc). This size is chosen because the wind-wind
interaction takes place within this distance for almost all models,
and the radius of observed WR bubbles is $R \leqslant 10$~pc
\citep{2000AJ....120.2670G}. One model is run on a slightly larger
grid ($750\times 1500$ cells) because the main wind-wind interaction
occurs further from the star, but the spatial resolution is the
same.

The transfer of the ionizing radiation is carried out by the method of
short characteristics \citep{1999RMxAA..35..123R} adapted for
spherical and cylindrical symmetry. We consider only a single point
source for the ionizing radiation (the star), which is positioned on
the symmetry axis. The hydrodynamics and radiative transfer are
coupled through the energy equation, where the source term depends on
the photoionization heating and the radiative cooling rate, and
through advection equations for the ions and neutrals. The ionization
balance equation takes into account both photoionization and
collisional ionization together with recombination. This basic code
has been extensively tested and applied to the evolution of \ion{H}{2}
regions in density gradients
\citep{{2005ApJ...627..813H},{2006ApJS..165..283A}}.

The radiative cooling function takes into account both cooling in
collisionally ionized gas (in the hot shocked wind bubble) and in the
photoionized gas in the \ion{H}{2} region where collisional ionization
of hydrogen is not an important contribution to the cooling
rate. Cooling rates are generated using the Cloudy photoionization
code \citep{1998PASP..110..761F} for appropriate stellar parameters
and used to generate a look-up table. The cooling rates extend down to
temperatures of 100~K but photoionization heating ensures that the
\ion{H}{2} region has a temperature of $\sim 10^4$~K.

The stellar wind is input to the grid through the
\citet{1985Natur.317...44C} thermal wind model, as a
volume-distributed source of mass and energy. This avoids the problem
of reverse shocks that can occur when there is a decrease in the ram
pressure of a stellar wind input in the form of a region of
high-density, high-velocity flow near the origin. The volume of the
stellar wind injection region is chosen to be sufficiently small so
that the stellar wind always reaches its terminal velocity and shocks
outside the source region, but sufficiently large so as to be a
reasonable geometrical representation of a sphere on a cylindrical
grid (in the case of the 2D numerical simulations).

The variation of the stellar mass-loss rate with time is taken from
publicly available stellar evolution models
\citep{{2003A&A...404..975M},{2006MNRAS.367..186E}}. The stellar wind
velocities are calculated using the tabulated stellar parameters
(mass, radius, effective temperature and surface abundances) and the
ionizing photon rates are obtained from the appropriate stellar
atmosphere models distributed with the Starburst 99 code
\citep{1999ApJS..123....3L}. All of this information is stored in
tables and interpolated as needed using the elapsed simulation time.

\subsection{Thermal Conduction}

We also run a set of models that includes the effects of electron
thermal conduction to assess the importance of this physical process
for the dynamics of stellar wind bubbles and their observable
properties such as X-ray emission \citep{1977ApJ...218..377W}. Thermal
conduction is taken to be a diffusion process, with the heat flux,
$\vec{q}$, given by
\begin{equation}
\vec{q} = - D \nabla T_{e},
\label{eq:diffq}
\end{equation}
where $D$ is the diffusion coefficient and $T_e$ is the electron
temperature \citep{1977ApJ...211..135C}. From
\citet{1962pfig.book.....S}, we have that the electron mean free path
can be written as
\begin{equation}
\lambda_{e} = 2.625 \times 10^{5} T_{e}^{2} / n_{e}/ \ln \Lambda , 
\label{eq:lam}
\end{equation}
where $n_{e}$ is the electron number density and $\ln\Lambda$ is the Coulomb
Logarithm. For a pure hydrogen plasma this can be approximated by
\begin{equation}
\ln \Lambda = 9.452 + 3/2 \ln T_{e} - \frac{1}{2} \ln n_{e} 
 \label{eq:CLam1}
\end{equation}
when $T_{e} \leq 4.2 \times 10^{5}$~K and, 
\begin{equation}
\ln \Lambda = 22.37 + \ln T_{e} - \frac{1}{2} \ln n_{e} 
\label{eq:CLam2}
\end{equation}
when $T_{e} \geq 4.2 \times 10^{5}$~K.

From \citet{2008A&A...489..173S}, the diffusion coefficient $D$ is then given by
\begin{equation}
D = 7.04 \times 10^{-11} \lambda_{e} n_{e} T_{e}^{1/2}, 
\label{eq:diffcoeff}
\end{equation}
where the numerical coefficient is the result of evaluating physical
constants \citep{1962pfig.book.....S}. In spherical symmetry, the
non-linear diffusion equation is
\begin{equation}
\frac{\partial e}{\partial t} = \rho c_{v} \frac{\partial T_e}{\partial t} = \frac{1}{r^{2}} \frac{\partial}{\partial r} \left( r^{2} D \frac{\partial T_e}{\partial r} \right),
\label{eq:sphdiffeq}
\end{equation}
and we solve this using a Crank-Nicholson-type scheme
\citep{1992nrfa.book.....P}.  Thermal conduction is also included in
the 2D simulations by solving the corresponding diffusion equation in
cylindrical symmetry.  The effect of conduction is included as an
update to the internal energy, $e$, once the diffusion equation has
been solved for the electron temperature, and we assume equipartition
between the electrons and ions.

When the electron temperature $T_{e}$ is high and the electron
density $n_{e}$ low, the electron mean free path $\lambda_{e}$ becomes
large compared to the size scale of the hot bubble, and the
diffusion approximation breaks down. This is the saturated heat flux
limit \citep{1977ApJ...211..135C} that can be described as
\begin{equation}
q_{\mathrm{sat}} = 1.72 \times 10^{-11} T_{e}^{3/2} n_{e}. 
\label{eq:qsat}
\end{equation}
Saturated conduction is treated by limiting the electron mean free
path in the following manner 
\begin{equation}
\lambda = \min \{ 0.244 \Delta r, 2.625 \times 10^{5} T_{e}^{2} / n_{e} \ln\Lambda \},
\label{eq:emfp}
\end{equation} 
where $\Delta r$ is the local grid spacing and the numerical constant
ensures that the conductive heat flux can never exceed the saturation
limit \citep{2008A&A...489..173S}. This formulation means that the
limiting flux can only be reached at the sharp edge of the hot bubble,
where $|\Delta T_e| \approx T_e$. A drawback of this method of
limiting the electron mean free path is that it depends explicitly on
the numerical resolution $\Delta r$ and so our results will be
affected by this. However, since all our 1D results are run at the
same numerical resolution, comparisons between simulations are
perfectly valid.

The temperature found from solving the diffusion equation for every
grid cell is used to update the internal energy of that cell once the
hydrodynamic and radiation steps have been completed.


\section{Stellar Evolution Models}
\label{sec:evol}
In this paper we use the publicly available stellar evolution models
for initial stellar masses of 40 and $60\, M_{\odot}$ from
\citet{2003A&A...404..975M}, hereafter MM2003, and the 40, 45, 50, 55,
and $60\, M_{\odot}$ initial mass models from STARS
\citep{{1971MNRAS.151..351E},{1995MNRAS.274..964P},{2004MNRAS.353...87E}}. Both
evolutionary models use the same nuclear reaction rates based on the
NACRE database \citep{1999NuPhA.656....3A}, and OPAL opacities
\citep{1996ApJ...464..943I}. At low temperatures, the opacities come
from \citet{1994ApJ...437..879A} and
\citet{2005ApJ...623..585F}. Differences between the codes such as
assumptions, approximations, and solution algorithms, lead to
differences in the post-main-sequence evolution of the massive stars.

We consider only the Solar metallicity models, which for MM2003 means
$X=0.705$ and $Y = 0.275$, while for STARS this is $X = 0.7$ and $Y =
0.28$ for $Z = 0.02$. In addition, MM2003 provide models including
stellar rotation, and so we investigate the differences between models
with and without rotation. The principal effects of an initial fast
stellar rotation are to increase the main-sequence lifetime of the
star by about $15$--$25\%$, and also to extend the lifetimes of the
stars in the Wolf-Rayet stage
\citep{2003A&A...404..975M}. Furthermore, the nature of the intense
period of post-main-sequence mass loss can change, for example, the
$60 M_{\odot}$ MM2003 model without rotation undergoes a short-lived
period of intense mass loss that can be interpreted as a Luminous Blue
Variable (LBV) phase, while the corresponding model with rotation
never becomes an LBV. Models from STARS do not include rotation.

We use the MM2003 models with an initial equatorial rotation velocity
of 300~km~s$^{-1}$. This rotation velocity is large enough such that
rotation effects on the stellar evolution are important, but not so
large that anisotropies in the stellar wind are important in the later
stages of stellar evolution. MM2003 find that, for an initial rotation
velocity of 300~km~s$^{-1}$, during the WR stage the ratio of the
stellar surface velocity to the break-up velocity is very small for
the 40 and 60$M_{\odot}$ stellar models (see their fig. 2) and so we
do not expect the stellar winds to become strongly anisotropic during
this stage \citep{1996ApJ...459..671I}. We do not take into account
the episodic and non-spherical nature of the wind during the most
intense phases of mass loss \citep{2010ASPC..425..247H}.

\begin{figure}[p]
\includegraphics[width=\linewidth]{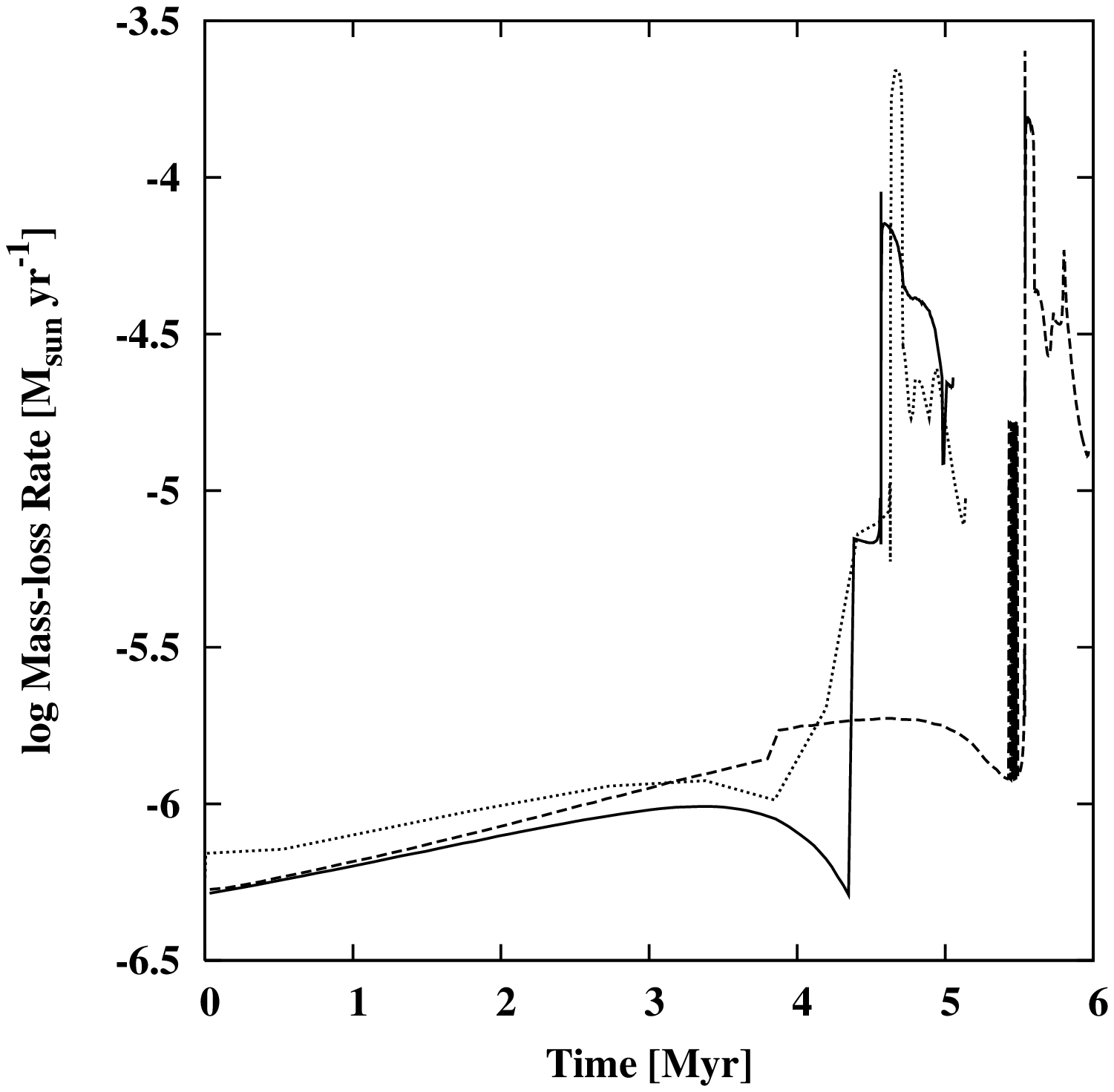}\\
\includegraphics[width=\linewidth]{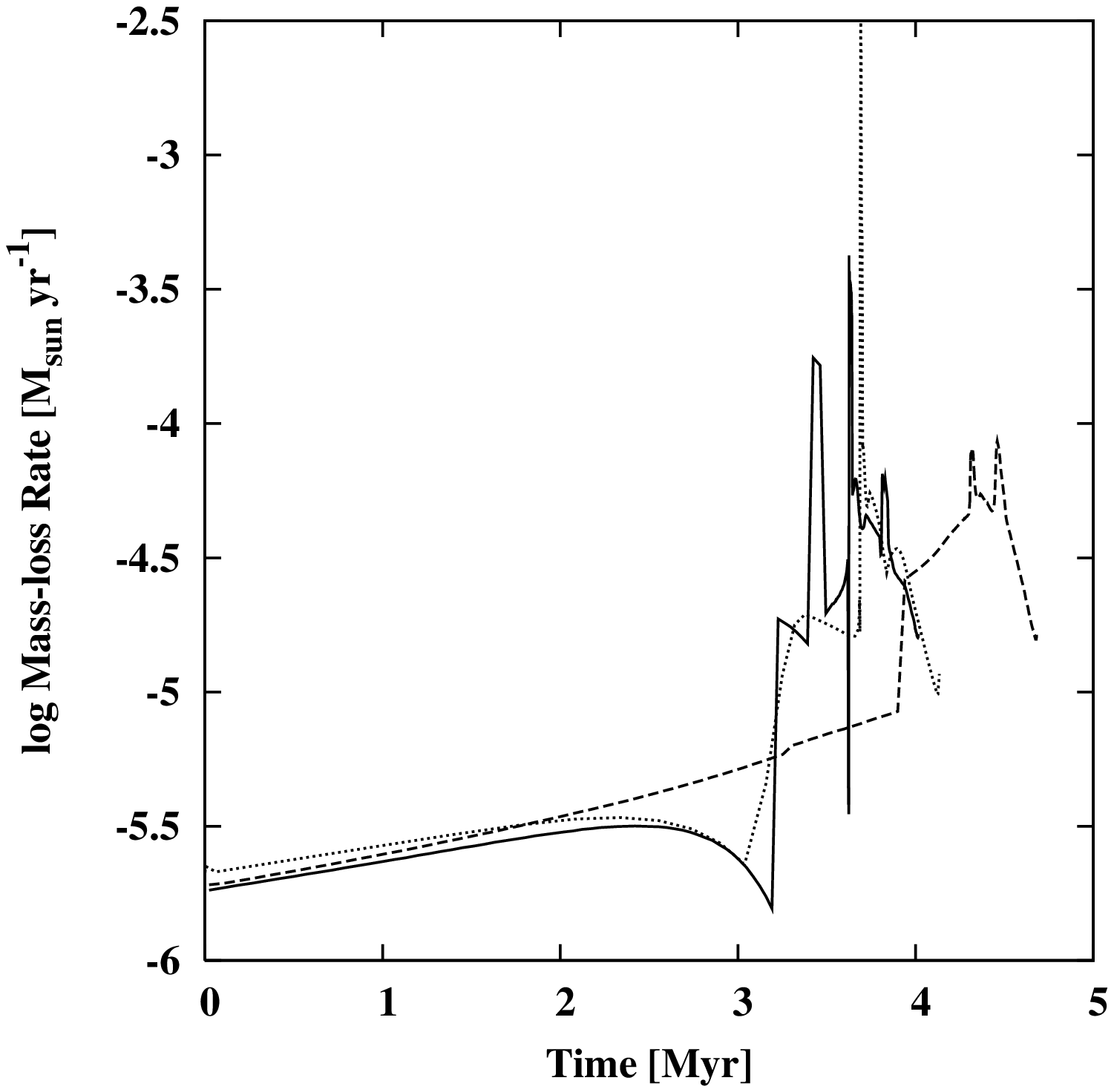}
\caption{Mass-loss rate for \textit{top panel} $40\, M_\odot$ and
  \textit{bottom panel} $60 M_{\odot}$ models. In each panel,
  \textit{solid line} --- MM2003 no rotation, \textit{dashed line} --
  MM2003 with rotation, \textit{dotted line} --- STARS.}
\label{fig:massloss}
\end{figure} 

Mass loss is very important in the evolution of massive stars. For the
mass-loss rates in pre-WR evolution, MM2003 use the theoretical
mass-loss rates of \citet{2001A&A...369..574V}, while STARS uses the
empirical mass-loss rates from \citet{1988A&AS...72..259J}. In the WR
phase, both evolution codes use the mass-loss rates proposed by
\citet{2000A&A...360..227N}. In Figure~\ref{fig:massloss} we compare
the mass-loss rates of the MM2003 and STARS models. Although the
details of the main-sequence mass loss is very similar in each case,
there are appreciable differences in the lifetimes of the different
post-main-sequence stages and the peak mass-loss rate. The MM2003
model with rotation has a far longer main-sequence lifetime.

\begin{figure}[p]
\includegraphics[width=\linewidth]{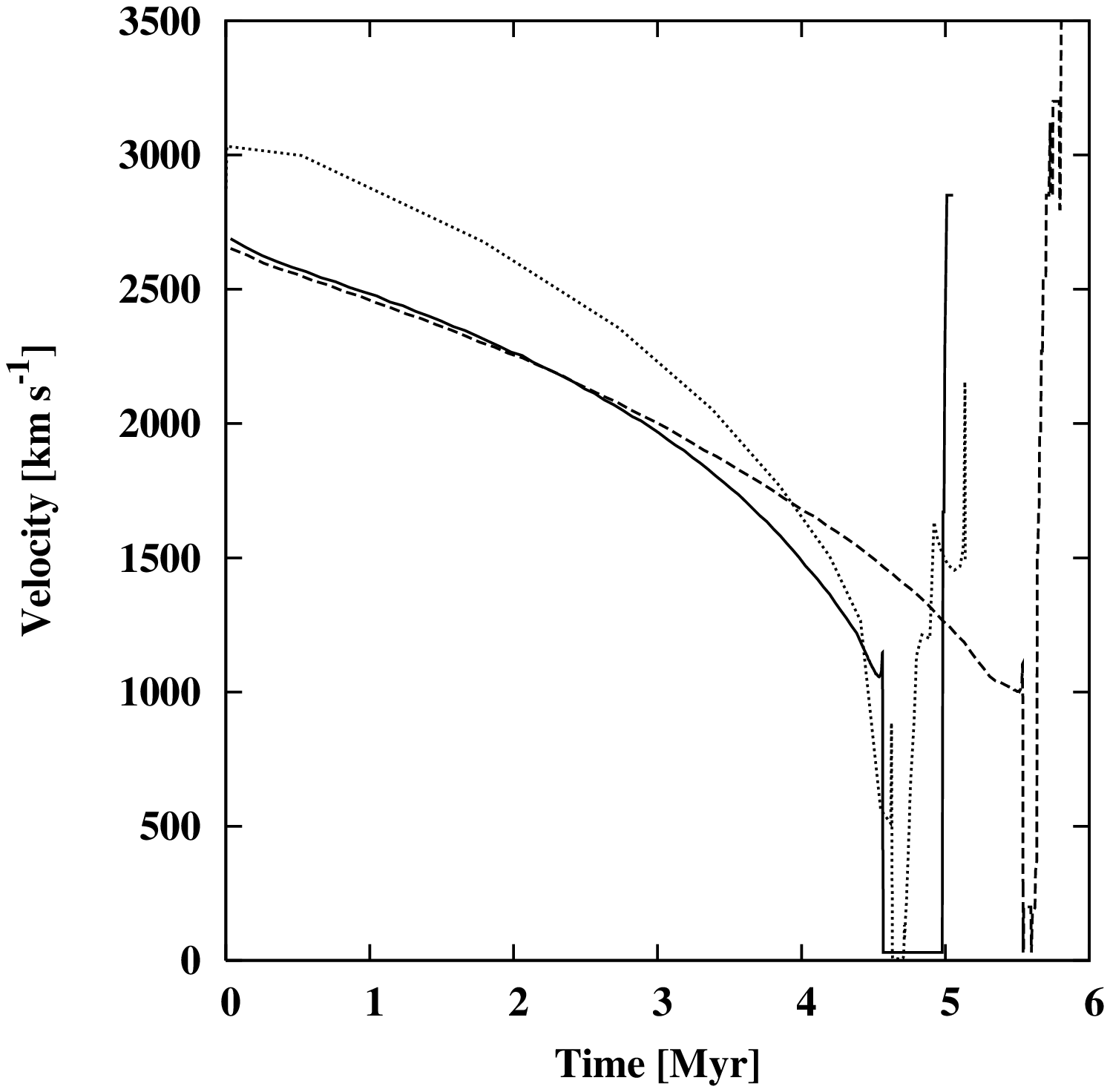}\\
\includegraphics[width=\linewidth]{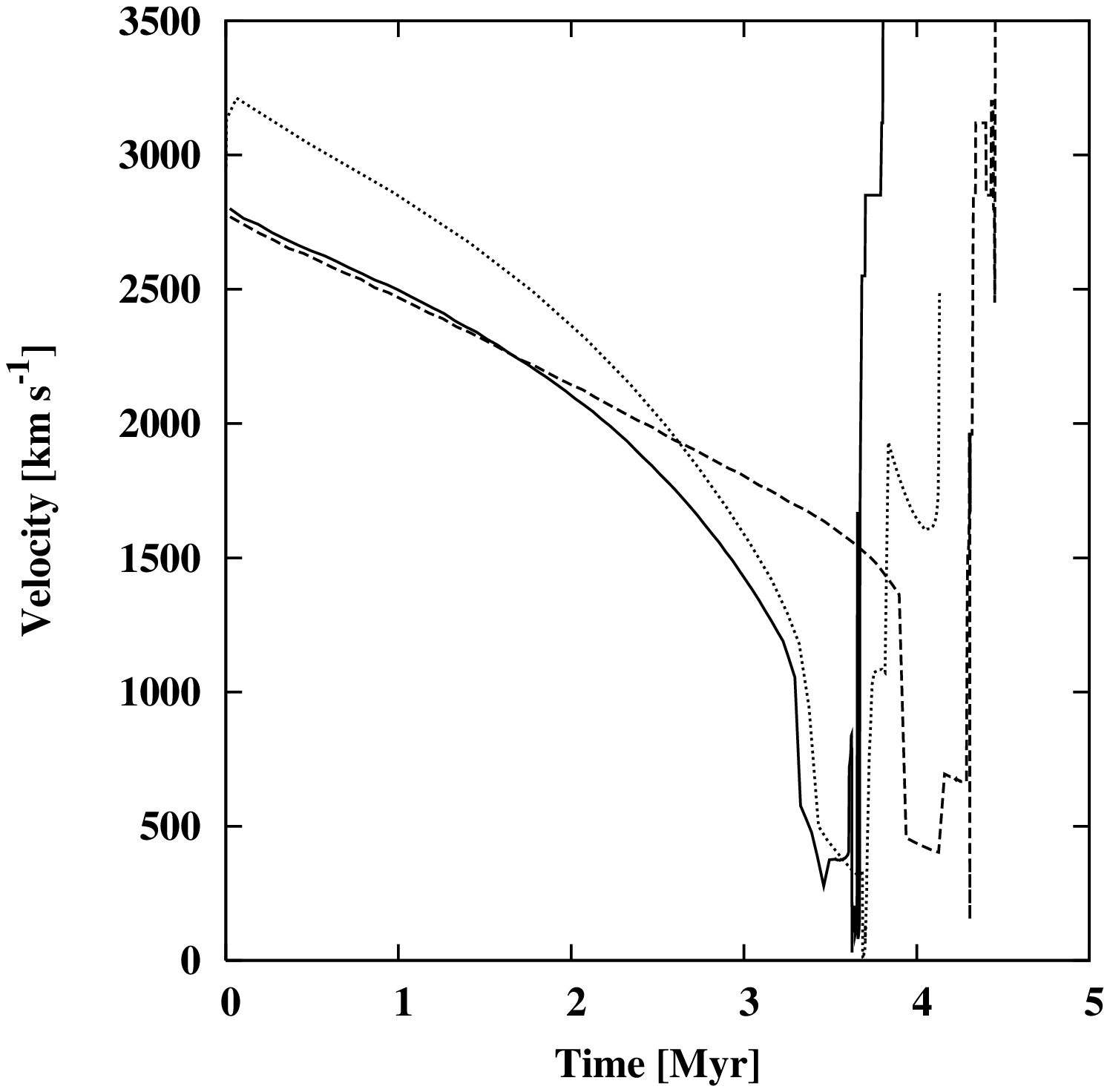}
\caption{Stellar wind velocity for \textit{top panel} $40\, M_\odot$ and
  \textit{bottom panel} $60 M_{\odot}$ models. In each panel the models are,
  \textit{solid line} --- MM2003 no rotation, \textit{dashed line} --
  MM2003 with rotation, \textit{dotted line} --- STARS.}
\label{fig:windvel}
\end{figure}

Our radiation-hydrodynamic simulations take the mass-loss rates as a
function of time directly from the stellar evolution models. We also
need the stellar wind velocities and ionizing photon rates for our
simulations, and these can be calculated using the values of the
stellar radius, effective temperature, and surface abundances from the
stellar evolution models. The stellar wind velocities for the MM2003
models are calculated as in \citet{1989A&A...219..205K} for the main
sequence phase, then we use 30~km~s$^{-1}$ for the red/yellow
supergiant wind ($T_{\mathrm{eff}} < 8\,000$~K) or 200~km~s$^{-1}$ for
the LBV wind ($8\,000 < T < 25\,000$~K and $\log\dot{M} > -3.9$), as
adopted in the Starburst 99 code \citep[][and
  updates]{1999ApJS..123....3L}. The wind velocities in the WR phase
(defined by surface hydrogen mass fraction $X < 0.4$ and
$T_{\mathrm{eff}} > 25\,000$~K) are taken from the tables of
\citet{2002MNRAS.337.1309S}. For the STARS models, the wind velocities
are listed as part of the stellar evolution tables, using the formulae
in \citet{2001A&A...369..574V} for the pre-Wolf-Rayet phase and those
of \citet{2000A&A...360..227N} for the Wolf-Rayet phase. We corrected
the tabulated values by the appropriate constant factors after noting
errors in the wind velocities. The resulting wind velocities are lower
in the RSG phase, even dropping as low as 5~km~s$^{-1}$, and are also
lower in the WR phase compared to the values we adopt for the MM2003
models. The stellar wind velocities are shown in
Figure~\ref{fig:windvel}. It is apparent that the WR stellar wind
velocities obtained from the \citet{2002MNRAS.337.1309S} tables are
extremely high, particularly in the final stages of stellar
evolution. These velocities were taken from the $v_\infty$-spectral
type calibrations of \citet{1990ApJ...361..607P}, where spectral type
is estimated from the stellar effective temperature and surface
chemical composition. The WR wind velocities obtained by using the
fits of \citet{2000A&A...360..227N} depend on the escape velocity of
the hydrostatic core, the luminosity and the surface abundances, and
are generally lower.

\begin{figure}[p]
\includegraphics[width=\linewidth]{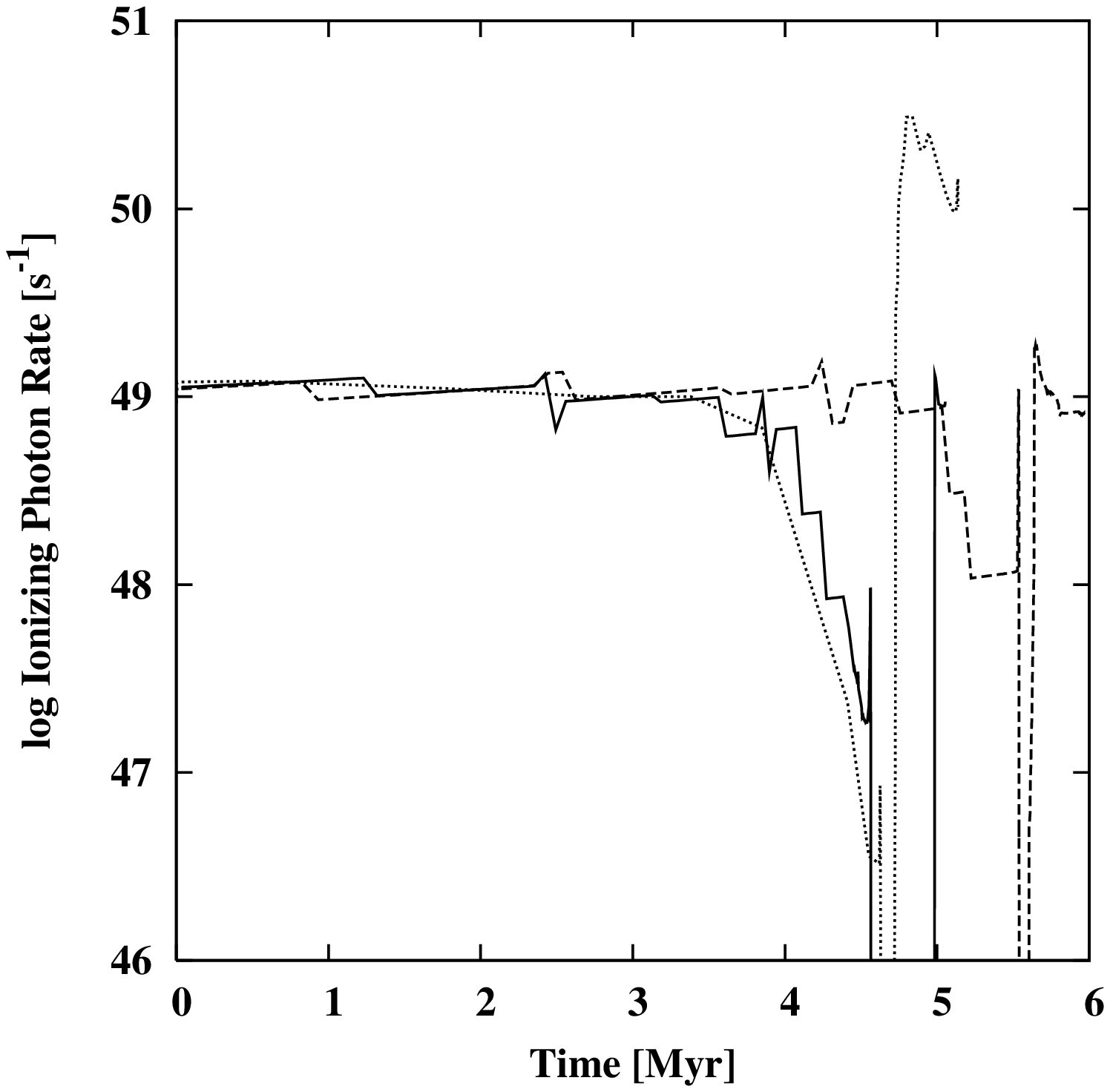}\\
\includegraphics[width=\linewidth]{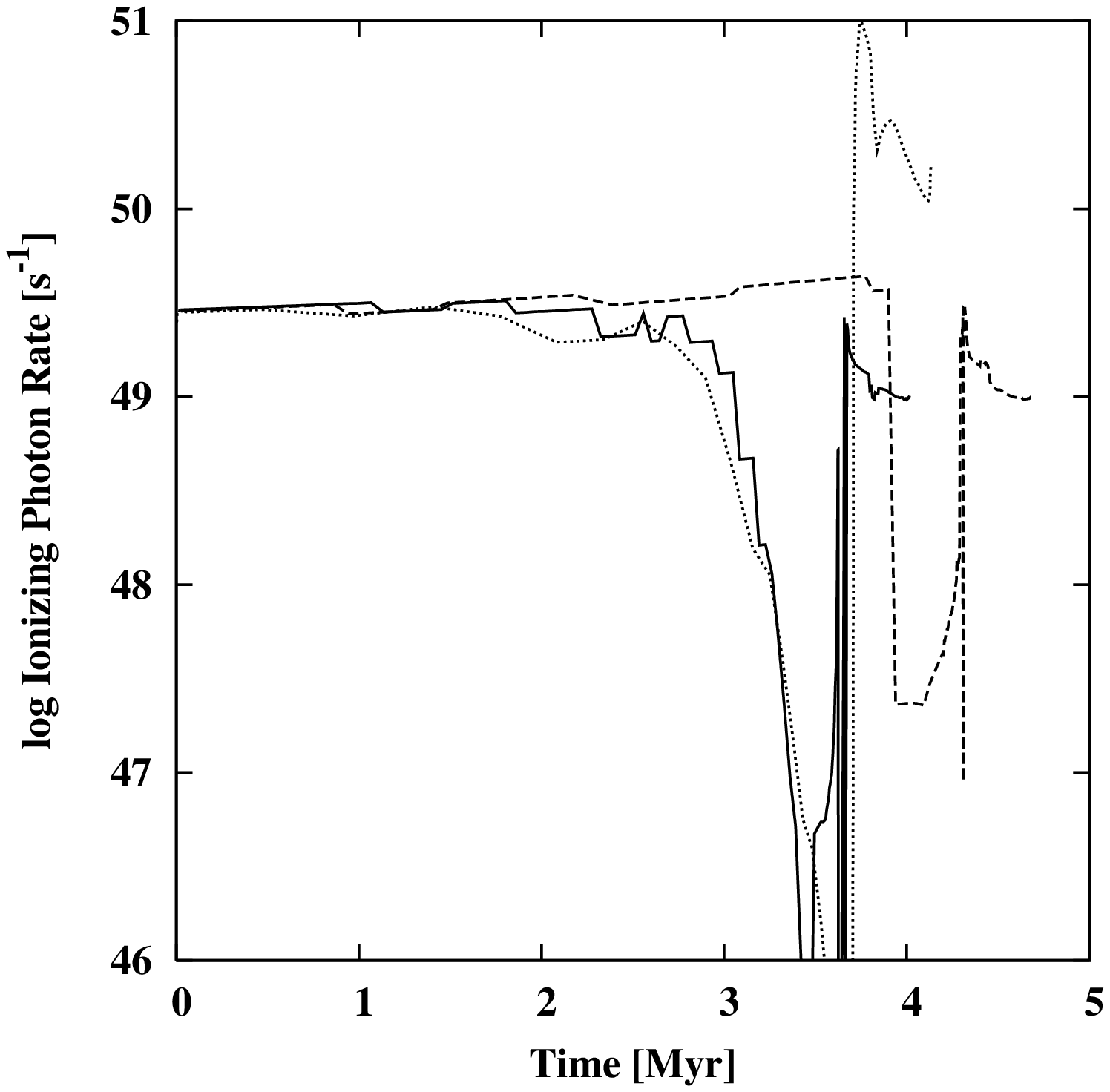}
\caption{Ionizing photon rates for \textit{top panel} $40\, M_\odot$ and
  \textit{bottom panel} $60 M_{\odot}$ models. In each panel the models are,
  \textit{solid line} --- MM2003 no rotation, \textit{dashed line} --
  MM2003 with rotation, \textit{dotted line} --- STARS.}
\label{fig:photonrate}
\end{figure}

The ionizing photon rate was obtained by using appropriate stellar
atmosphere models and integrating over the ionizing flux. For both
MM2003 and STARS models, we used the stellar atmosphere tables from
Starburst 99 \citep{1999ApJS..123....3L}, concretely the atmosphere
models of \citet{2001A&A...375..161P} for the main-sequence stage, the
stellar library of \citet{1997A&AS..125..229L} for the post-main
sequence stage, and for the WR stage we use the models of
\citet{1998ApJ...496..407H}, as described by
\citet{2002MNRAS.337.1309S}. The resultant ionizing photon rates as a
function of time are shown in Figure \ref{fig:photonrate}.

\begin{figure}[p]
\includegraphics[width=\linewidth]{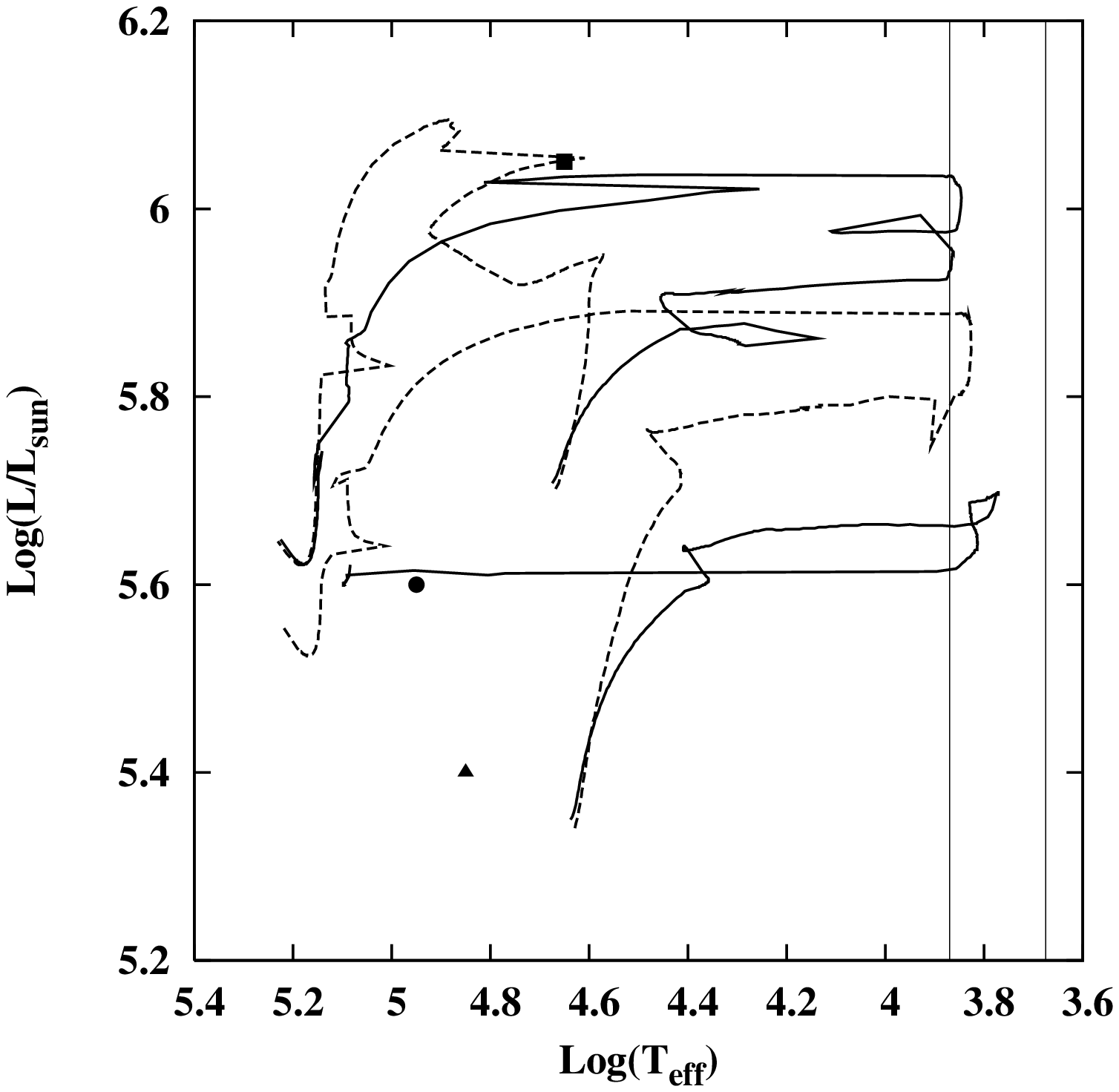}\\
\includegraphics[width=\linewidth]{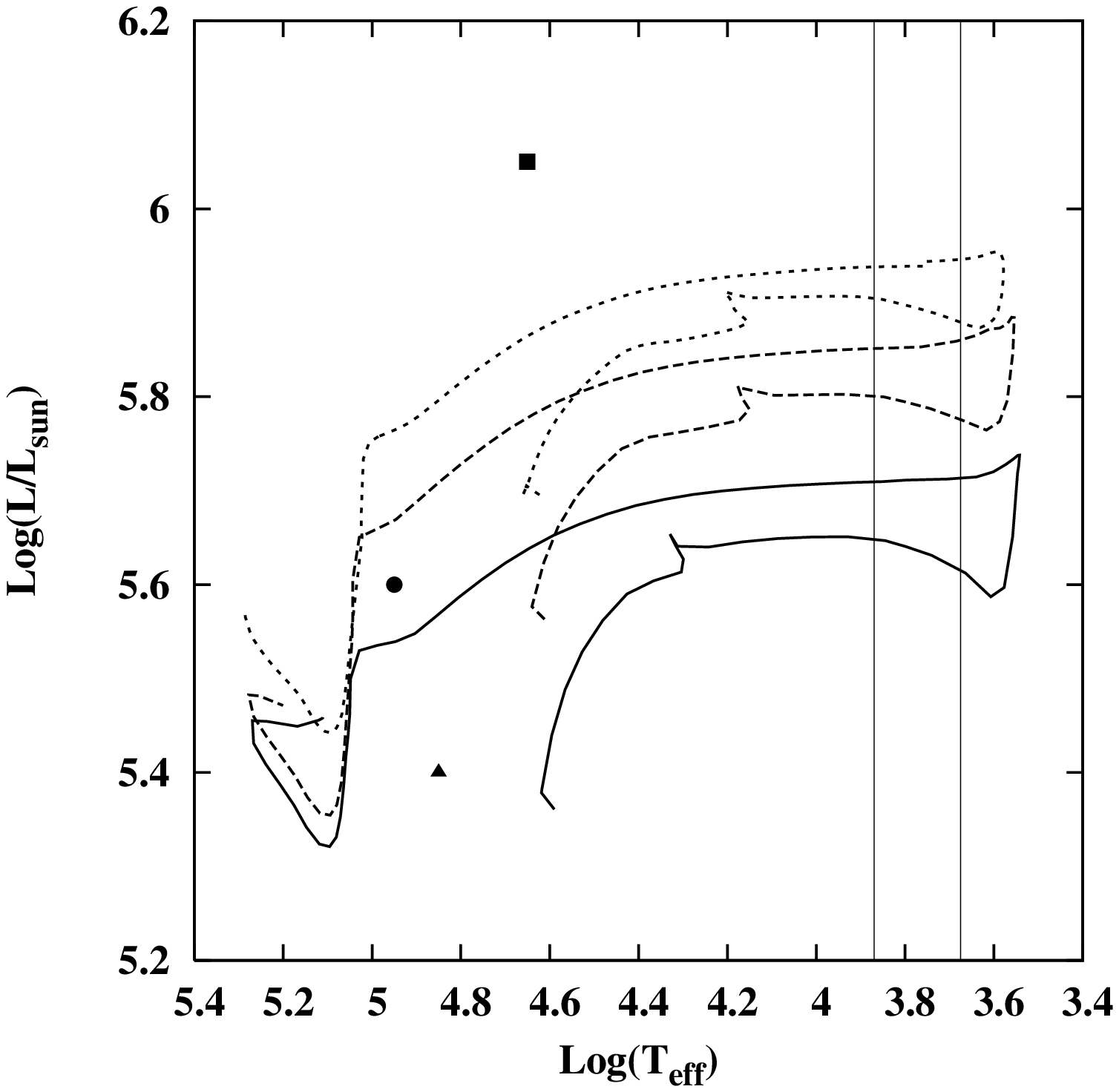}
\caption{Evolutionary tracks in the $T_\mathrm{eff}$--luminosity
  plane. \textit{Top panel}: MM2003 models. Solid lines indicate no
  stellar rotation, dashed lines indicate models with rotation, for
  $60 M_\odot$ (upper) and $40\, M_\odot$ (lower) initial masses.
  \textit{Bottom panel:} STARS models. Solid line -- $40\, M_\odot$,
  dashed line -- $50 M_\odot$, dotted line -- $60 M_\odot$. In each
  panel, the circle, square and triangle represent the values for
  WR\,6, WR\,40 and WR\,136 from \protect\citet{2006A&A...457.1015H},
  respectively. The region delimited by the thin vertical lines
  indicates the effective temperature range for which the stars would
  be considered yellow supergiants (4800 -- 7500~K).}
\label{fig:track}
\end{figure}
In Figure~\ref{fig:track} we plot the evolutionary tracks for the
$40\, M_\odot$ and $60\, M_\odot$ MM2003 and STARS models together
with data points that represent the properties of the WR stars WR~6,
WR~40, and WR~136 from \citet{2006A&A...457.1015H}. These are the
central stars of the three WR wind-blown bubbles S\,308, RCW\,58, and
NGC\,6888, respectively. The initial masses of these WR stars have
been controversial. We can see that the WN4 star WR\,6 falls exactly
on the evolutionary track of the $40\,M_{\odot}$ no-rotation model
from MM2003 but slightly below the corresponding model with rotation,
while it lies between the $40$ and $50 M_{\odot}$ STARS models. The
WN8 star WR\,40 seems to fit the $60\,M_{\odot}$ models from MM2003,
but apparently would require a much higher mass model from
STARS. Finally, the WN6 star WR\,136 would indicate an initial mass
lower than $40\,M_{\odot}$ in both sets of models. However, there are
large uncertainties in the luminosities of these stars due to
variability and distance determinations.

\section{Results}
\label{sec:results}
We present results of our one and two-dimensional simulations. The 1D
spherically symmetric simulations are used to follow the formation of
the main-sequence stellar wind bubble and \ion{H}{2} region and to
have a general idea of the evolution of the circumstellar medium once
the star has left the main sequence. Two-dimensional, cylindrically
symmetric ($r,z$) simulations are used to study the interaction
between the fast WR wind and the slow, dense RSG wind, where
instabilities are expected to break the swept-up shell into
clumps. The starting point for the 2D simulations are the results of
the spherically symmetric simulations at the end of the RSG phase,
which we remap onto the cylindrical grid.
 
\subsection{1D Results}
\label{subsec:1dresults}
\subsubsection{Main-sequence stage}
\label{subsubsec:MS_stage}
We start from an initial condition of a cold, uniform, neutral medium
with density $n_0 = 100$~cm$^{-3}$ and temperature $T = 100$~K. The
photoionized (\ion{H}{2}) region forms first. The photoionized gas is
heated to $T \sim 10^4$~K and begins to expand into the surrounding
medium at $v_\mathrm{if} \gtrsim 10$~km~s$^{-1}$. The conditions at
the ionization front require a shock to form in the neutral medium
ahead of it, which sweeps up and compresses the neutral gas. Inside
the \ion{H}{2} region, the highly supersonic stellar wind forms a
two-shock structure: an outer shock sweeps up, heats and compresses
the photoionized gas, and an inner shock brakes and heats the stellar
wind. The two regions have the same pressure but different densities
and are separated by a contact discontinuity. The outer shocked region
can cool down quickly to the temperature of the \ion{H}{2} region due
to the high densities in the swept-up material, but its temperature is
maintained at $10^{4}$~K by photoionization. The shocked stellar wind,
on the other hand, has a very high temperature ($T > 10^7$~K) and low
density and so does not cool efficiently. A hot bubble forms inside
the \ion{H}{2} region, which is in turn surrounded by a slowly
expanding shell of swept-up neutral material. These structures are
well known from textbooks \citep[e.g.,][]{1997pism.book.....D} and the
slow time evolution of the stellar parameters does not cause any
noticeable modifications\footnote{Two-dimensional simulations of the
  early stages of stellar wind bubbles inside \ion{H}{2} regions show
  more complicated structures. Rapid cooling in the material swept up
  by the outer wind shock forms clumps, which then trigger the
  shadowing instability, leading to complex low-velocity
  ($<20$~km~s$^{-1}$) kinematics in the photoionized gas
  \citep{2003ApJ...594..888F,2006ApJ...638..262F,2006ApJS..165..283A}.}.
The evolution of a stellar wind bubble inside an \ion{H}{2} region is
different to that when ionization is not taken into account
\citep[e.g.,][]{2004RMxAC..22..136V}. Once the initial Str\"omgren
region has been formed, the pressure in the \ion{H}{2} region is much
higher than that in the surrounding ISM. Moreover, although to begin
with the pressure in the hot, shocked, stellar wind bubble is higher
than that in the photoionized gas, pressure equilibrium with the
\ion{H}{2} is reached after a short time and, thereafter, the pressure
across the bubble structure is regulated by both the \ion{H}{2} region
and the stellar wind together.

\begin{figure*}
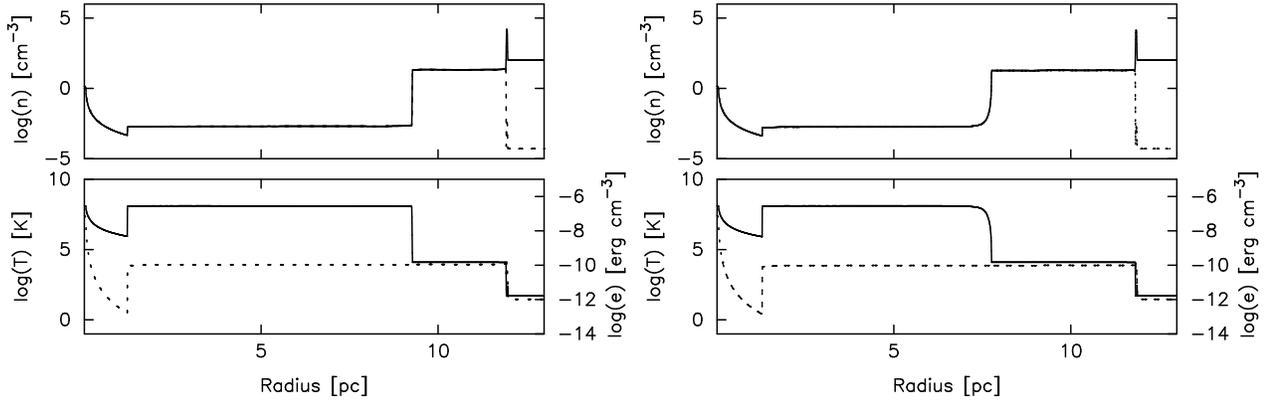

\includegraphics[width=0.5\linewidth]{fig5a.ps}~
  \includegraphics[width=0.5\linewidth]{fig5b.ps}\\
\caption{Density and temperature at $10^{6}$~yrs of evolution
  for the 40$M_{\odot}$ model from STARS. \textit{Left-hand panels}:
  model without thermal conduction, \textit{right-hand panels}: model
  with thermal conduction. The density plots show total density (solid
  line) and ionized density (dashed line). The temperature plots also
  show the internal energy, $e=p/[\gamma - 1]$ (dashed line).}
\label{fig:1DMS}
\end{figure*}

To illustrate the main-sequence evolution, in Figure~\ref{fig:1DMS} we
show the results from a 40$M_{\odot}$ STARS model for simulations with
and without thermal conduction after $t=10^{6}$~yrs. The hot, shocked
wind bubble, \ion{H}{2} region, and neutral shell are cleary
identified.  The bubble radius in the simulation with thermal
conduction is slightly smaller than that without conduction due to
increased radiative losses in the hot bubble. These enhanced radiative
losses in the hot, stellar wind bubble reduce the pressure here and
enable the \ion{H}{2} region to expand back towards the star, and the
phoionized gas has slightly lower density in the thermal conduction
simulations that in those without conduction as a result.

\begin{deluxetable}{lcccc}
\tablewidth{0pt} 
\tablecaption{Radii and masses of the outer \ion{H}{1} shell at the end of MS phase}
    \tablehead{ \multicolumn{1}{l}{Mass} & \multicolumn{1}{c}{Model} &
      \multicolumn{1}{c}{$R_{\mathrm{HI}}$} &
      \multicolumn{1}{c}{$M_{\mathrm{ism}}$} &
      \\ \multicolumn{1}{l}{$M_{\odot}$} & \multicolumn{1}{c}{} &
      \multicolumn{1}{c}{pc} & \multicolumn{1}{c}{$M_{\odot}$} }
    \startdata 40 & MM2003 N & 24.6 & $2.0 \times 10^5$ \\ 40 & MM2003
    R & 27.5 & $2.8 \times 10^5$ \\ 40 & STARS & 29.0 & $3.3 \times
    10^5$ \\ 60 & MM2003 N & 25.6 & $2.2 \times 10^5$ \\ 60 & MM2003 R
    & 29.4 & $3.4 \times 10^5$ \\ 60 & STARS & 29.4 & $3.4 \times
    10^5$ \\ \enddata
\label{tab:radii}
\end{deluxetable}

For our particular models, we find that by the end of MS stage, the
radius of the region affected by the massive star is 25--30~pc (see
Table~\ref{tab:radii}) and the mass of swept-up neutral material is
$>10^{5}M_{\odot}$. The exact size of the bubble depends mainly on the
length of the main-sequence stage (e.g., models with rotation have
longer main sequences than those without) and the stellar wind
velocity. Higher stellar wind velocities lead to higher pressures in
the shocked wind bubbles and therefore faster expansion speeds (e.g.,
the STARS models have higher main-sequence stellar wind velocities,
see Fig.~\ref{fig:windvel}). We remark that bubbles expanding into
denser media will, of course, be smaller, since on dimensional grounds
the external radius of a stellar wind bubble in an uniform medium of
density $n_{o}$ is $R \sim (\dot{M_{W}}V_{w}^{2} / n_{0})^{1/5}
t^{3/5}$ where $\dot{M_{W}}$ is the mass-loss rate and $V_{W}$ the
stellar wind velocity \citep[see e.g.][]{1997pism.book.....D}, while
that of an expanding \ion{H}{2} region follows $R_{\mathrm{i}} \sim
(S_{*}/\alpha_{\mathrm{B}}n_{0}^{2})^{3/7} c_{\mathrm{II}}^{4/7}
t^{4/7}$, where $\alpha_{\mathrm{B}}$ and $c_{\mathrm{II}}$ are the
recombination coefficient and the sound speed in the ionized gas,
respectively, and $S_{*}$ is the stellar ionizing photon rate. Both of
these relations depend inversely on the density
\citep{1978ppim.book.....S}.
\begin{figure*}
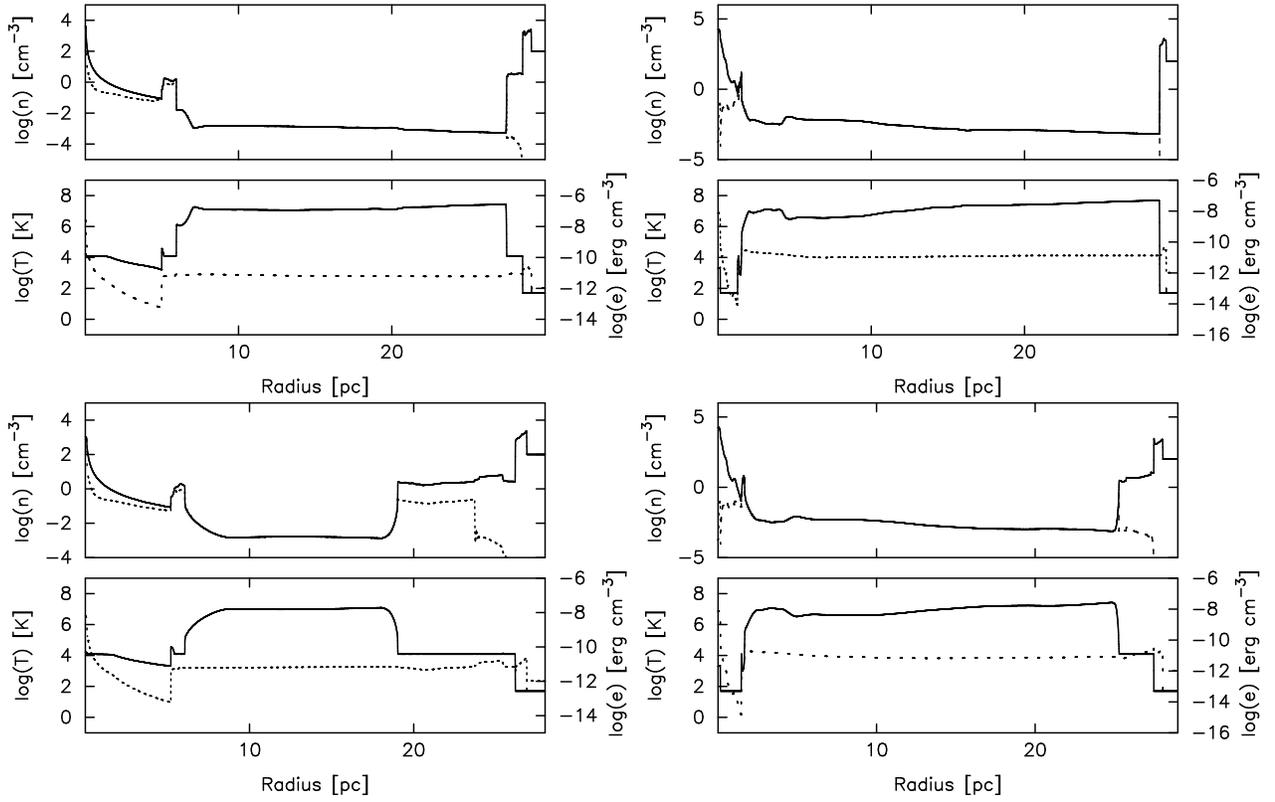

  \includegraphics[width=0.5\linewidth]{fig6a.ps}~
  \includegraphics[width=0.5\linewidth]{fig6b.ps}\\
  \includegraphics[width=0.5\linewidth]{fig6c.ps}~
  \includegraphics[width=0.5\linewidth]{fig6d.ps}\\
\caption{Density and temperature at the end of the RSG phase for the
  $40\, M_\odot$ models. \textit{Left-hand panels}: MM2003 models,
  \textit{right-hand panels}: STARS models. \textit{Top row}: models
  without thermal conduction, \textit{bottom row}: models with thermal
  conduction. The density plots show total density (solid line) and
  ionized density (dashed line). The temperature plots also show the
  internal energy, $e=p/[\gamma - 1]$ (dashed line).}
\label{fig:1D40}
\end{figure*}
\begin{figure*}
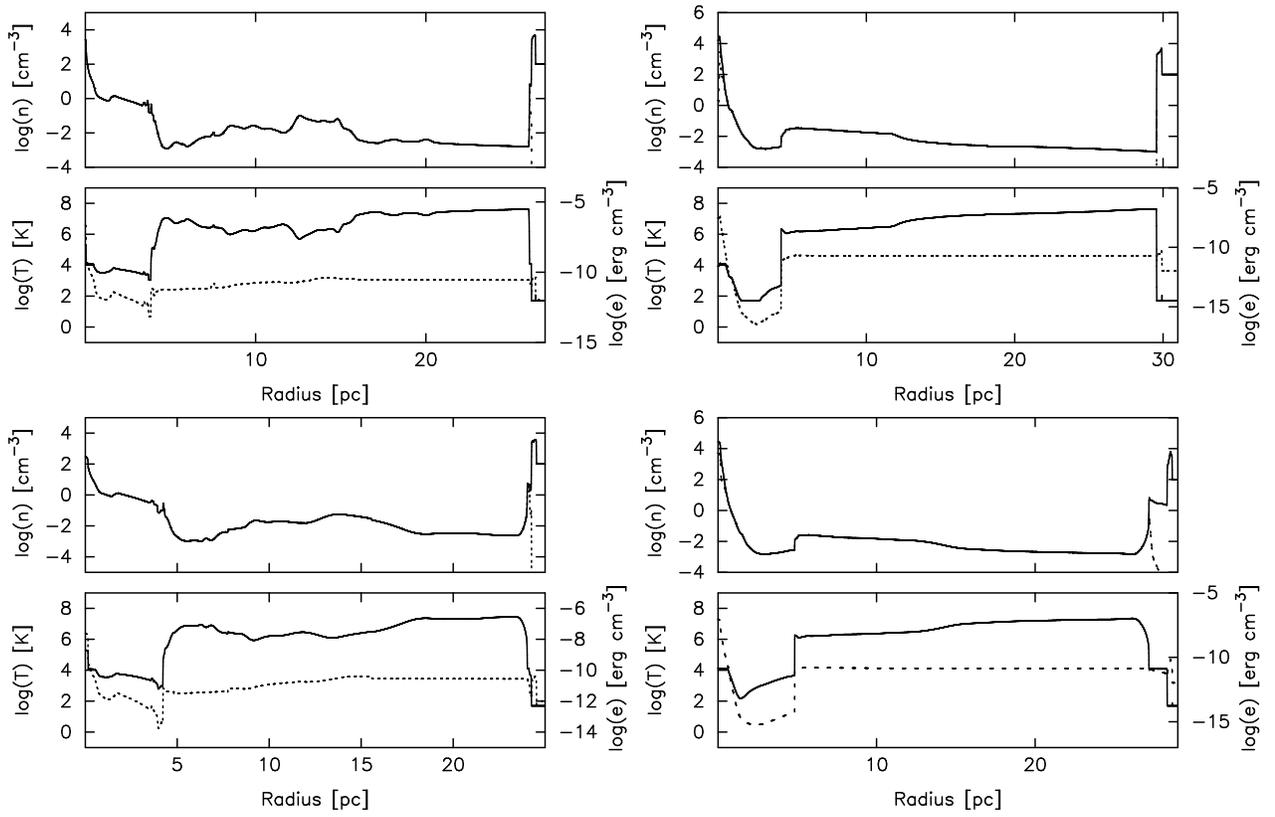

  \includegraphics[width=0.5\linewidth]{fig7a.ps}~
  \includegraphics[width=0.5\linewidth]{fig7b.ps}\\
  \includegraphics[width=0.5\linewidth]{fig7c.ps}~
  \includegraphics[width=0.5\linewidth]{fig7d.ps}\\
\caption{Same as Fig.~\protect\ref{fig:1D40} but for the $60\,M_\odot$ models.}
\label{fig:1D60}
\end{figure*}

\subsubsection{Post-main-sequence stage}
\label{subsubsec:PMS_stage}
During the RSG phase the star loses a large amount of its mass in a
slow, dense wind (see Fig.~\ref{fig:massloss}). This wind is
subsonic with respect to the hot wind bubble left by the main-sequence
stellar wind and so piles up at the bottom of a $r^{-2}$ density
gradient into a thin dense shell without forming a two-shock pattern.

In Figure~\ref{fig:1D40} we show the density, temperature and internal
energy (i.e. thermal pressure, since $e=p/[\gamma - 1]$) as a function
of radius at the end of RSG phase for the $40\,M_\odot$ models. We
compare the results using the MM2003 stellar evolution model (without
rotation) with those of STARS. For the simulations without thermal
conduction, we see an inner region of RSG wind, which has a $r^{-2}$
density profile and temperature $T < 10^4$~K, with a high central
density and bounded by a dense shell of partially ionized
material. For the MM2003 model, this region has a radius of $\sim
6$~pc, while for the STARS model the radius is only $\sim 2$~pc, the
shell is much narrower and the densities are higher. This is because
the RSG phase is much shorter in the STARS models, even though roughly
the same amount of stellar mass is lost as in the MM2003 models. The
ram pressure of the RSG wind is less than the thermal pressure of the
hot, shocked bubble produced during the main-sequence stage. As a
result, there is some back filling as the hot diffuse gas pushes back
towards the star. The lack of ionizing photons at this stage (see
Fig.~\ref{fig:photonrate}), and enhanced opacity due to the RSG wind
and neutral shell, leads to recombinations in the outer \ion{H}{2}
region at radius $R \sim 27$~pc. This is most pronounced for the
MM2003 model, since the inner absorbing RSG shell contains more
neutral atoms than the shell formed in the STARS case. When thermal
conduction is included in the simulations, evaporation of cold
material from both the RSG shell and the outer \ion{H}{2} region
enhances the density and thus the cooling rate in the hot shocked wind
bubble. In the MM2003 case, this results in a much smaller hot bubble
($\sim 19$~pc as opposed to $\sim 27$~pc in the simulation without
conduction). The circumstellar medium in the STARS simulations is less
affected by thermal conduction, possibly due to the shorter timescales
in the RSG phase.

In Figure~\ref{fig:1D60} we show the corresponding circumstellar
structures for the $60\,M_\odot$ models at the end of the RSG
phase. The MM2003 and STARS results look quite different. This is
because the MM2003 model experiences several episodes of enhanced mass
loss, whereas the STARS model undergoes a single intense outburst. The
ram pressure of the slow wind in the MM2003 case fluctuates and this
leads to acoustic waves propagating from the inner edge of the hot
bubble outwards. In Figure~\ref{fig:1D60} the waves have yet to reach
the outer edge of the bubble, but once they do, they will reflect from
the dense neutral shell and pass back through the bubble and reflect
backwards and forwards through the hot shocked gas, forming localized
heated and compressed regions where oppositely directed waves shock
against each other. There is no clearly defined shell of RSG material
in the $60\,M_\odot$ models, because the duration of the most intense
period of mass loss is very short and just before the onset of the WR
phase (see Fig.~\ref{fig:massloss}), hence the dense material has not
had time to move far away from the star before the fast wind starts.

We are particularly interested in the radii of the dense, neutral RSG
shells formed in these models. These shells will be impacted and
broken up by the fast WR winds and so have important consequences for
the observables (X-ray and optical emission) of the WR wind-blown
bubbles. The shells are most evident in the $40\,M_\odot$ models,
reaching radii of $\sim 6$~pc and $\sim 2$~pc for the MM2003 and STARS
models, respectively, by the end of the RSG phase. These distances
reflect the duration of the RSG stage in each case: $5.6 \times
10^5$~yrs for the MM2003 model as opposed to $3.1 \times 10^5$~yrs for
the STARS model. Also, the wind velocities in the STARS model are
allowed to fall below 30~km~s$^{-1}$ and in general are lower that
those of the MM2003 models (see Fig.~\ref{fig:windvel}). We have not
presented results for the MM2003 models with stellar rotation because
they are qualitatively similar to the results for the other models and
result in a dense neutral shell being formed at $\sim 2$~pc, since the
RSG phase in this model only lasts $1.0 \times 10^5$~yrs.  We continue
the simulations in 1D until the end of the WR phase but we do not show
these models because we are interested in the formation of
instabilities during the fast wind-slow wind interaction, which do not
occur in 1D.

\subsection{2D Results}
\label{subsec:2dresults}
At the end of the RSG phase, we remap our 1D, spherically symmetric
distributions of density, momentum and energy onto a 2D axisymmetric
grid using a volume-weighted averaging procedure to ensure
conservation of these quantities. The 2D grid is, of computational
necessity, coarser than the 1D grid, and we also zoom in on the inner
$\sim 10$~pc of the simulation, since this is where the hydrodynamic
interaction between the fast WR wind and the slow RSG wind will take
place. Also, the observed WR bubbles all have radii 10~pc or less. We
continue to follow the radiation-hydrodynamic evolution of the
circumstellar medium using the stellar wind parameters and the
tabulated ionizing photon rate as a function of time shown in
Figures.~\ref{fig:massloss}, \ref{fig:windvel}, and
\ref{fig:photonrate}. For all 2D simulations, we present only one
quadrant of the calculation, since the other is the same due to
symmetry.
\subsubsection{$40 M_{\odot}$ models}
\begin{figure*}[p]
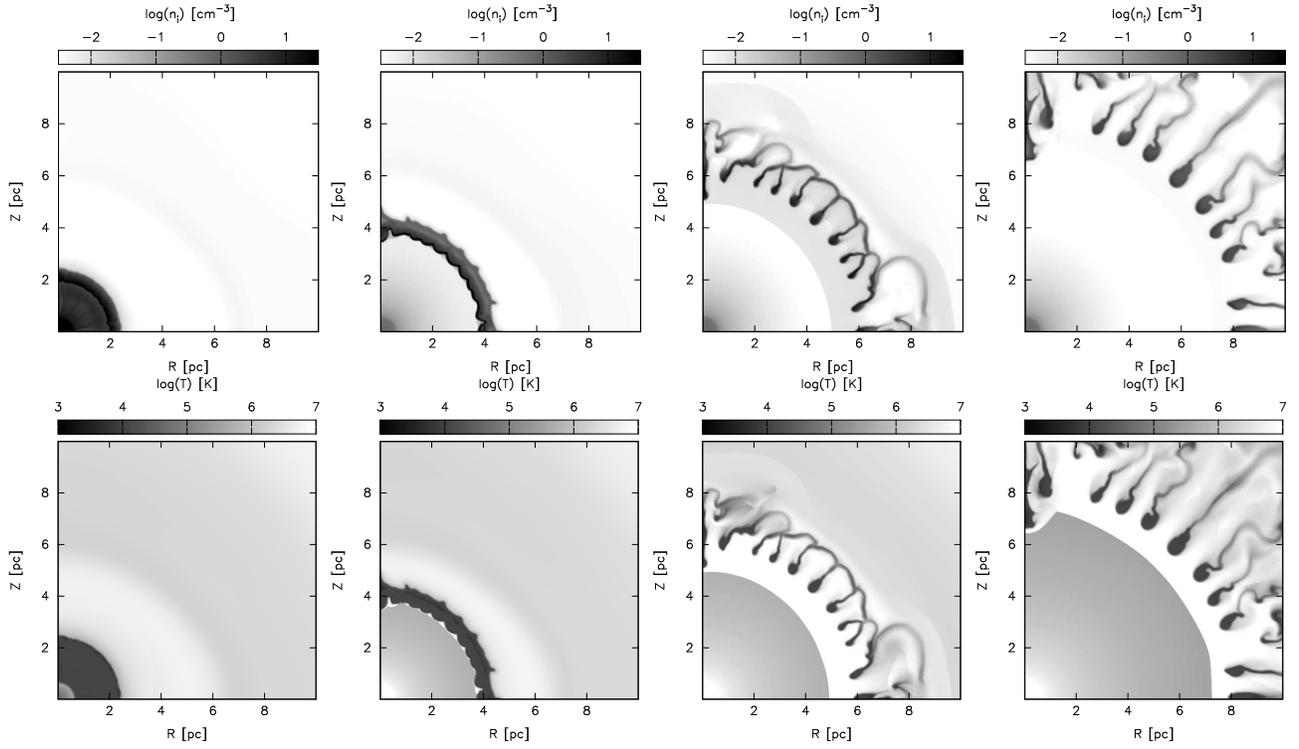

  \includegraphics[width=0.25\linewidth]{fig8a.ps}~
  \includegraphics[width=0.25\linewidth]{fig8b.ps}~
  \includegraphics[width=0.25\linewidth]{fig8c.ps}~
  \includegraphics[width=0.25\linewidth]{fig8d.ps}\\
  \includegraphics[width=0.25\linewidth]{fig8e.ps}~
  \includegraphics[width=0.25\linewidth]{fig8f.ps}~
  \includegraphics[width=0.25\linewidth]{fig8g.ps}~
  \includegraphics[width=0.25\linewidth]{fig8h.ps}
\caption{Wind-wind interaction for the $40\,M_{\odot}$ STARS model
  without thermal conduction. From left to right the evolution times
  are $39800, 65900, 86000,$ and $104000$~yrs after the onset of the
  WR wind. Top panels show ionized number density and bottom panels
  show temperature, both on logarithmic scale.}
\label{fig:2D40_snapshots1}
\end{figure*}

\begin{figure*}[p]
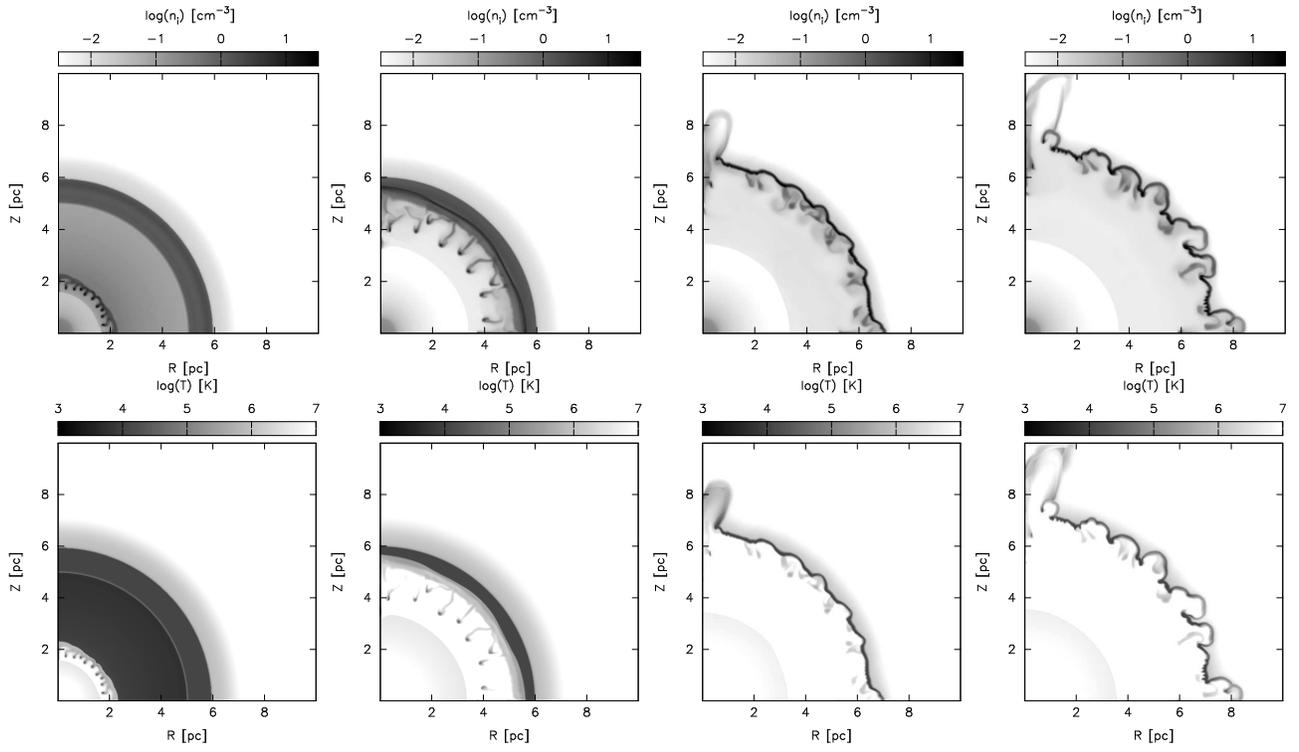

  \includegraphics[width=0.25\linewidth]{fig9a.ps}~
  \includegraphics[width=0.25\linewidth]{fig9b.ps}~
  \includegraphics[width=0.25\linewidth]{fig9c.ps}~
  \includegraphics[width=0.25\linewidth]{fig9d.ps}~\\
  \includegraphics[width=0.25\linewidth]{fig9e.ps}~
  \includegraphics[width=0.25\linewidth]{fig9f.ps}~
  \includegraphics[width=0.25\linewidth]{fig9g.ps}~
  \includegraphics[width=0.25\linewidth]{fig9h.ps}\\
\caption{Same as Figure \ref{fig:2D40_snapshots1} but for the
  $40\,M_{\odot}$ MM2003 model without stellar rotation and without
  thermal conduction. From left to right the evolution times are
  $10600, 22600, 32650,$ and $36600$~yrs after the onset of the WR
  wind.}
\label{fig:2D40_snapshots2}
\end{figure*}

\begin{figure*}[p]
  \includegraphics[width=0.25\linewidth]{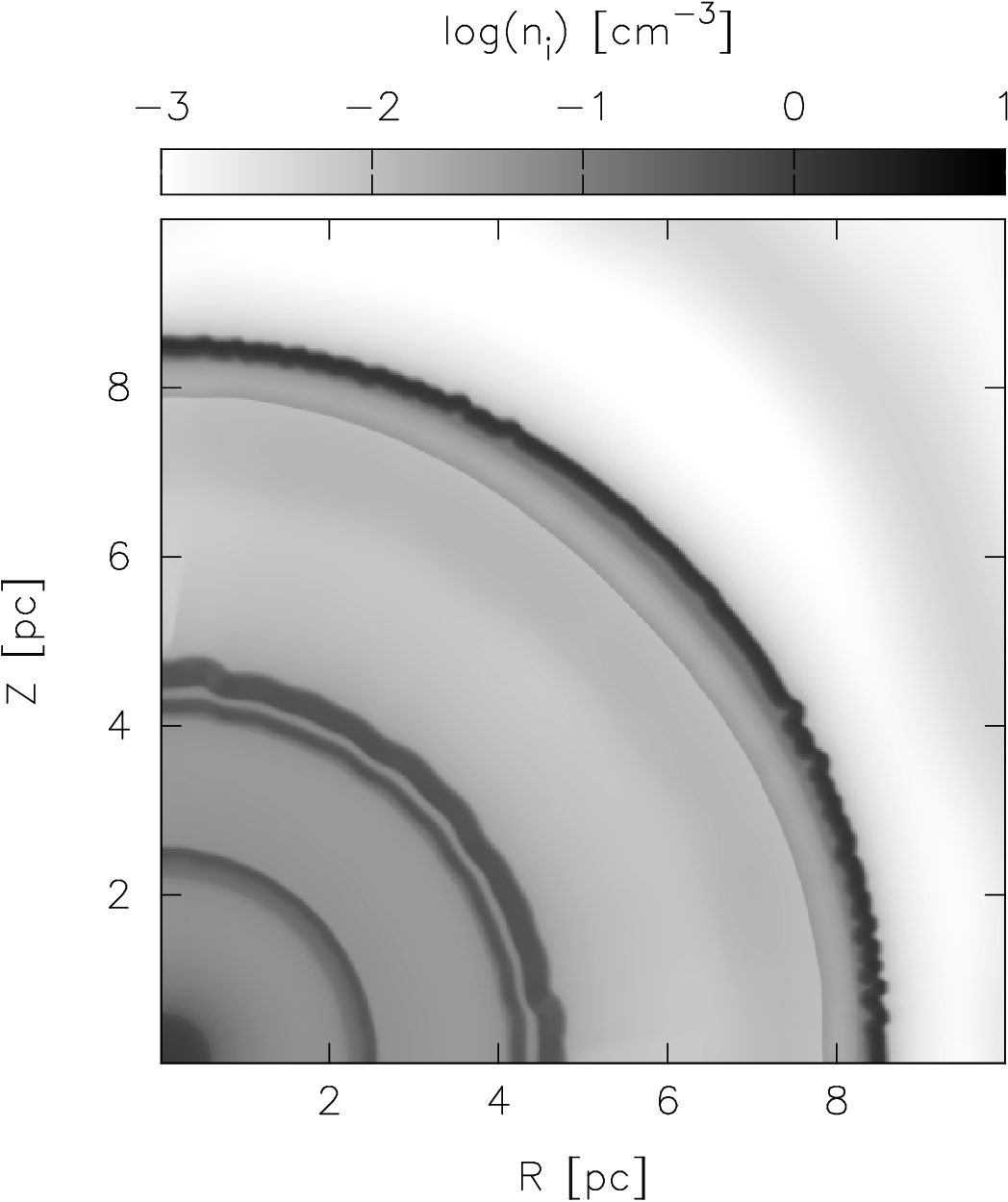}~
  \includegraphics[width=0.25\linewidth]{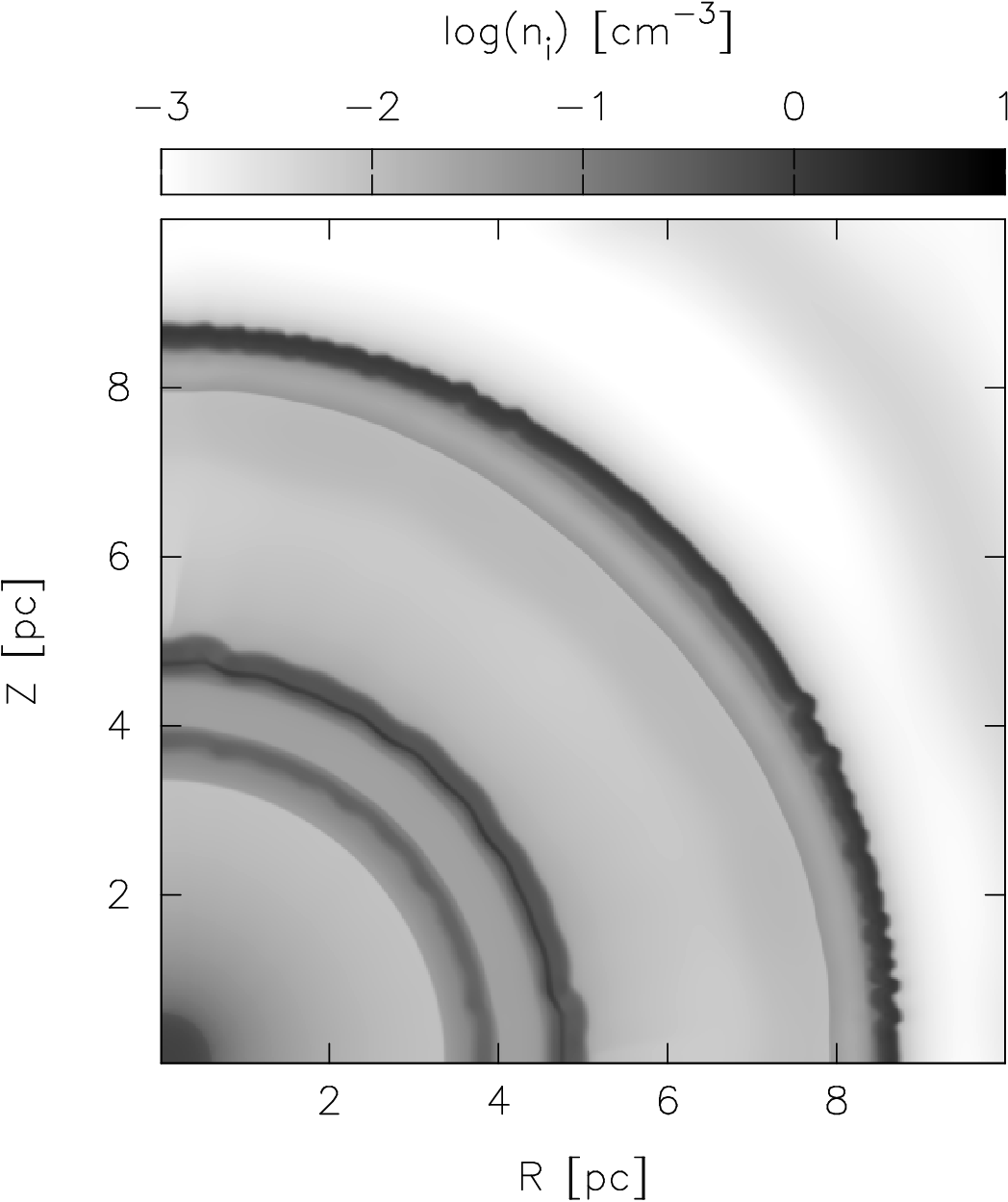}~
  \includegraphics[width=0.25\linewidth]{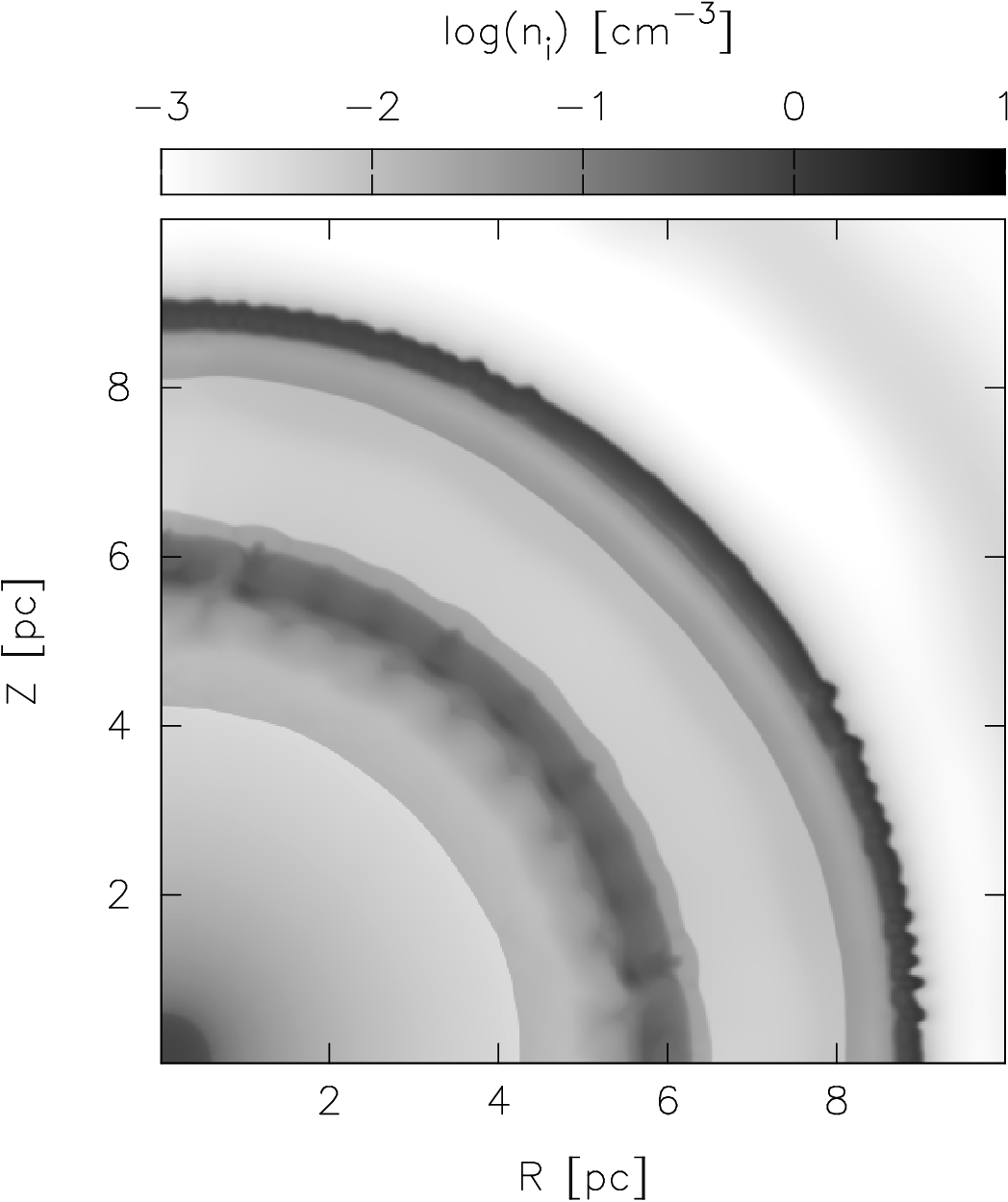}~
  \includegraphics[width=0.25\linewidth]{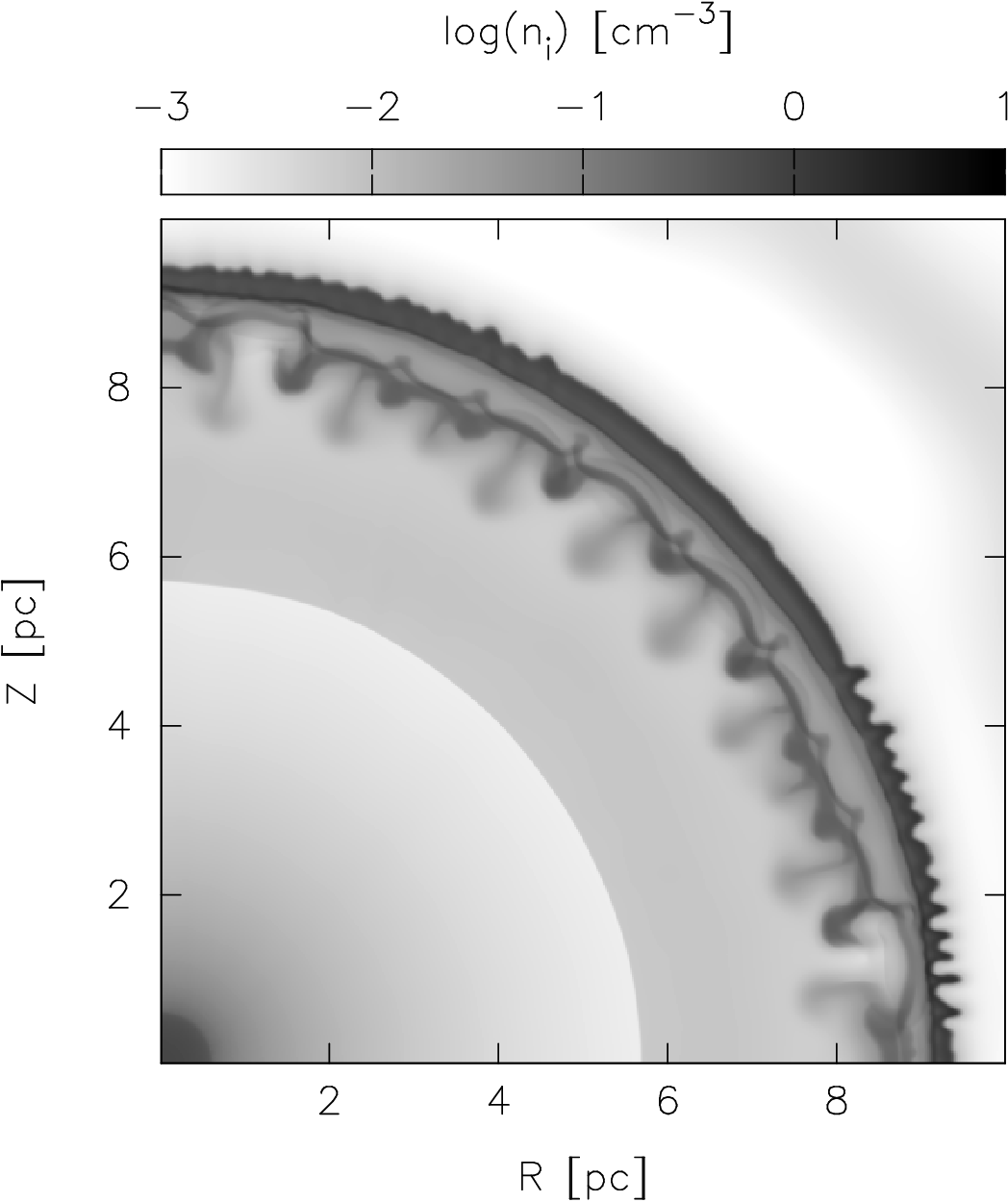}\\
  \includegraphics[width=0.25\linewidth]{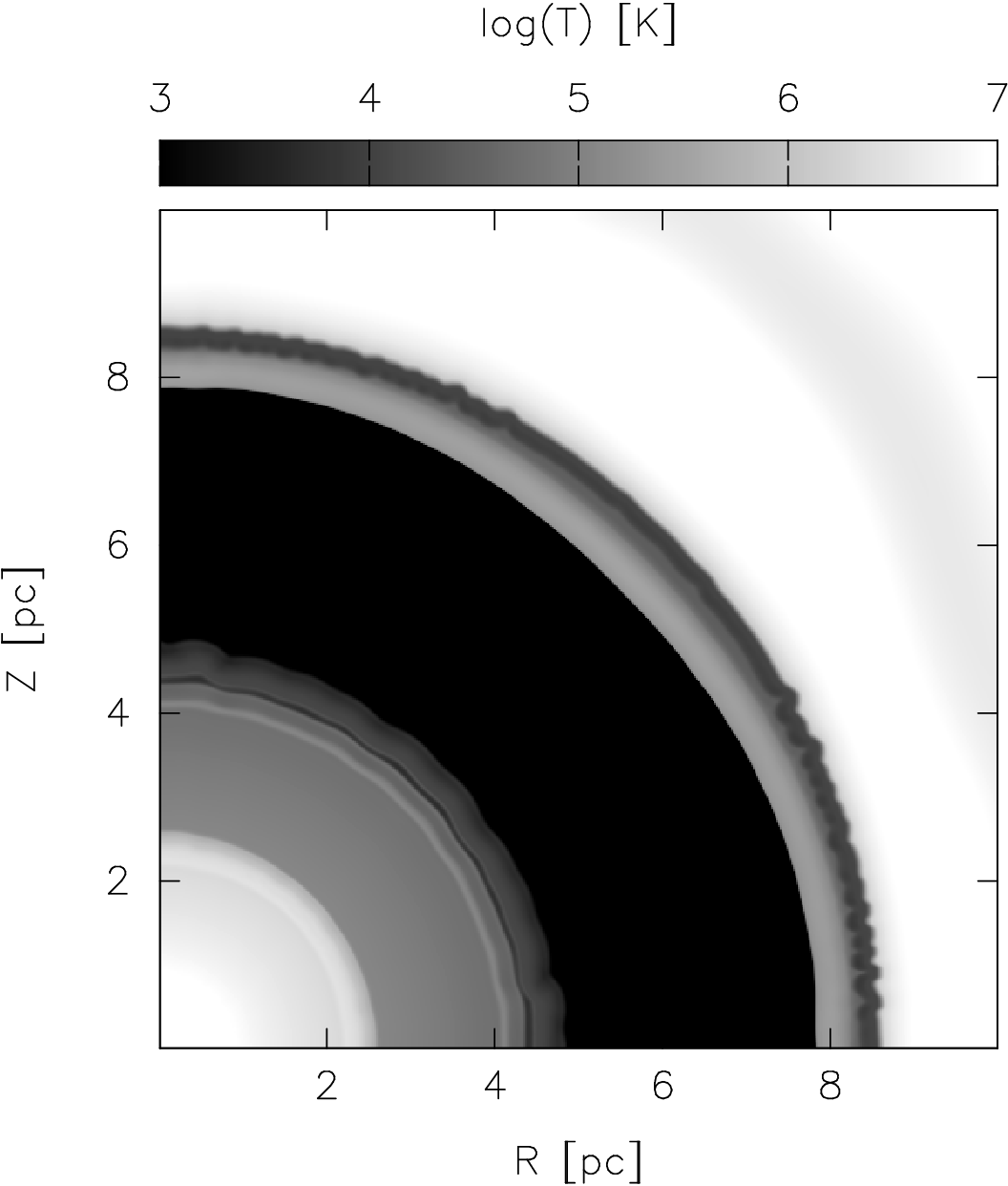}~
  \includegraphics[width=0.25\linewidth]{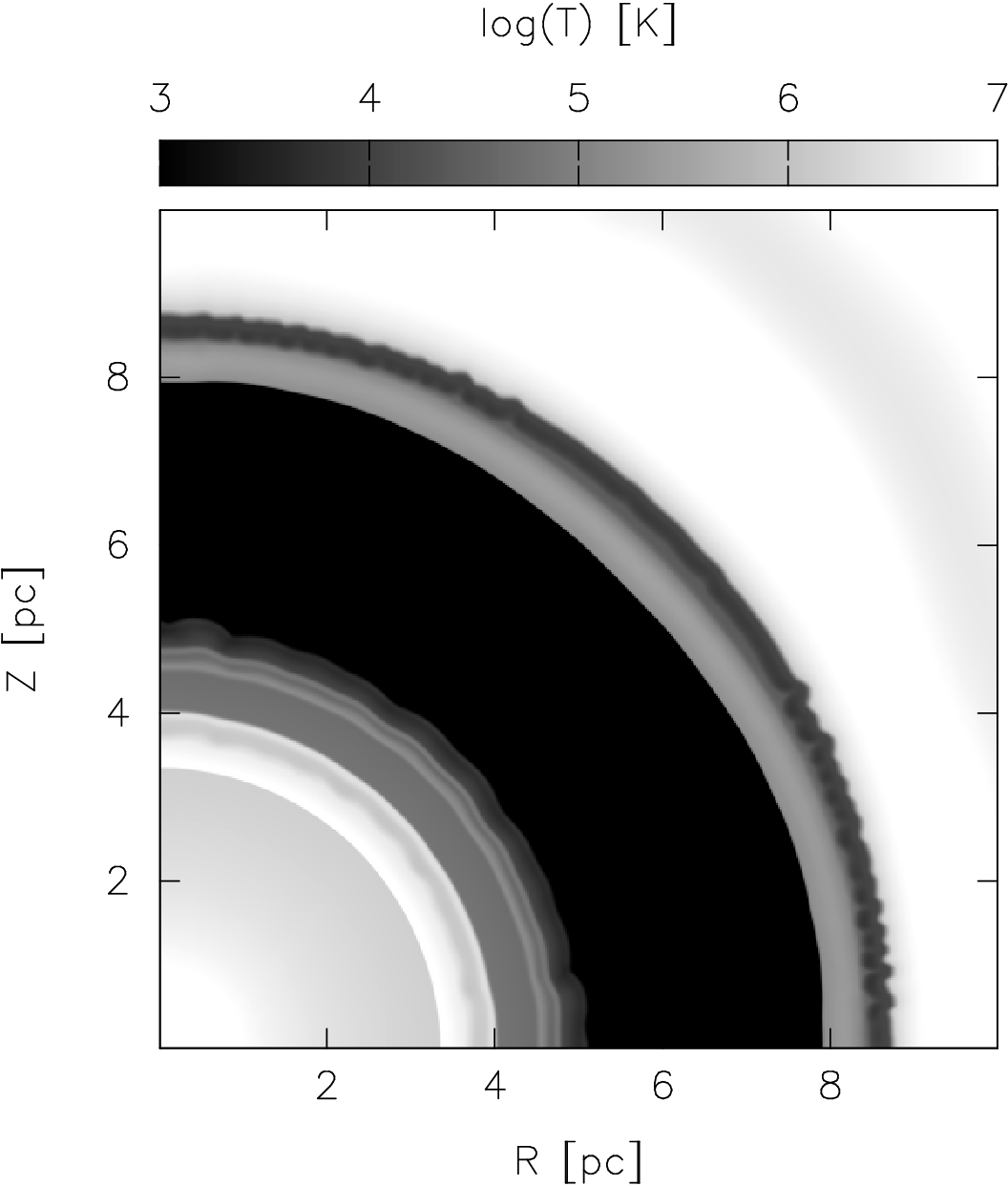}~
  \includegraphics[width=0.25\linewidth]{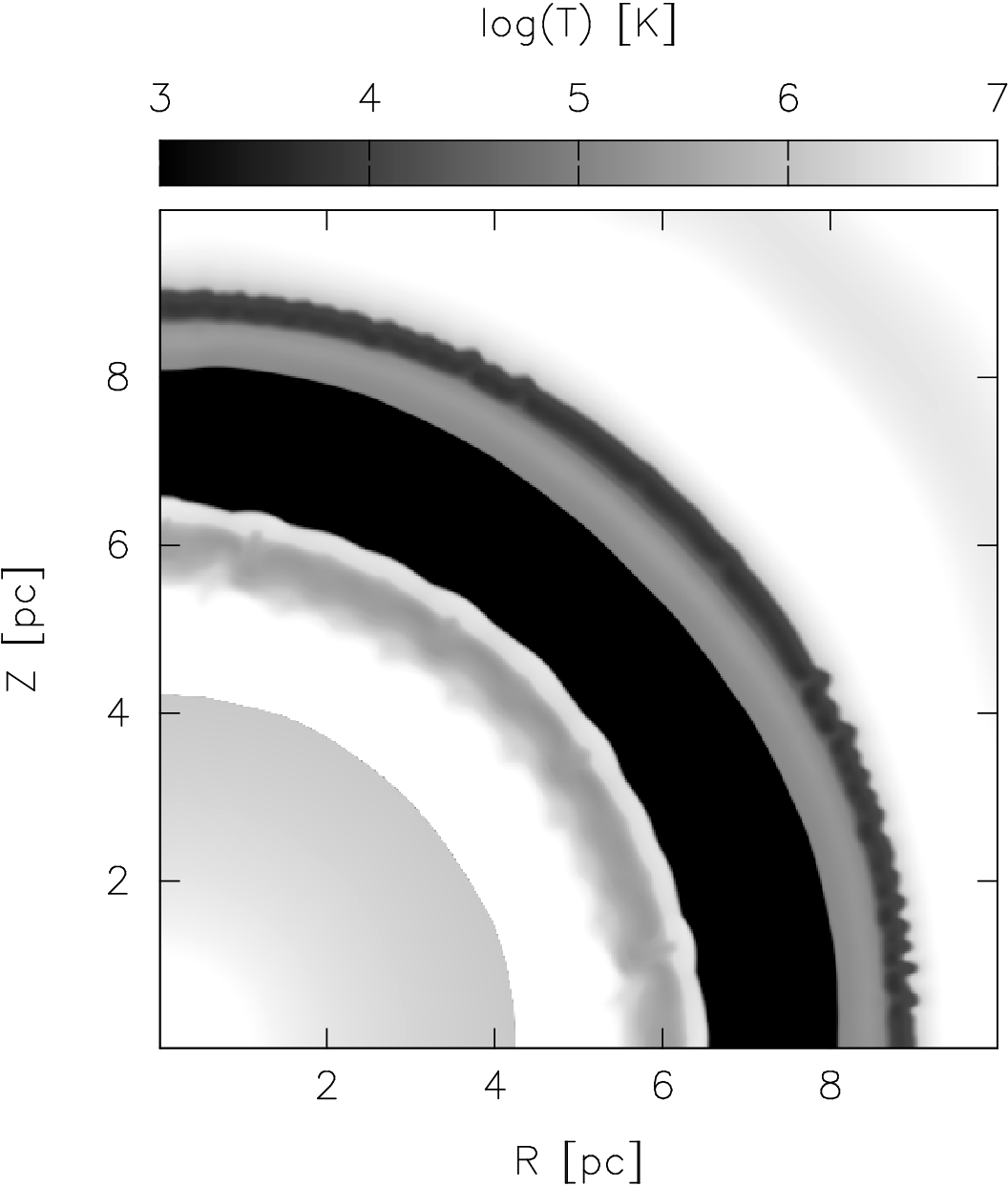}~
  \includegraphics[width=0.25\linewidth]{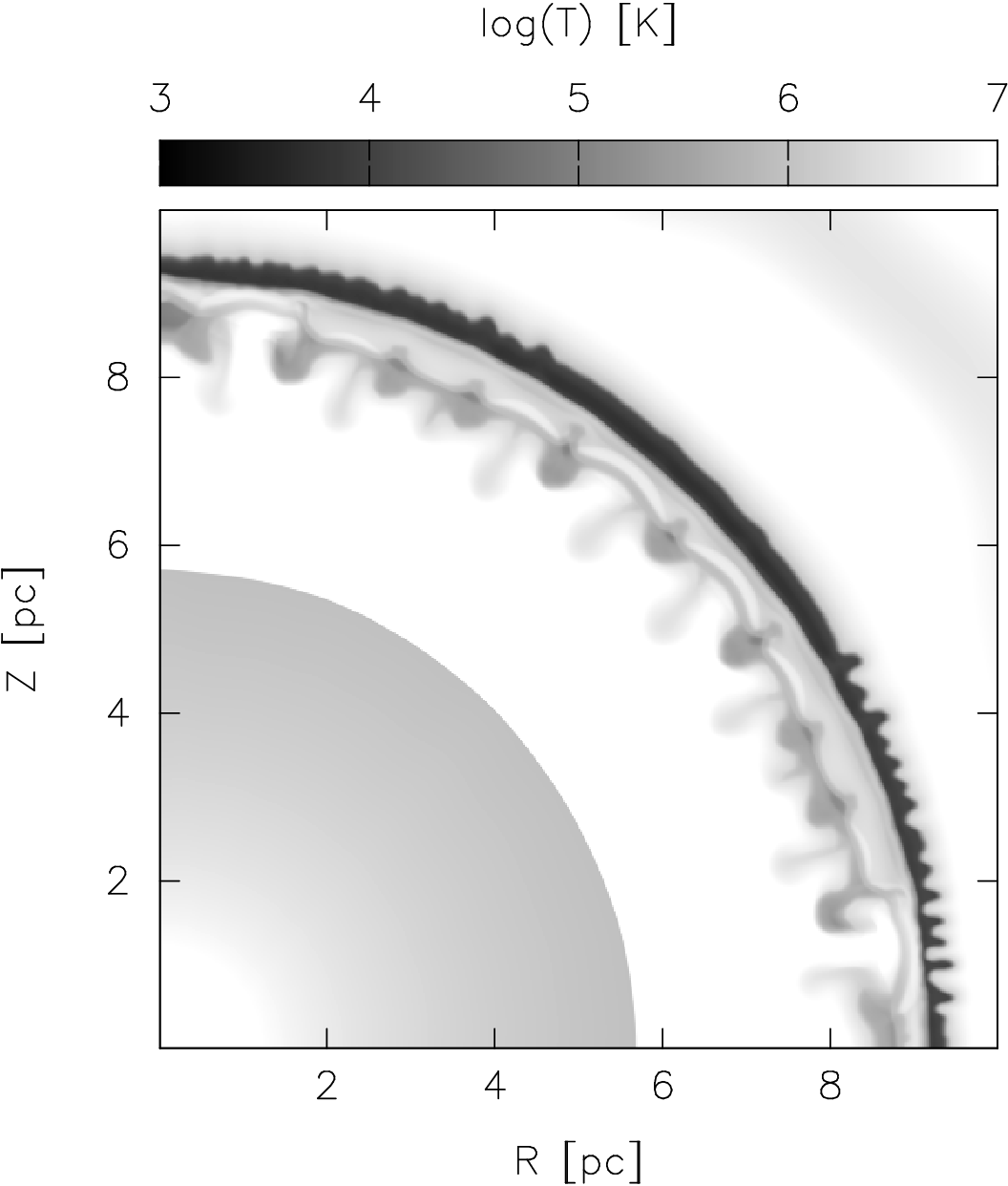}
\caption{Same as Figure \ref{fig:2D40_snapshots1} but for the
  $40\,M_{\odot}$ MM2003 model with stellar rotation without thermal
  conduction. From left to right the evolution times are $6800, 8800,
  12800,$ and $16900$~yrs after the onset of the WR wind.}
\label{fig:2D40_snapshots3}
\end{figure*}
In Figure~\ref{fig:2D40_snapshots1} we present grayscale images for
the wind-wind interaction for the $40\,M_{\odot}$ STARS model without
thermal conduction at four different times corresponding to when the
interaction region has reached distances 2, 4, 6, and 8~pc from the
central star. These times are, respectively, $39800, 65900, 86000,$
and $104000$~yrs after the onset of the WR wind. In this figure, we
see the disruption of the RSG shell by the fast wind and the formation
of clumps. The RSG material consisted initially of a thin, dense shell
of partially ionized material located $\sim 2$~pc from the star. The
fast WR wind accelerates down the density gradient and slams into the
dense shell. The WR wind is shocked and heated by this interaction
while the dense shell is shocked and compressed. Vishniac
instabilities of the linear thin-shell variety break the shell into
dense clumps, which then travel radially outwards
\citep{1983ApJ...274..152V,1989ApJ...337..917V}. The clumps have a
cometary form, with the dense, cold ($10^4$~K photoionized) head
pointing towards the central star and tails extending several parsecs
radially outwards, showing that the heads are traveling more slowly
than the diffuse material that surrounds them. Interactions between
clump material and the shocked fast wind create a pattern of multiple
interacting shocks, which heat the ablated clump gas to X-ray
temperatures. For this particular case the RSG shell breaks up at
$\sim5$~pc (see Fig.~\ref{fig:2D40_snapshots1}).

Figure~\ref{fig:2D40_snapshots2} shows the results obtained when the
stellar wind parameters come from the MM2003 $40\, M_{\odot}$ model
without rotation and without conduction. The wind-wind interaction
region is again shown at 2, 4, 6 and 8~pc from the central star,
corresponding to $10600, 22600, 32650,$ and $36600$~yrs after the
onset of the WR wind. For this model, the RSG dense wind material,
traveling at $\sim 30$~km~s$^{-1}$, is in the form of a broad shell
about 6~pc from the star at the start of the WR stage. The interaction
of the WR wind with this shell occurs only a few times $10^3$~yrs
after the onset of the fast wind. Compared to the STARS model, the
MM2003 models experience Rayleigh-Taylor instabilities, which disrupt
the accelerating WR shell before it interacts with the main shell of
RSG material, forming small clumps at $<2$~pc. The main interaction of
the WR shell with the RSG shell occurs at 6~pc and compresses the RSG
material into a very thin, dense shell, which then corrugates and
starts to break up at around 8~pc due to the linear thin-shell
instability.

Figure~\ref{fig:2D40_snapshots3} shows the results from MM2003 model
with stellar rotation but without conduction at $6800, 8800, 12800,$
and $16900$~yrs after the onset of the WR wind. In this model the
interaction takes place much further from the star and the shell of
RSG material is not so dense, so the interaction does not produce such
violent instabilities and the shell remains essentially intact, being
swept up and compressed by the fast wind but not suffering the
complete disruption seen in the STARS case discussed above. For this
model, the intense mass-loss phase has a relatively short duration,
$\sim 10^5$~yrs, and there are several mass-loss episodes before and
after the main ejection event. Rather than a RSG stage, the extremely
strong mass loss ($\dot{M}_\mathrm{w} \sim
10^{-3.6}M_{\odot}$~yr$^{-1}$), together with the fact that the
stellar effective temperature is relatively high ($T_\mathrm{eff} \sim
10^4$~K) for part of this period, indicate a Luminous Blue Variable
stage, with stellar wind velocities $V_\mathrm{w} >
200$~km~s$^{-1}$. The densest wind material travels quite a large
distance from the star before the onset of the WR stage and so the
interaction with the fast wind, when it occurs, does not lead to
complete breakup of the shell.

\begin{figure*}[p]
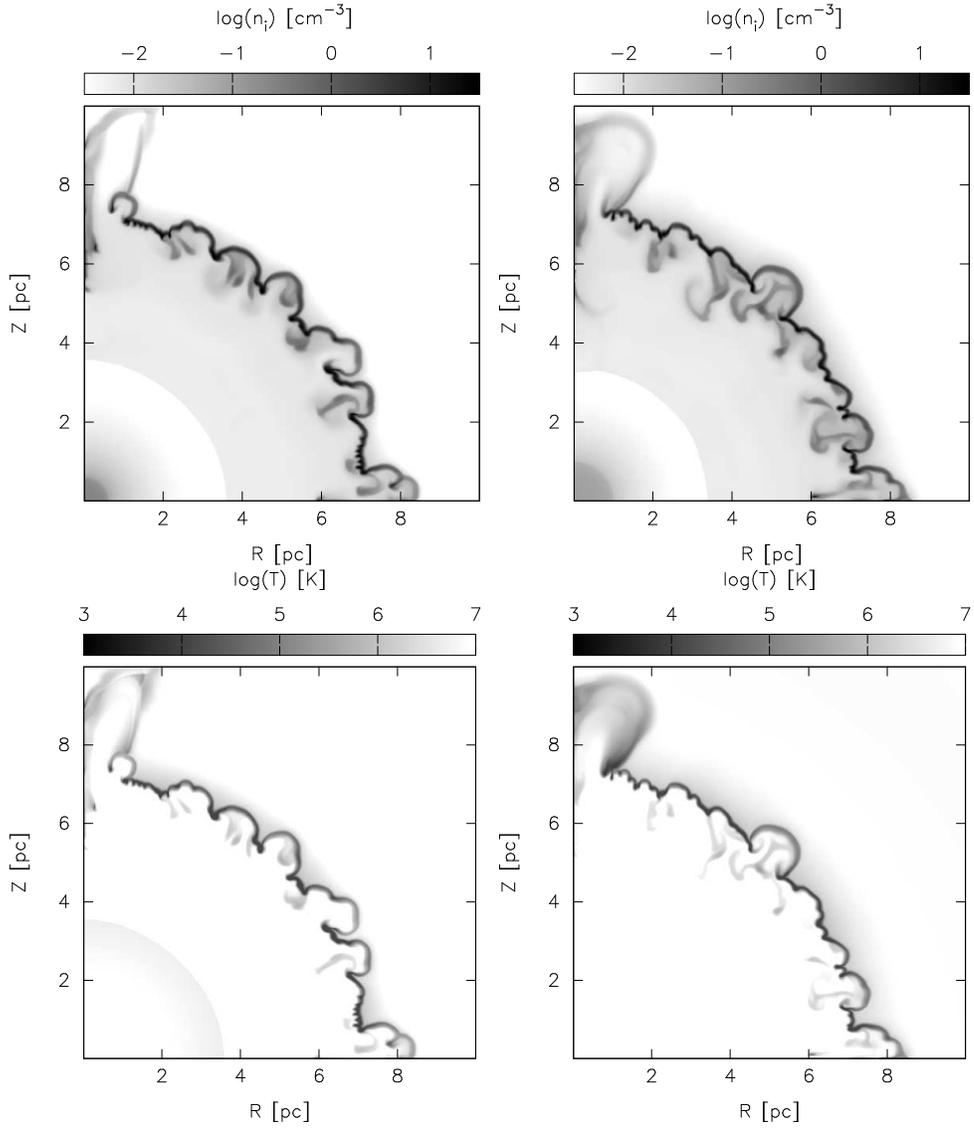

\centering
\includegraphics[width=0.38\linewidth]{fig11a.ps}~ 
\includegraphics[width=0.38\linewidth]{fig11b.ps} \\
\includegraphics[width=0.38\linewidth]{fig11c.ps}~ 
\includegraphics[width=0.38\linewidth]{fig11d.ps} \\
\caption{Ionized density (Top panels) and temperature (bottom panels)
  during the interaction of the fast WR wind with the slow, dense RSG
  wind for the $40\,M_\odot$ MM2003 model without stellar
  rotation. \textit{Left panel}: model without thermal conduction
  \textit{Right panel}: with thermal conduction. Both numerical
  results are at $36600$~yrs after the the onset of the WR stage.}
\label{fig:2D40_2}
\end{figure*}

We find that thermal conduction has little discernible effect either
on the density or the temperature structure of the 2D simulations and
the results are essentially the same as those shown in Figures
\ref{fig:2D40_snapshots1}, \ref{fig:2D40_snapshots2} and
\ref{fig:2D40_snapshots3}. In Figure \ref{fig:2D40_2} we present one
set of results including thermal conduction to illustrate this. In
this figure we show the results of the $40\,M_{\odot}$ MM2003
simulations (without rotation) at 36600~yrs after the onset of the WR
wind. There are slight differences between the two sets of results,
but the short timescales of the wind-wind interaction stage compared
to cooling timescales in the $10^{6}$~K gas mean that there is no
major effect on the hydrodynamics due to thermal conduction during
this stage.

We have run further set of STARS simulations to investigate the
effects of the radius of the wind injection zone and the grid
resolution on the number and distribution of clumps formed during
shell disruption. Increasing the grid resolution to 800$\times$1600
cells while maintaining the same number of cells in the wind injection
region (radius 30 cells) produced the same results as those presented
here. Decreasing the size of the wind injection region (radius 10
cells) did produce fewer clumps but increasing the radius of the wind
injection region to 50 cells gave the same results as in Figure
\ref{fig:2D40_snapshots1}. The clumps formed in our simulations have
diameters between 15 and 50 cells and are photoionized, hence are
fully resolved.

\subsubsection{$60 M_{\odot}$ models}
\begin{figure*}[p]
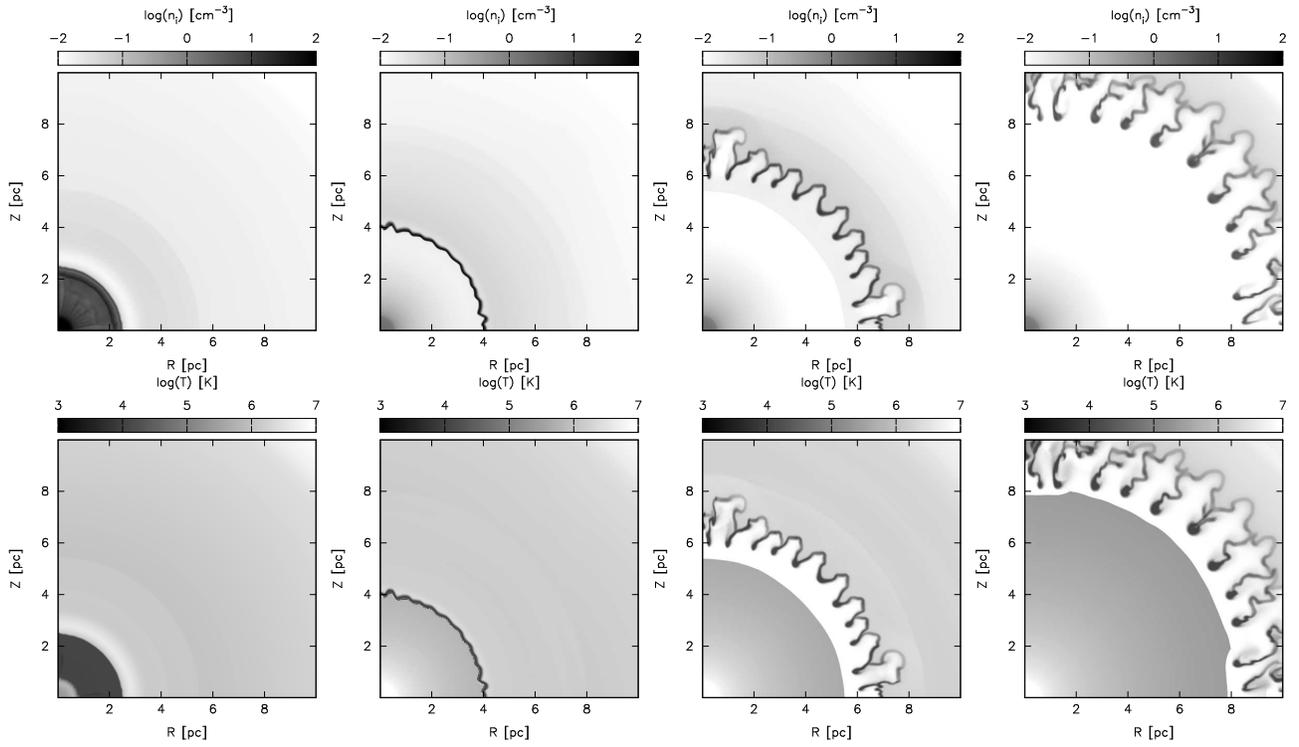

  \includegraphics[width=0.25\linewidth]{fig12a.ps}~
  \includegraphics[width=0.25\linewidth]{fig12b.ps}~
  \includegraphics[width=0.25\linewidth]{fig12c.ps}~
  \includegraphics[width=0.25\linewidth]{fig12d.ps}\\
  \includegraphics[width=0.25\linewidth]{fig12e.ps}~
  \includegraphics[width=0.25\linewidth]{fig12f.ps}~
  \includegraphics[width=0.25\linewidth]{fig12g.ps}~
  \includegraphics[width=0.25\linewidth]{fig12h.ps}\\
\caption{Wind-wind interaction for the $60\,M_{\odot}$ STARS model
  without thermal conduction. From left to right the evolution times
  are $27150, 43200, 57200,$ and $71250$~yrs after the onset of the WR
  wind.}
\label{fig:2D60_snapshots1}
\end{figure*}
\begin{figure*}[p]
  \includegraphics[width=0.25\linewidth]{fig13a.ps}~
  \includegraphics[width=0.25\linewidth]{fig13b.ps}~
  \includegraphics[width=0.25\linewidth]{fig13c.ps}~
  \includegraphics[width=0.25\linewidth]{fig13d.ps}\\
  \includegraphics[width=0.25\linewidth]{fig13e.ps}~
  \includegraphics[width=0.25\linewidth]{fig13f.ps}~
  \includegraphics[width=0.25\linewidth]{fig13g.ps}~
  \includegraphics[width=0.25\linewidth]{fig13h.ps}\\
\caption{Same as Figure~\ref{fig:2D60_snapshots1} but for the
  $60\,M_{\odot}$ MM2003 model without stellar rotation and without
  thermal conduction. From left to right the evolution times are
  $6800, 14800, 24800,$ and $30900$~yrs after the onset of the WR
  wind.}
\label{fig:2D60_snapshots2}
\end{figure*}
\begin{figure*}[p]
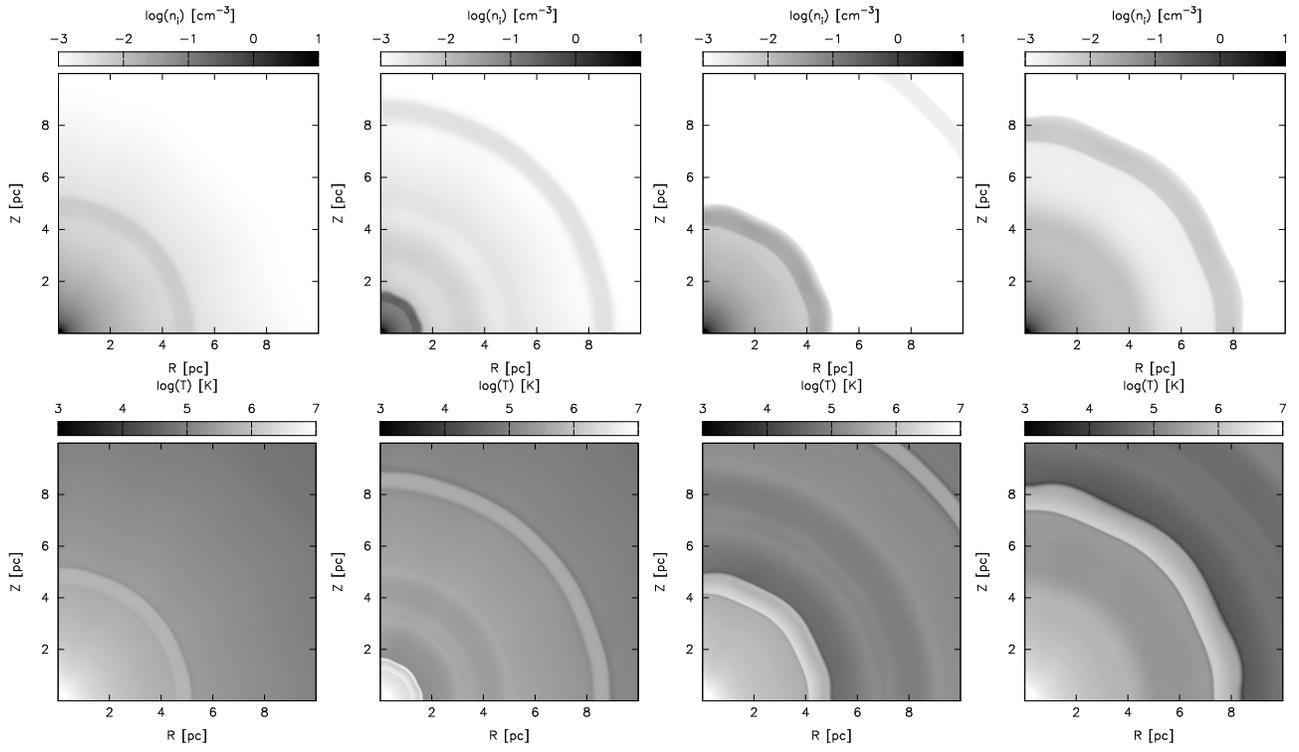

  \includegraphics[width=0.25\linewidth]{fig14a.ps}~
  \includegraphics[width=0.25\linewidth]{fig14b.ps}~
  \includegraphics[width=0.25\linewidth]{fig14c.ps}~
  \includegraphics[width=0.25\linewidth]{fig14d.ps}\\
  \includegraphics[width=0.25\linewidth]{fig14e.ps}~
  \includegraphics[width=0.25\linewidth]{fig14f.ps}~
  \includegraphics[width=0.25\linewidth]{fig14g.ps}~
  \includegraphics[width=0.25\linewidth]{fig14h.ps}\\
\caption{Same as Figure~\ref{fig:2D60_snapshots1} but for the
  $60\,M_{\odot}$ MM2003 model with stellar rotation and without
  thermal conduction. From left to right the evolution times are
  $19450, 21450, 23450,$ and $25500$~yrs after the onset of the WR
  wind.}
\label{fig:2D60_snapshots3}
\end{figure*}
In Figure~\ref{fig:2D60_snapshots1} we show the corresponding results
for the $60\, M_\odot$ STARS model without conduction. Again, we show
the wind-wind interaction at 2, 4, 6, and 8~pc, corresponding to
evolution times of $27150, 43200, 57200,$ and $71250$~yrs after the
onset of the WR wind. The dense RSG (or LBV) material is swept up into
a thin shell by the fast WR wind. Linear thin-shell instabilities are
induced in the swept-up material and break the shell into many cold
($< 10^4$~K), dense ($> 1$~cm$^{-3}$) clumps which drift outwards,
embedded in the hot, shocked WR wind. The morphologies of the
$40\,M_{\odot}$ and $60\,M_{\odot}$ STARS models are very similiar.

For the MM2003 $60\,M_{\odot}$ model without rotation, there are
several episodes of enhanced mass loss during the RSG/LBV stage and
even in the WR stage there are short-lived outbursts (see
Fig.~\ref{fig:massloss}). This leads to the formation of successive
shells of material traveling with different velocities away from the
star shown in Figure~\ref{fig:2D60_snapshots2}. The most intense
mass-loss episode shown in Figure~\ref{fig:massloss} is followed by a
reduction in mass-loss rate and increase in wind velocity (see
Fig.~\ref{fig:windvel}), which sweeps up, compresses and accelerates
the ejected material. This leads to the formation of Rayleigh-Taylor
instabilities, which disrupt the shell once it has reached $\sim
10$~pc from the star. The clumpy shell is traveling quite quickly and
soon leaves the region of interest ($R < 10$~pc). Other, less dense,
shells follow, which suffer instabilities but do not contain enough
material to form dense, cold clumps. In
Figure~\ref{fig:2D60_snapshots2} we show the wind-wind interaction and
the formation of different shells and shock patterns as a result of
the variability of the mass-loss and wind velocity up to 30,900~yrs
after the onset of the WR wind.

For the MM2003 $60\,M_{\odot}$ model with rotation, from
Figure~\ref{fig:massloss} we note that there is no period of
particularly enhanced mass loss. In this case, therefore, although the
RSG is swept up by the faster WR wind, the swept-up material is not
particularly dense and does not form a thin, dense shell nor any
instabilities or clumps. In Figure~\ref{fig:2D60_snapshots3} we show
the density and temperature in the evolving CSM up to the point when
the swept-up material has reached a radius of 8~pc from the star,
which occurs 25,500~yrs after the onset of the WR stage.

\subsection{Comparisons with previous works}
Differences between our results and the work of previous authors can
be attributed to several factors: the stellar evolution models, the
numerical method, and the initial conditions and physics that is
included in the numerical simulations. In this section we discuss each
of these aspects in turn.

Much of the previous numerical work on the evolving circumstellar
media around massive stars has used the same $35\,M_\odot$ and
$60\,M_\odot$ stellar evolution models as
\citet{1996A&A...305..229G,1996A&A...316..133G}, which are those
developed by \citet{1994A&A...290..819L}. These models were used in
their full time-dependent form by
\citet{1996A&A...305..229G,1996A&A...316..133G},
\citet{2003ApJ...594..888F,2006ApJ...638..262F}, and
\citet{2007ApJ...667..226D}.  For both stellar masses considered, the
stars evolve to be red supergiants ($T_\mathrm{eff} < 4800$~K). One
difference between these models and the ones we use in this paper is
the stellar wind velocity. In the
\citet{1996A&A...305..229G,1996A&A...316..133G} models, the stellar
wind velocity remains high, above 2500~km~s$^{-1}$, throughout the
main-sequence stage, whereas in all of our models we see a significant
decrease from an initial $\sim 3000$~km~s$^{-1}$ to $\sim
1000$~km~s$^{-1}$. This affects both the pressure and size of the hot,
shocked wind bubble at the end of the main-sequence stage. In the case
of their $60\,M_\odot$ model, there is a ``H-rich WN'' stage
immediately after the main-sequence stage, in which more than
$10\,M_\odot$ of material is lost by the star at high velocity over
the space of a million years. None of the models we have studied have
a similar evolutionary stage. This mass is moving too fast to remain
close to the star during the subsequent evolutionary stages, and thus
does not contribute to structure formation in the circumstellar
medium. Other authors have used the STARS stellar evolution models
\citep{2006MNRAS.367..186E}, where the only differences between that
work and ours is that we have corrected the stellar wind velocities
(upwards) by missing numerical factors, and those of
\citet{1992A&AS...96..269S}, for $40\,M_\odot$ and $60\,M_\odot$
\citep{2005A&A...444..837V,2007A&A...469..941V}, which are very
similar to the MM2003 models without rotation.  However,
\citet{2005A&A...444..837V,2007A&A...469..941V} approximates the
evolution with a three-constant-winds scenario, and so does not follow
the episodic ejections of material which are a main characteristic of
the post-main-sequence evolution of stars with initial masses $\gtrsim
60\,M_\odot$ \citep{2010ASPC..425..247H}. By following the full time
variation of the stellar wind parameters, we are able to study the
formation and interaction of multiple shells in the circumstellar
medium in the LBV and WR stages.

All of the numerical simulations by previous authors use well-tested
hydrodynamical schemes and so we do not expect major differences to
arise from differences between codes. The codes used include ZEUS
\citep{1996A&A...305..229G,1996A&A...316..133G,2005ApJ...631..435R,2005A&A...444..837V,2006MNRAS.367..186E,2007A&A...469..941V,2008A&A...488L..37C,2009A&A...506.1249P},
VH1 \citep{2007ApJ...667..226D} and a code due to
\citet{1995CoPhC..89...29Y} and \citet{1996A&A...315..555Y}
\citep{2003ApJ...594..888F,2006ApJ...638..262F}. Some authors choose
spherical coordinates for their two-dimensional models
\citep[e.g.][]{1996A&A...305..229G,1996A&A...316..133G,2005A&A...444..837V,2007ApJ...667..226D,
  2007A&A...469..941V}, while others use cylindrical $r-z$ coordinates,
as we do \citep[e.g., ][]{2003ApJ...594..888F,2006ApJ...638..262F}. Each
coordinate system has its advantages and disadvantages. In 2D
spherical coordinates, the cell sizes increase with increasing radius,
while in $r-z$ coordinates it is more difficult to represent a
spherical wind-injection zone.

One key ingredient, which can strongly affect the results of the
numerical simulations, is the inclusion (or not) of radiative transfer
and the modeling of the \ion{H}{2} region that will develop around the
star during the main-sequence and WR stages. Of the previous numerical
work we have mentioned, only
\citet{2003ApJ...594..888F,2006ApJ...638..262F},
\citet{2005A&A...444..837V,2007A&A...469..941V} and
\citet{2008A&A...488L..37C} (for a $12\,M_\odot$ star) include the
effect of ionizing photons on the evolution of the circumstellar
medium. Both \citet{2005A&A...444..837V,2007A&A...469..941V} and
\citet{2008A&A...488L..37C} use a simple Str\"omgren radius approach
to the radiative transfer, whereby gas is either ionized or neutral
depending on whether the ionizing photons reach it or not (it is not
clear whether these authors take collisional ionization into
account). \citet{2003ApJ...594..888F,2006ApJ...638..262F} uses a
ray-tracing method and includes collisional ionization. Our radiative
transfer method (see \S~\ref{sec:general}) allows us to follow gas at
intermediate stages of ionization. For example, once the RSG stage begins,
the ionizing photons do not reach the outer radii of the nebula, but
the recombination timescale is of order $10^{5}/n$~yrs and thus the gas
does not recombine immediately. This has important consequences for
the pressure across the bubble. In our results, even once the ionizing
photons have switched off, we find roughly constant pressure right
across the bubble, from the inner edge of the hot, shocked wind
through the recombining \ion{H}{2} region and the outer neutral shell,
since acoustic waves smooth out the pressure distribution as the
\ion{H}{2} region slowly recombines. However, the results presented by
\citet[][Figure~4]{2004RMxAC..22..136V} and
\citet[][Figure~1]{2005A&A...444..837V} show a drop in pressure in the
ex-\ion{H}{2} region where the number of particles has suddenly been
reduced by half due to instantaneous recombination. This will lead to
the formation of spurious shells of material as shock waves propagate
into the depressurized region from both sides.

The expansion of a stellar wind bubble into an \ion{H}{2} region
cannot simply be modeled by a bubble expanding into a uniform medium
at $10^4$~K
\citep{1996A&A...305..229G,1996A&A...316..133G,2005ApJ...631..435R},
since the density of the photoionized gas in the \ion{H}{2} region
becomes lower with time as the expansion proceeds. That is, not only
is the stellar wind bubble expanding, but the \ion{H}{2} region is
expanding as well.

In our 2D models we see the Rayleigh-Taylor and linear thin-shell
instabilities reported by other authors
\citep{1996A&A...305..229G,1996A&A...316..133G,2003ApJ...594..888F,2006ApJ...638..262F,2007ApJ...667..226D}
for the wind-wind interaction stage. The details very much depend on
the stellar evolution model being used. Because we do not model the
main-sequence stage in 2D we do not see the \ion{H}{2} shadowing
instability reported by
\citet{2003ApJ...594..888F,2006ApJ...638..262F} and
\citet{2006ApJS..165..283A}.

We remark that the initial uniform ambient medium density of our
models is $n_0 = 100$~cm$^{-3}$, which is higher than that used by all
the previous authors we have mentioned. The justification for our
value is that a massive star will live a large fraction of its
lifetime in the vicinity of the molecular cloud where it was born. As
the \ion{H}{2} region around the star evolves, the density of the
photoionized gas falls with time. For example, electron densities in
the Orion nebula (age about one million years) range from
$100$~cm$^{-3}$ \citep{1993A&AS...98..137F} to $>1000$~cm$^{-3}$
\citep{1959ApJ...129...26O}, while those in the Carina nebula (which
harbors several high-mass main-sequence stars and post-main-sequence
stars) vary between $\sim 30$~cm$^{-3}$ and $\sim 350$~cm$^{-3}$
\citep{2002A&A...382..610M}. By the end of the main sequence, the
density in our photoionized gas has fallen to $<10$~cm$^{-3}$. The
stellar wind bubble evolves within this changing environment, with the
position of the inner wind shock regulated by the pressure in the
\ion{H}{2} region. Models in which the external medium is initially
less dense
\citep[e.g., ][]{2003ApJ...594..888F,2006ApJ...638..262F,2005A&A...444..837V,2007A&A...469..941V}
produce larger, more diffuse \ion{H}{2} regions and consequently
larger stellar wind bubbles than in our models. Simulations with a
lower density medium and without radiative transfer produce large
stellar wind bubbles with thin outer shells of swept-up material
\citep{2007ApJ...667..226D} instead of broad shells of photoionized
gas.

\subsection{Energy evolution}
\label{subsec:energy}
\begin{figure*}[p]
  \includegraphics[width=0.5\linewidth]{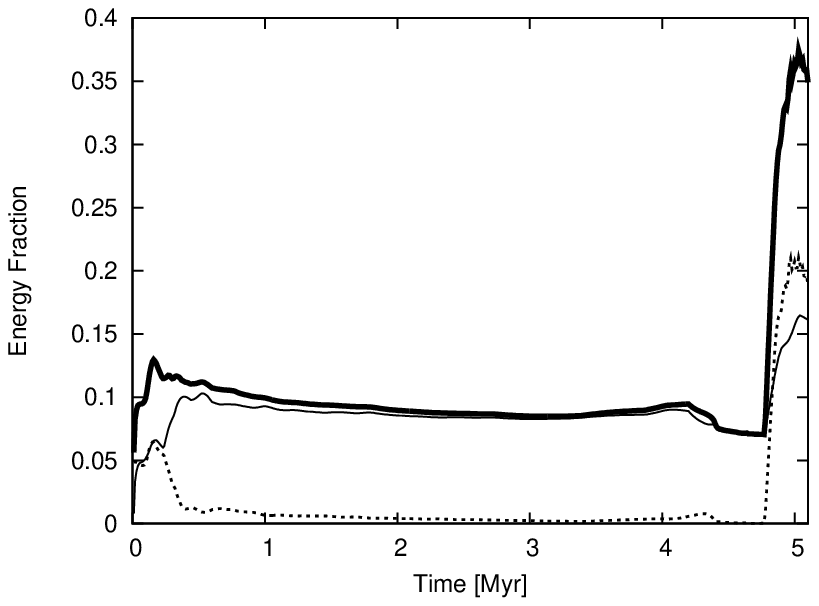}~
  \includegraphics[width=0.5\linewidth]{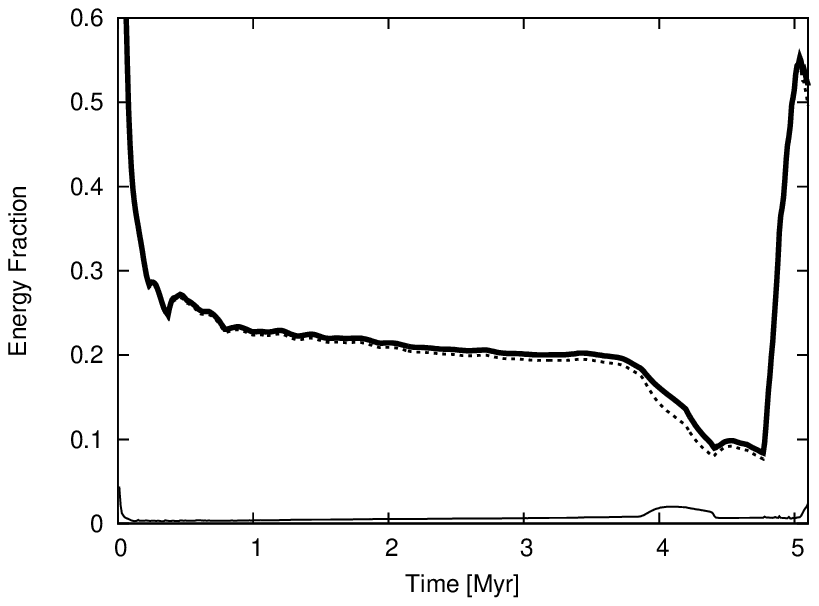}\\
  \includegraphics[width=0.5\linewidth]{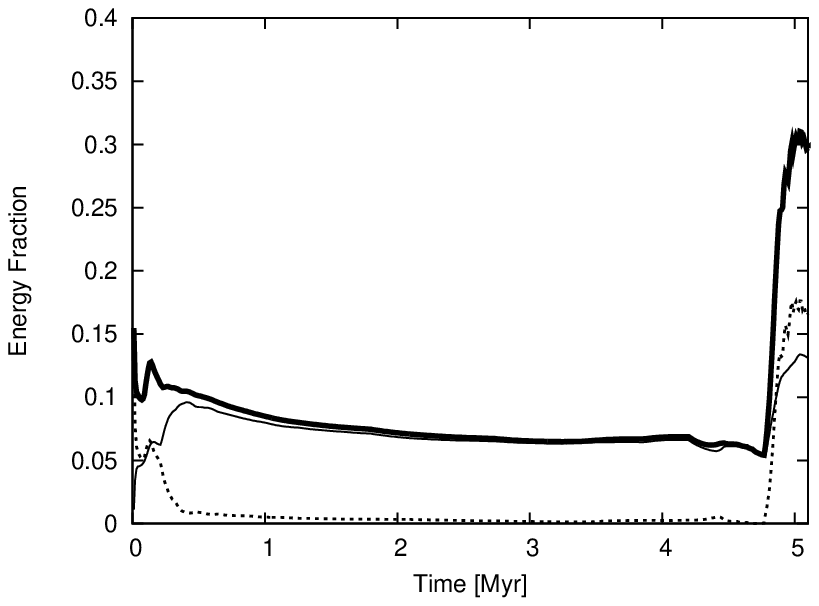}~
  \includegraphics[width=0.5\linewidth]{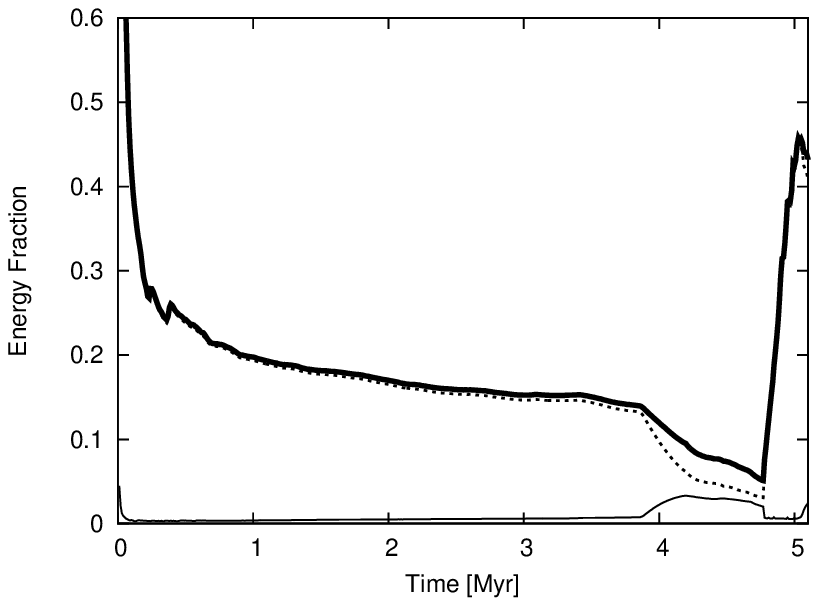}\\
\caption{Kinetic and thermal energy as fractions of the
  time-integrated wind
  mechanical energy as a function of time for the $40\,M_\odot$ STARS
  models. \textit{Left panels}: Gas kinetic energy as a fraction of
  total wind mechanical energy. \textit{Right panels}: Gas thermal
  energy. \textit{Solid thick lines}: Total gas kinetic (thermal)
  energy. \textit{Solid thin lines}: Neutral gas kinetic (thermal)
  energy. \textit{Dotted lines}: Ionized gas kinetic (thermal)
  energy. \textit{Top panels:} simulations without thermal conduction,
  \textit{bottom panels:} simulations with thermal conduction.}
\label{fig:energy}
\end{figure*}

In Figure~\ref{fig:energy} we show the evolution of neutral and
ionized gas kinetic and thermal energy fractions with respect to the
total integrated stellar wind mechanical energy, $E_\mathrm{w}(t) =
\int \dot{M}_\mathrm{w} V_\mathrm{w}^2/2 \, dt$, resulting from the
numerical simulations from the $40\,M_{\odot}$ STARS models with and
without thermal conduction. We do not include the energy of the
stellar ionizing photons in this calculation because nearly all of
the interstellar UV energy will be radiated away in forbidden-line
processes (see \citealp{1997pism.book.....D}, Chapter 7). At the
beginning of the main-sequence stage, the energy fraction profiles
show some variations due to the formation and expansion of the initial
\ion{H}{2} region around the star \citep{2003ApJ...594..888F,
  2007dmsf.book..183A}\footnote{Note that
  \citet{2003ApJ...594..888F,2006ApJ...638..262F} do include the
  stellar UV energy in their calculations, which leads to lower gas
  thermal and kinetic energy fractions by more than an order of
  magnitude.}. For most of the main-sequence stage, the energy
fractions of both kinetic and thermal energy are roughly constant
\citep{2003ApJ...594..888F,2006ApJ...638..262F}, with the thermal
energy being lower in the model with conduction due to enhanced
cooling in the hot bubble because the material evaporated from the dense
outer shell increases the density, and hence the cooling rate, in the
outer part of the bubble. It is evident that the outer neutral shell
dominates the kinetic energy fraction of these models, while the hot,
shocked bubble dominates the thermal energy.

At the end of the main-sequence stage, the neutral
shell remains coasting along but the hot bubble suffers a drastic drop in
pressure when the (subsonic with respect to the hot bubble) RSG wind
starts. This can clearly be seen in Figure~\ref{fig:energy} where,
starting at about $4\times 10^6$~yrs, the thermal energy fraction dips
while the kinetic energy fraction remains roughly constant. With the
onset of the fast WR wind, the thermal energy increases enormously
once the high velocity gas shocks and thermalizes into a new
pressure-driven hot bubble. In contrast, the kinetic energy becomes
shared between the ionized and neutral components. This is due to
photoionization of part of the outer neutral shell and the
repressurized hot bubble rejuvenating the expansion of the outer
swept-up material.

Between them, the thermal and kinetic energy fractions add up to
substantially less than 1, particularly during the main-sequence
stage. The remaining energy is lost as radiation both from the hot
bubble and from the \ion{H}{2} region and neutral shell which
surround the wind bubble.

\section{Comparision with observations}
\label{sec:comparison}
\begin{deluxetable}{lcccccccc}
\tablewidth{0pt} 
\tablecaption{Stellar and nebular parameters from selected WR wind-blown
  bubbles} \tablehead{ \multicolumn{1}{l}{WR
    bubble} & 
  \multicolumn{1}{c}{Distance\tablenotemark{a}} & 
  \multicolumn{1}{c}{Radius\tablenotemark{b}}   & 
  \multicolumn{1}{l}{$\log M_{\odot}$\tablenotemark{c}}     & 
  \multicolumn{1}{c}{$V_{\infty}$\tablenotemark{c}}        & 
  \multicolumn{1}{c}{$\log$~N/O\tablenotemark{d}}          & 
  \multicolumn{1}{c}{$N_{\mathrm{H}}$\tablenotemark{e}}     & 
  \multicolumn{1}{c}{$T$\tablenotemark{e}}                & 
  \multicolumn{1}{c}{$L_{\mathrm{X}}$\tablenotemark{e}} \\ 
  \multicolumn{1}{l}{}                                     & 
  \multicolumn{1}{c}{kpc}                                  & 
  \multicolumn{1}{c}{arcmin}                               & 
  \multicolumn{1}{l}{$M_{\odot}$~yr$^{-1}$}                  & 
  \multicolumn{1}{c}{km~s$^{-1}$}                           & 
  \multicolumn{1}{c}{}                                     & 
  \multicolumn{1}{c}{$10^{21}$~cm$^{-2}$}                    & 
  \multicolumn{1}{c}{K}                         & 
  \multicolumn{1}{c}{$10^{34}$~erg~s$^{-1}$} } 
\startdata 
 S\,308    & 1.5$\pm$0.3  & 19.4   & -4.12 & 1720    & 0.22    & 1.1  & $1.1\times10^{6}$ & 1.2  \\ 
 NGC\,6888 & 1.8$\pm$0.5  & 8.6    & -4.02 & 1605    & 0.27    & 3.1  & $1.3\times10^{6}$ & 3    \\ 
 RCW\,58\tablenotemark{f}  & 3.0   & 4.9  & \nodata & 975 & -0.30 & \nodata & \nodata & \nodata \\ \enddata
\tablenotetext{a}{Distances were obtained from:
  \citet{2003IAUS..209..415C} -- WR\,6~(S\,308),
  \citet{1986ApJ...303..239A} -- WR\,136~(NGC\,6888), and
  \citet{1982ApJ...254..578C} -- WR\,40~(RCW\,58).}
\tablenotetext{b}{\citet{2000AJ....120.2670G}. For NGC\,6888 this is the semi-major axis.}
\tablenotetext{c}{\citet{1990ApJ...361..607P}.}
\tablenotetext{d}{\citet{1992A&A...259..629E}.}
\tablenotetext{e}{Obtained from soft X-ray observations
  \citep{2003ApJ...599.1189C,2005ApJ...633..248W}.}
\tablenotetext{f}{RCW\,58 has not been detected in X-ray observations
  \citep{2005A&A...429..685G}.}
\label{tab:nebulae}
\end{deluxetable}

\subsection{Optical observations}
\label{subsec:optical}
\begin{figure}[p]
\includegraphics[width=\linewidth]{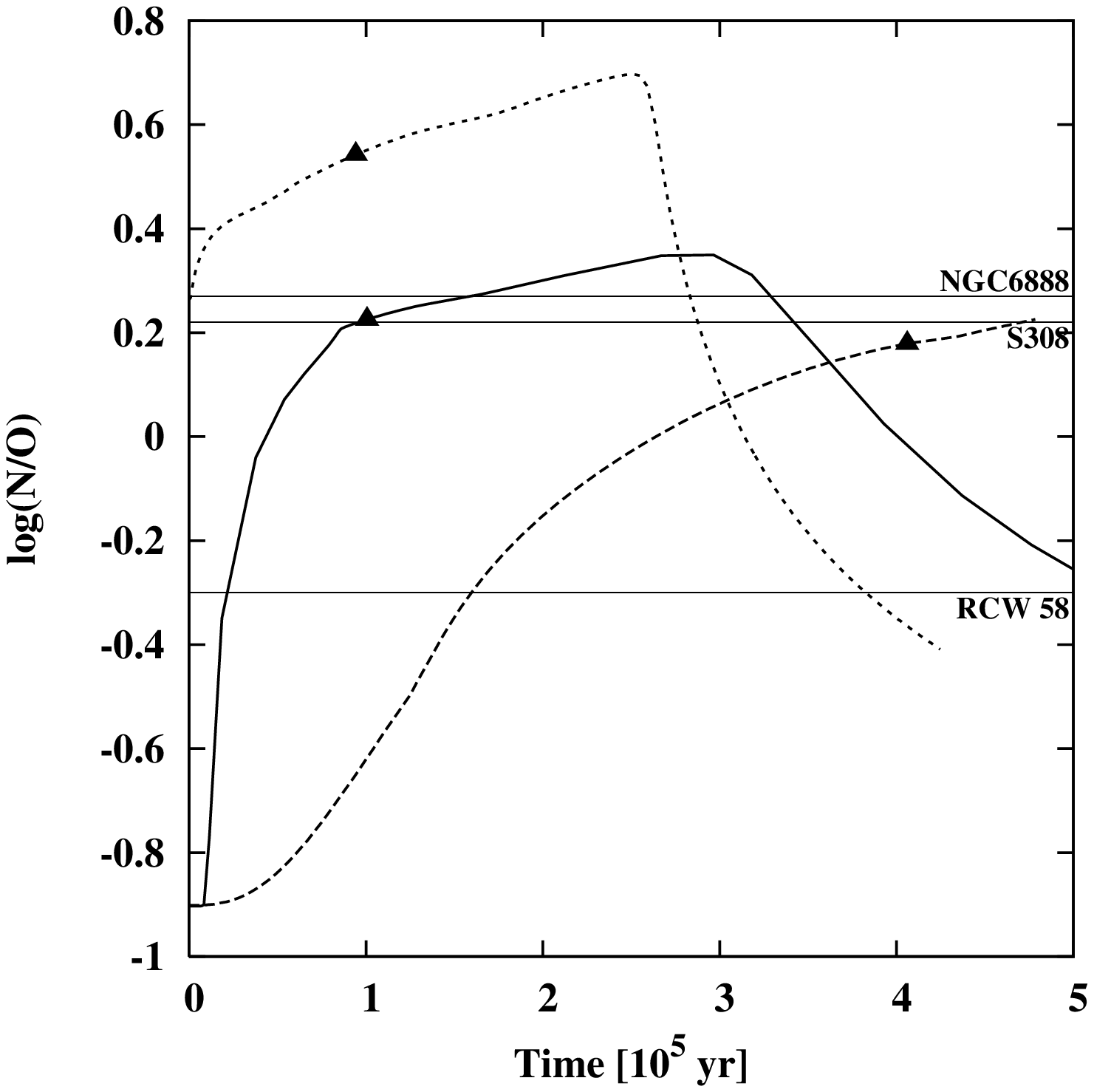}\\
\includegraphics[width=\linewidth]{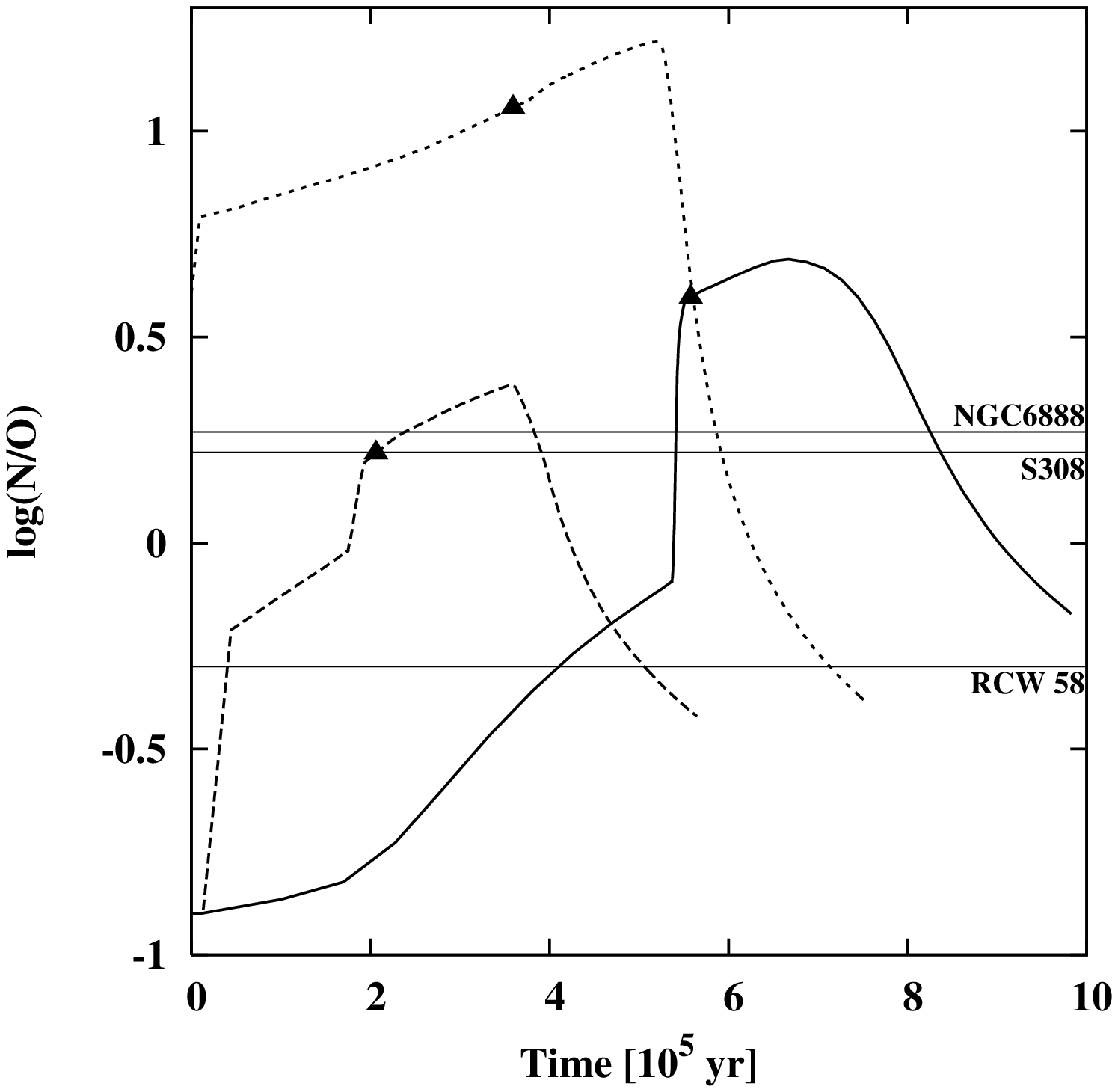}
\caption{Nitrogen-to-oxygen abundance ratio in the circumstellar
  nebula as a function of time since the end of the main-sequence
  stage for the stellar evolutionary models used in this
  work. \textit{Top panel}: $40\, M_{\odot}$ models, \textit{bottom
    panel}: $60\, M_{\odot}$ models. In each panel the solid line is
  for the STARS stellar evolution models, the dashed line is for the
  MM2003 models without rotation, while the dotted line is for the
  MM2003 models with stellar rotation. Also shown are thin horizontal
  lines corresponding to the spectroscopically determined values for
  the three Wolf-Rayet ring nebulae S\,308, NGC\,6888 and RCW\,58 from
  \protect\citet{1992A&A...259..629E}. The triangles on each line mark
  the point where each stellar model enters the Wolf-Rayet phase.}
\label{fig:abun}
\end{figure}

Our simulated WR wind bubbles bear morphological similarity to optical
images of observed WR nebulae such as RCW\,58 and S\,308
\citep{2000AJ....120.2670G}. Instabilities in the swept-up RSG
material shell lead to the formation of dense, cold clumps, which can
be compared to the H$\alpha$ knots, for example, in images of
RCW\,58. \citet{2000AJ....120.2670G} divide a sample of WR bubbles
into morphological types depending on the offset of the H$\alpha$ and
[\ion{O}{3}] emission. Type I has no measurable offset between the
H$\alpha$ and [\ion{O}{3}] fronts, Type II morphology has an H$\alpha$
front trailing close behind an [\ion{O}{3}] front of similar shape,
while for Type III the H$\alpha$ trails far behind a faint [\ion{O}{3}]
front and can differ appreciably in shape. Type IV morphology has a
faint [\ion{O}{3}] front with no H$\alpha$ counterpart. Under this
classification system, RCW\,58 is Type III, while S\,308 is Type
II. While the H$\alpha$ is a recombination line produced in gas with
temperatures $\sim 10^4$~K, [\ion{O}{3}] comes from hotter gas $>
10^4$~K due to either photoionization by harder UV photons or
collisional excitation behind shock waves. In our simulations, the
dense clumps that result from the instabilities which break up the
swept-up shell are cold ($T < 10^4$~K) and are photoionized on the
side closest to the star. These clumps will therefore be sources of
H$\alpha$ emission and, moreover, lag behind the remainder of the
shocked shell, which has temperatures $T > 10^4$~K and could be a
source of [\ion{O}{3}] emission. The images presented in section
\ref{subsec:2dresults} were chosen to facilitate comparison with the
ring nebula S\,308, which has an optical radius of $\sim 8$~pc.

The initial abundances in the stellar evolution models from MM2003 and
STARS are Solar \citep{1989GeCoA..53..197A} but towards the end of the
RSG phase the stellar surface abundances begin to change, becoming
enriched with nitrogen in particular. This surface material is
expelled from the star in the form of stellar winds during the period
of enhanced mass loss and can be expected to enrich the circumstellar
nebula. Indeed, spectroscopic abundance determinations of Wolf-Rayet
ring nebulae \citep{1992A&A...259..629E} show that the
nitrogen-to-oxygen ratio, $\log(\mathrm{N}/\mathrm{O})$, can even
become positive 
in this circumstellar medium, while the Solar value is
$\log(\mathrm{N}/\mathrm{O}) = -0.9$ (see Table~\ref{tab:nebulae}).

In Figure~\ref{fig:abun} we
plot the nitrogen-to-oxygen abundance ratio for stellar material
expelled since the onset of enhanced mass loss for the different
stellar evolution models used in this paper, using the model stellar
surface abundances as being representative of the stellar wind
chemical composition. These are mass-weighted
average values over this time period and do not include material
expelled during the main-sequence stage. The main-sequence stellar
wind material is spread throughout an extended, diffuse bubble of
radius up to some tens of parsecs, while the RSG and Wolf-Rayet wind material
will form a circumstellar nebula around the Wolf-Rayet star. We
compare our calculated abundance ratios to those determined for the
three Wolf-Rayet ring nebulae S\,308, NGC\,6888 and RCW\,58 by
\citet{1992A&A...259..629E} and listed in Table~\ref{tab:nebulae}. It
is clear from these figures that 
rotation significantly affects the abundances at the stellar surface,
and hence in the stellar wind, throughout the main-sequence stage so
that by the start of the enhanced mass-loss period the
nitrogen-to-oxygen ratio is already positive. For the evolution models
without rotation, the abundances at the start of the RSG stage are
still Solar, since there has been no mixing with enriched material from
the stellar core up to this point. 

From Figure~\ref{fig:abun} we see that the abundance ratio for RCW\,58
can only come from material expelled in the early stages of enhanced
mass loss. The S\,308 and NGC\,6888 values could be reproduced by
various models, except that of the MM2003 $40\,M_{\odot}$ without
rotation. For the STARS models, however, these abundances are
achieved only once the Wolf-Rayet stage has been entered and WR wind
material has been added to the circumstellar nebula. For the MM2003
models with rotation, the nitrogen-to-oxygen abundance ratio in the
circumstellar medium is predicted to be far higher than the observed
values throughout the RSG and WR stages. These findings are broadly
consistent with the classification of the central star of RCW\,58 as
WNL, while the central stars of S\,308 and NGC\,6888 are WNE. 

\subsection{X-ray observations}
\label{subsec:xray}
\begin{figure}[p]
\includegraphics[width=\linewidth]{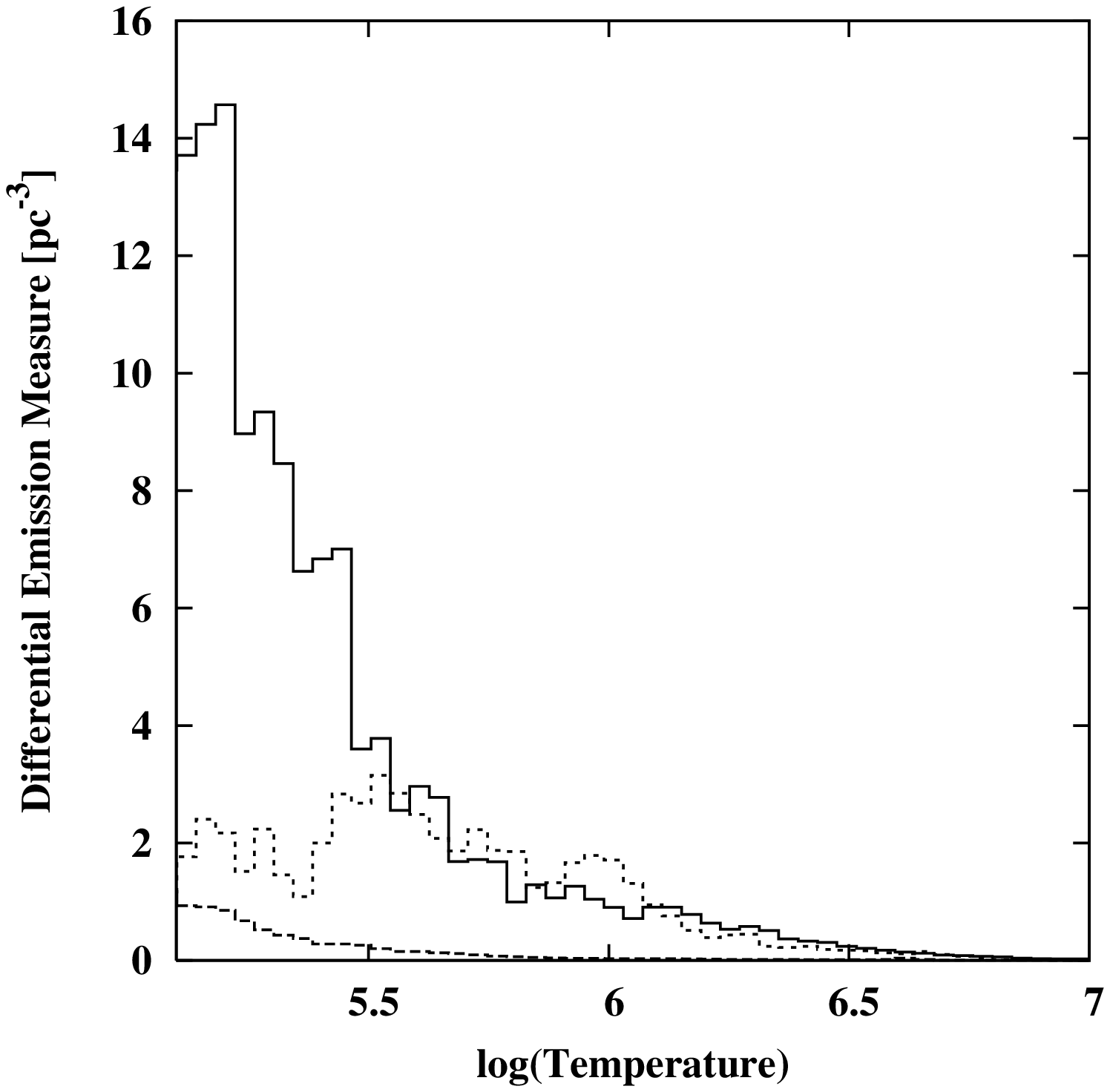}\\
\includegraphics[width=\linewidth]{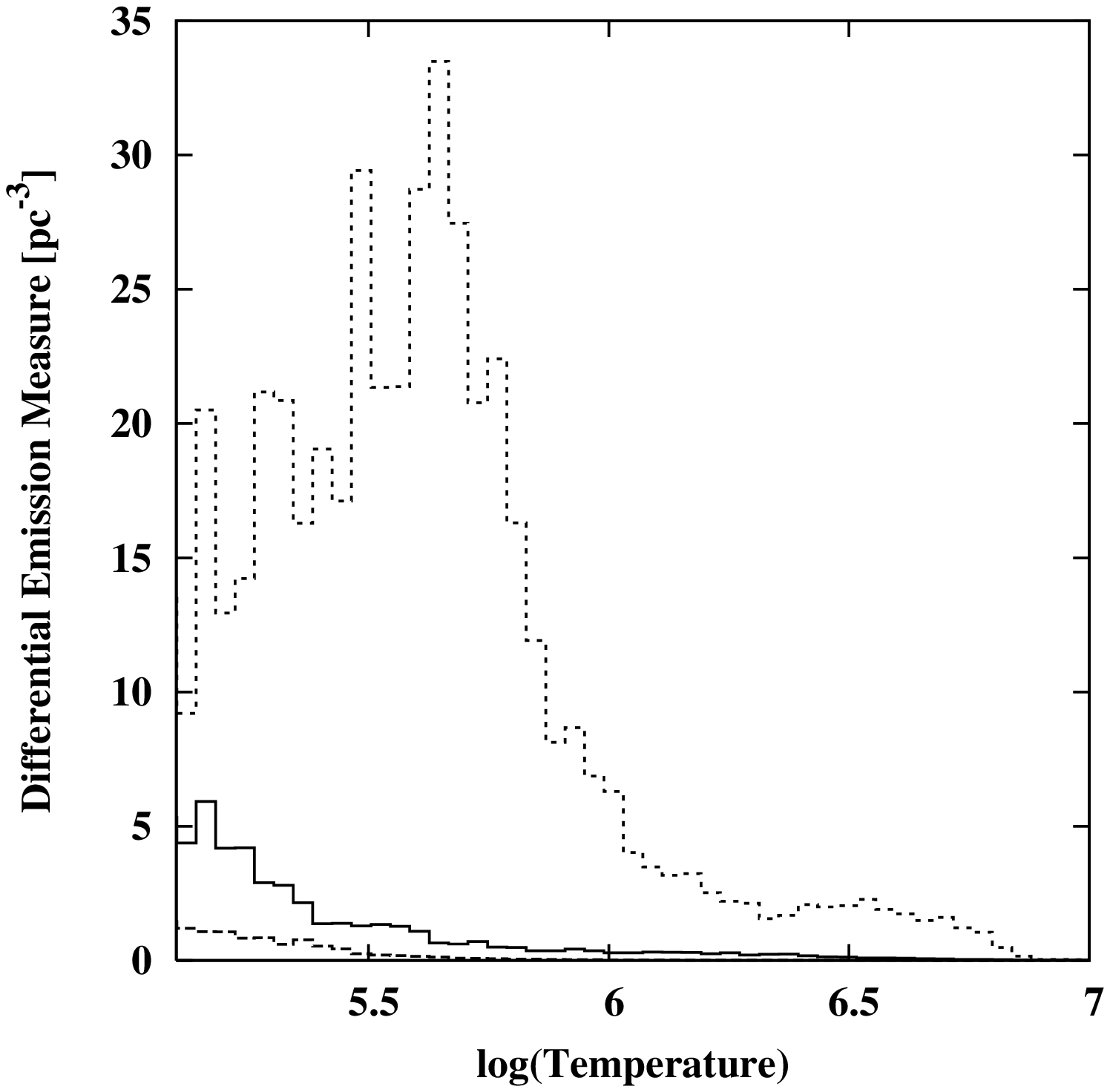}
\caption{Differential emission measure for the models presented in
  Figs.~\protect\ref{fig:2D40_snapshots1}, \ref{fig:2D40_snapshots2},
  and \ref{fig:2D40_snapshots3} when the wind-wind interaction is at
  8~pc. \textit{Top panel}: models without conduction, \textit{bottom
    panel}: models with conduction at similar points in their
  evolution. In each panel, \textit{dashed line} --- STARS model,
  \textit{solid line}: MM2003 without stellar rotation, \textit{dotted
    line}: MM2003 with rotation. The temperature bin width is 0.04~dex
  in each case.}
\label{fig:dem}
\end{figure}

There are around 200 WR stars in our Galaxy, and only two of them have
nebulae detected in diffuse X-rays: S\,308 and NGC\,6888
\citep{1988Natur.332..518B,1994A&A...286..219W,1999A&A...343..599W,2003ApJ...599.1189C,2011ApJ...728..135Z}.
The X-ray emission is very soft and the diffuse emission from S\,308
can be fit with a one temperature component model with ($T= 1.1\times
10^6$~K), enhanced nitrogen abundance and X-ray luminosity
$L_\mathrm{X} \simeq 1.2 \times 10^{34}$~erg~s$^{-1}$ in the
0.25--1.5~keV energy band, assuming a distance of 1.8~kpc
\citep{2003ApJ...599.1189C}.  NGC\,6888 \textit{ASCA} and
\textit{ROSAT} X-ray observations are best fit by a two-component
model with the dominant component at $1.3\times 10^6$~K and the weaker
component at $8 \times 10^6$~K and a total X-ray luminosity
$L_\mathrm{X} \simeq 3\times 10^{34}$~erg~s$^{-1}$ in the energy range
0.4--2.4~keV, assuming a distance of 1.8~kpc
\citep{{1994A&A...286..219W},{1998LNP...506..425W},{2005ApJ...633..248W}}.
The recent \textit{SUZAKU} observations of NGC\,6888
\citep{2011ApJ...728..135Z} confirm that ninety percent of the
observed X-ray emission is in the soft 0.3--1.5~keV energy range. In
both bubbles, the X-ray emission is internal to the optical
[\ion{O}{3}] emission. Such soft X-ray emission indicates that the
observations will be strongly affected by the absorbing column
density, $N_\mathrm{H}$, along the line of sight. This could be a
reason for the non-detection of other wind-blown WR bubbles such as
RCW\,58, since these objects are found in the plane of the Galaxy
where $N_\mathrm{H}$ will be greatest. The two bubbles with detected
X-ray emission are relatively close by whereas WR\,40, the central
star of RCW\,58, is more than twice as distant (see Table~\ref{tab:nebulae}).

\subsubsection{X-ray simulations}
\label{subsubsec:xray_simulations}
From our numerical simulations, both in one and two dimensions, we can
calculate the expected X-ray luminosity for any energy range at any
time in the simulation. To do this efficiently, we use the temperature
and density in each cell of the hydrodynamic simulation to construct
the array of Differential Emission Measures (DEM), i.e. the emission
measure (E.M.  = $\int{n_{e} n_{\mathrm{i}} dV}$) that corresponds to
each bin of a preassigned temperature array for a given numerical
simulation. Note that we mask out the unshocked thermal wind in this
procedure. The total X-ray spectrum is then found by
summing the spectra of each temperature bin and the X-ray luminosity
is then found by summing over the required energy range. This method
is much more 
efficient than summing the spectrum (or luminosity) of each individual
hydrodynamic cell, since we find that 100 temperature bins are
perfectly adequate, while the numerical simulations consist of
$\sim6000$~cells for the one-dimensional models and $>10^5$~cells for
the two-dimensional models.  In Figure~\ref{fig:dem} we show the
corresponding DEM histograms for the $40\,M_{\odot}$ results presented
in Figures.~\protect\ref{fig:2D40_snapshots1},
\ref{fig:2D40_snapshots2}, and \ref{fig:2D40_snapshots3} when the
wind-wind interaction is at 8~pc, together with their counterparts
that include thermal conduction.

The first thing to notice about Figure~\ref{fig:dem} is that most of
the mass has temperatures $T < 10^6$~K. This means that the spectra will
be very soft and the absorbing column of neutral hydrogen towards such
objects will play a very important role in what is observed. We see no
systematic difference between models without thermal conduction and
their equivalent models with conduction beyond the differences that
can be attributed to slightly different volumes of emitting gas. The
main difference we see is between models with complete, if distorted,
shells (those of MM2003) and those where the shell has already broken
up into clumps (those of STARS). The models with clumps have a much
smaller differential emission measure for the hot gas ($T >
10^{5.5}$~K) than those where the whole shell is being shocked,
because the hot gas in these cases is more diffuse and can flow around
the clumps and out of the computational domain.

Once we have the DEM arrays, the X-ray spectra are calculated assuming
collisional ionization equilibrium (CIE) and the ion fractions of
\citet{1998A&AS..133..403M}. The line spectrum uses the APED database
of CIE line intensities \citep{2001ApJ...556L..91S}, while the
continuum is calculated with our own code, taking into account
free-free, free-bound and two-photon emission. The chemical abundances
can be varied according to the requirements of the models.

Typical electron densities in the X-ray emitting gas for our
simulations are $n_e \sim 0.1$~cm$^{-3}$ and temperatures are in the
range 1--$4\times10^6$~K (see Fig.~\ref{fig:2D40_snapshots1} through
\ref{fig:2D60_snapshots3} and \ref{fig:dem}). For these parameters, we can
assess the closeness of the different elements (C, N, O, Ne, etc.) to
ionization equilibrium using Figure~1 of
\citet{2010ApJ...718..583S}. At $10^6$~K, for nitrogen \citep[the main
  X-ray spectral characteristic of S\,308, ][]{2003ApJ...599.1189C} to
be in ionization equilibrium requires $n_e t > 8\times
10^{11}$~cm$^{-3}$s$^{-1}$, whereas at $4\times 10^6$~K, ionization
equilibrium for nitrogen requires $n_e t > 1\times
10^{11}$~cm$^{-3}$s$^{-1}$. For our typical electron density, these
correspond to timescales of 253,000~yrs and 32,000~yrs
respectively. Thus, it could be argued that a non-equilibrium
ionization treatment is required for the X-rays in the shocked stellar
wind interaction region, but this is beyond the scope of this paper.

\begin{figure}[p]
\includegraphics[width=\linewidth]{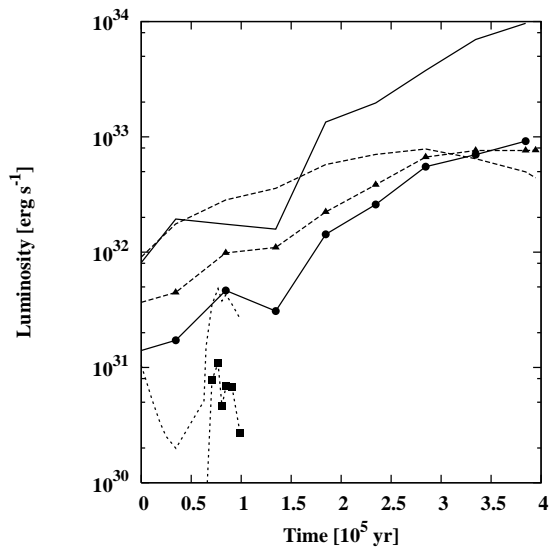}
\caption{Variation of X-ray luminosity with time in the WR phase for
  the STARS $40 
  M_\odot$ model for 1D and 2D
  simulations, assuming Solar abundances. \textit{Solid line}: 1D simulation with thermal
  conduction, \textit{dashed line}: 1D simulation without conduction,
  \textit{dotted line}: 2D simulation without conduction. In each case
  the simple lines represent the soft-band emission (0.25--1.5~keV)
  while the lines with symbols represent the hard X-ray emission (1.5--7~keV).}
\label{fig:lum1D2}
\end{figure}
\begin{figure}[p]
\includegraphics[width=\linewidth]{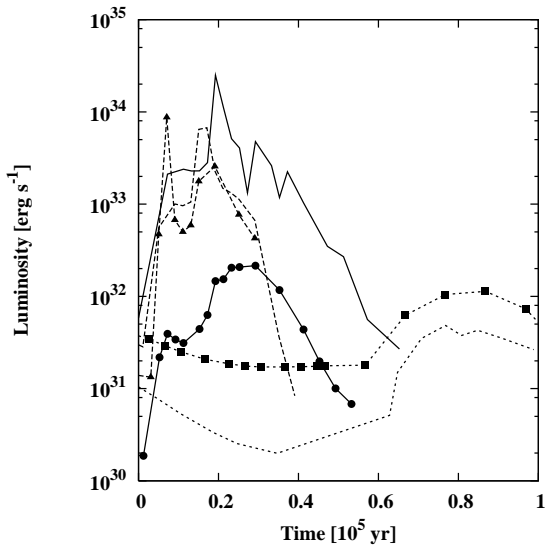}
\caption{Variation of soft-band (0.25--1.5~keV) X-ray luminosity with
  time in the WR phase for three sets of $40\,M_{\odot}$ 2D
  models, assuming Solar abundances. \textit{Solid 
  line}: MM2003 with no rotation, \textit{dashed line}: MM2003 with
  rotation, \textit{dotted line}: STARS. In each case the simple lines
  represent the model without thermal conduction, while the lines with
  symbols represent the models with conduction.}
\label{fig:lum2D3}
\end{figure}
In Figures~\ref{fig:lum1D2} and \ref{fig:lum2D3} we show the X-ray
luminosity as a function of time in the WR phase for the $40\,
M_{\odot}$ models, assuming Solar abundances
\citep{1989GeCoA..53..197A}. Figure~\ref{fig:lum1D2} plots the
soft-band (0.25--1.5~keV) and hard-band (1.5--7~keV) X-ray emission
for the 1D STARS models with and without thermal conduction, together
with the STARS 2D model emission with thermal conduction. From this
figure we see that the 1D models produce more intense X-ray emission,
by an order of magnitude, than the 2D model. This can be understood by
referring to Figure~\ref{fig:2D40_snapshots1} where we see that the RSG shell
breaks up into clumps and the low-density, hot, shocked WR wind flows
around them. The X-ray emission starts to diminish after
$\sim$~70,000~yrs because the hot gas flows off the 10~pc 2D grid. 

In the 1D models, the WR wind is confined by the shell of RSG
material, which cannot break up because of the geometrical
restriction. The soft X-ray emission of the two 1D models is very
similar until $1.5\times 10^5$~yrs, when the thermal conduction at the
edge of the hot bubble starts to play a dominant role, raising the
density and lowering the temperature of the hot gas here and thereby
increasing the contribution of the soft X-rays. In the 1D
model without conduction, the gradual increase in the stellar wind
velocity raises the temperature of the hot, shocked gas and finally
results in the hard X-rays being the dominant component. Note that,
for the 1D models, we are summing the X-ray contribution from the
whole computational grid (except the thermal wind injection region):
the contributions from the diffuse, hot, shocked main-sequence wind to
the soft and hard X-rays are, respectively, $2.1 \times
10^{31}$~erg~s$^{-1}$ and $2.9 \times 10^{31}$~erg~s$^{-1}$ for the
model without conduction and $1.6 \times 10^{31}$~erg~s$^{-1}$ and
$2.1 \times 10^{31}$~erg~s$^{-1}$ for the model with thermal
conduction (see Fig.~\ref{fig:1D40}).

Figure~\ref{fig:lum2D3} shows the complete time evolution of the soft
X-ray emission from all of the $40\, M_{\odot}$ 2D models, assuming
Solar abundances: MM2003 with and without stellar rotation, and STARS,
both with and without thermal conduction. Unfortunately, in all cases,
the hot gas moves off the 2D computational grid (radius 10~pc) after a
few tens of thousands of years, and this is what causes the reduction
in the soft X-ray luminosity. However, we comment that this is a
larger size scale than either of the X-ray observed WR wind
bubbles. Our results are very varied. The STARS model, in which, as
commented above, the shell of RSG material rapidly breaks up into
clumps, has the lowest soft X-ray luminosity by an order of
magnitude. In the MM2003 models, although the swept-up shells suffer
instabilities, they do not break up into clumps before they move off
the grid. The differences between the models with and without thermal
conduction should be attributed mainly to differences in the position
(and volume) of the wind interaction region due to the effect of
thermal conduction on the pressure in the main-sequence hot bubble, and
only partly due to the local effect of thermal conduction on the
interaction region between the hot WR wind and the swept-up RSG
material (see Fig.~\ref{fig:2D40_2}). Only three of our models attain
X-ray luminosities above $10^{33}$~erg~s$^{-1}$: the MM2003 model
without stellar rotation for the case without conduction, and the
MM2003 models with stellar rotation both with and without
conduction. Recall that the estimated total 0.2--1.5~keV band
luminosity of S\,308 is $1.2 \times 10^{34}$~erg~s$^{-1}$ (see
Table~\ref{tab:nebulae}) from analysis of the NW quadrant of this
object \citep{2003ApJ...599.1189C}, although a recent study of the
full set of observations of S\,308 suggests that the total luminosity
is actually $\sim5\times10^{33}$~erg~s$^{-1}$ (see our companion paper
Toal\'{a} et al., in preparation).

\section{Discussion}  
\label{sec:discuss}
\subsection{Comments on the stellar evolution models and their
  consequences for the circumstellar medium}
\label{subsec:comevol}
\begin{deluxetable}{lrrrrrrrr}
\tablewidth{0pt}
\tablecaption{Mass loss and duration of yellow supergiant and red
  supergiant stages}
\tablehead{
\colhead{Model} & \colhead{$M_\mathrm{FMS}$\tablenotemark{a}} & \colhead{$M_\mathrm{IWR}$\tablenotemark{b}} & \colhead{$t_\mathrm{YR}$\tablenotemark{c}} &\colhead{$\Delta M_\mathrm{YR}$\tablenotemark{d}} & \colhead{$t_\mathrm{YB}$\tablenotemark{e}} & \colhead{$\Delta M_\mathrm{YB}$\tablenotemark{f}} & \colhead{$t_\mathrm{RG}$\tablenotemark{g}} & \colhead{$\Delta M_\mathrm{RG}$\tablenotemark{h}}\\ 
\colhead{}      & \colhead{$M_{\odot}$} & \colhead{$M_{\odot}$} &
\colhead{yr}  & \colhead{$M_{\odot}$}  & \colhead{yr} &
\colhead{$M_{\odot}$} & \colhead{yr} & \colhead{$M_{\odot}$}
}
\startdata
MM2003&&&&&&&\\
$40\, M_{\odot}$ N\tablenotemark{i} & 36.5 & 15.7 & 12100 &  0.765  & 391500 &  18.40 & \nodata & \nodata \\
$40\, M_{\odot}$ R & 32.8 & 22.2 &  3420 &  0.335  &  54200 &   8.00 & \nodata & \nodata \\
$60\, M_{\odot}$ N & 51.5 & 26.7 &  4410 &  1.050  &  10300 &   3.21 & \nodata & \nodata \\
$60\, M_{\odot}$ R & 44.4 & 32.2 & \nodata &  \nodata  &  \nodata &  \nodata & \nodata & \nodata \\
STARS&&&&&&&\\
$40\, M_{\odot}$ N & 35.5 & 16.7 &  785  & 0.042  &   3210  &  0.282 & 77600 & 15.8  \\
$45\, M_{\odot}$ N & 39.2 & 18.6 &  724  & 0.069  &   2690  &  0.455 & 31100 & 16.1  \\
$50\, M_{\odot}$ N & 42.9 & 20.4 &  718  & 0.119  &   3400  &  0.896 & 16400 & 16.1  \\ 
$55\, M_{\odot}$ N & 46.4 & 22.5 &  802  & 0.244  &   6570  &  3.310 &  8320 & 13.3  \\
$60\, M_{\odot}$ N & 48.9 & 23.8 &  809  & 0.325  &   7730  &  4.210 &  6020 & 12.0  \\
\enddata
\tablenotetext{a}{Mass at the end of the main-sequence stage.}
\tablenotetext{b}{Mass at the beginning of the Wolf-Rayet stage.}
\tablenotetext{c}{Time spent as a yellow supergiant heading redwards.}
\tablenotetext{d}{Mass lost as a yellow supergiant heading redwards.}
\tablenotetext{e}{Time spent as a yellow supergiant heading bluewards.}
\tablenotetext{f}{Mass lost as a yellow supergiant heading bluewards.}
\tablenotetext{g}{Time spent as a red giant.}
\tablenotetext{h}{Mass lost as a red giant.}
\tablenotetext{i}{N indicates models without stellar rotation, R
  indicates models with an initial rotation rate of 300~km~s$^{-1}$ at the stellar equator.}
\label{tab:masslost}
\end{deluxetable}
The stellar evolution tracks shown in Figure~\ref{fig:track} indicate
that none of the MM2003 models, 40 or $60\, M_{\odot}$, with or without
stellar rotation actually become red supergiants, since the minimum
stellar surface temperature attained by these models is above 4800~K.
Rather, they become yellow supergiants. The STARS models depicted in
Figure~\ref{fig:track} all become red supergiants. In
Table~\ref{tab:masslost} we list the duration and mass lost in the
yellow and red supergiant phases of evolution, separating the yellow
supergiants into those heading to the red and those heading back to
the blue. We use the temperature range $4800 < T_\mathrm{eff} <
7500$~K to define the yellow supergiants \citep{2009ApJ...703..441D}.

The yellow supergiant stage is a particularly interesting stage of
stellar evolution, since statistical surveys of nearby galaxies
suggest that it should be short lived \citep{2009ApJ...703..441D}. For
the STARS stellar evolution models, this is indeed the case, with the
yellow supergiant phase lasting at most a few thousand
years. Furthermore, the most important mass loss occurs during the red
supergiant phase, when the stellar wind velocities are lowest ($<
30$~km~s$^{-1}$, although mass loss in the hotter, bluer phase becomes
more important as the initial stellar mass increases). For the MM2003
models, on the other hand, there is no red supergiant stage, and the
yellow supergiant stage lasts tens, or even hundreds, of thousands of
years. For the MM2003 $40\,M_{\odot}$ model without rotation, the most
important mass loss occurs as a yellow supergiant heading
bluewards. However, for the other MM2003 models, the most important
mass loss does not even occur in the yellow supergiant phase, but in a
hotter, bluer evolutionary state.

The different periods of mass loss have consequences for the
circumstellar medium around the massive star. Intense mass loss in a
short-lived red supergiant phase will result in a slow-moving dense
shell of material close to the star; mass-loss spread out over
hundreds of thousands of years in a yellow supergiant phase will lead
to an extended region of expanding circumstellar medium. Mass lost
predominantly in hotter, bluer evolutionary stages may form
fast-moving shells of material. When the fast WR wind interacts with
the circumstellar medium, different structures will be produced
depending on where the bulk of the circumstellar mass is and how fast
it is moving \citep{{1996A&A...305..229G},{1996A&A...316..133G}}.

From our results, we see that individual dense clumps are formed when
the fast wind interacts with dense, slow-moving material close to the
star, such as is the case for the STARS 40 and $60\,M_{\odot}$ models
(see Figs.~\ref{fig:2D40_snapshots1} and
\ref{fig:2D60_snapshots1}). This is because the linear thin-shell
instabilities have time to break up the shell before it has traveled
too far from the star. The clump densities and temperatures are such
that the clumps will be photoionized and we would expect H$\alpha$
emission in this case. When the circumstellar medium is moving more
quickly, then the main fast wind-slow wind interaction occurs further
from the star and, due to the lower densities (because of the
geometrical dilution), the instabilities take longer to break up the
swept-up shell. This is what we see in our MM2003 models
(Figs.~\ref{fig:2D40_snapshots2}, \ref{fig:2D40_snapshots3},
\ref{fig:2D60_snapshots2}). In this case, the densities of the clumps
and filaments are lower and their temperatures are higher, so we would
expect neutral material only at the beginning of the interaction, when
the shell is densest.

\subsection{Direct comparision with individual nebulae}
\label{subsec:dircomp}
In Table \ref{tab:nebulae} we present a summary of the observed and
derived properties of three WR bubbles S\,308, NGC\,6888, and RCW\,58 and
their central stars WR\,6, WR\,136, and WR\,40. In this section, we
compare these properties and other details with the results of our
numerical simulations.

\subsubsection{S\,308}
\label{sssec:comps308}
\begin{figure}[p]
\includegraphics[width=\linewidth]{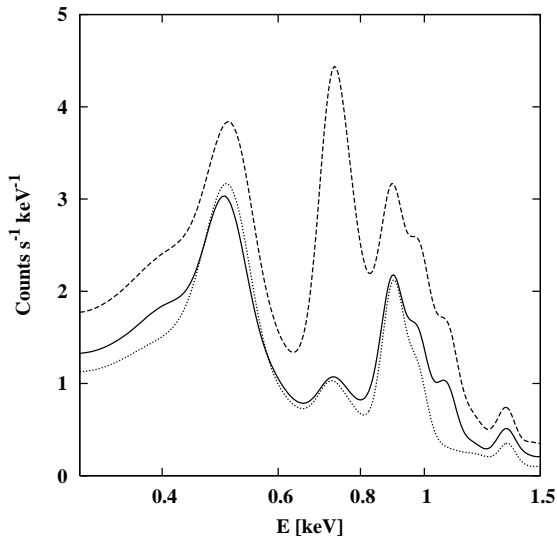}
\caption{X-ray spectra for the $40\,M_{\odot}$ 2D MM2003 model without
  rotation but with thermal conduction, corresponding to the DEM of
  Fig.~\protect\ref{fig:dem}. \textit{Solid line}: Modified,
  nitrogen-rich abundances, \textit{dashed line}: Solar abundances,
  \textit{dotted line}: Single temperature model ($T= 3.6\times10^{6}$~K)
  with nitrogen-rich abundances. All spectra take into account
  absorption by a neutral column density of $N_\mathrm{H} = 1.1 \times
  10^{21}$~cm$^{-2}$ and are convolved with the EPIC/pn instrumental
  response matrices.}
\label{fig:spec}
\end{figure}
The wind-blown bubble S\,308 around the WN4 star WR\,6 is surprisingly
spherical in [\ion{O}{3}] images, apart from a ``blowout'' region to
the northwest. The H$\alpha$ emission from this object shows a
filamentary structure, and \citet{2000AJ....120.2670G} classified it
as a Type~II bubble. The optical radius of S\,308 is $9 \pm 2$~pc at a
distance of $1.5 \pm 0.3$~kpc \citep{2003ApJ...599.1189C}. Our figures
in \S~\ref{subsec:2dresults} were chosen such that the main features
of the swept-up shell are roughly at this distance. It is important to
point out that we are not trying to fit the wind parameters from the
central star to the observed values, we are only choosing hydrodynamic
results that approximately reproduce the morphology using the wind
parameters obtained from the most appropriate of our stellar models.

The stellar evolution tracks shown in Figure~\ref{fig:track} suggest
that the central star, WR\,6, would be well fit by any one of the
MM2003 $40\,M_{\odot}$ models with or without stellar rotation, or by
a STARS $45\,M_{\odot}$ model \citep{2006A&A...457.1015H}. Our
numerical simulations show that the STARS $45\,M_{\odot}$ model (not
shown, but very similar morphologically to the $40\,M_{\odot}$ model)
would produce a Type III bubble (H$\alpha$ clumps well separated from
an [\ion{O}{3}] shell). On the other hand, the MM2003 $40\,M_{\odot}$
model without rotation would appear to produce structures where dense,
photoionized gas exists just inside a warmer, collisionally excited
(i.e., shocked) region, which could be interpreted as a Type II
bubble. Model MM2003 with rotation breaks up the shell too far from
the star and at the point depicted in
Figure~\ref{fig:2D40_snapshots3}\,(when the wind-wind interaction is
at 8~pc), the H$\alpha$ emission and [\ion{O}{3}] emission would be
coincident.

Spectroscopic analysis of S\,308 shows the nebula to have enhanced
nitrogen abundances \citep{1992A&A...259..629E}. Our simple estimation
of the evolution of the nebular abundances with time (see
Fig.~\ref{fig:abun}), made by time-averaging the total ejected mass of
the different elements in the post-main-sequence phase, where the
relative abundances are given by those at the stellar surface at a
given time, suggests that both the STARS model and the MM2003
$40\,M_{\odot}$ model with rotation could reproduce the observed
abundances. However, the MM2003 model without rotation would not
achieve the nebular abundances until the very end of its evolution.

Another test of the models is to compare the predicted X-ray
luminosity and spectrum with observations. As can be seen from
Figure~\ref{fig:dem}, our simulations produce large amounts of gas
with temperatures $T \leq 10^6$~K, whereas observations are generally
fit by a single temperature component, which is used to estimate the
total unabsorbed luminosity. For S\,308, \citet{2003ApJ...599.1189C}
estimate a total X-ray luminosity in the 0.25--1.5~keV band of
$L_\mathrm{X} = 1.2 \pm 0.5 \times 10^{34}$~erg~s$^{-1}$ by
extrapolating their single temperature component ($T \sim 1.1 \times
10^6$~K) results from the northwest quadrant assuming a spherical
bubble. \citet{2003ApJ...599.1189C} find that the spectrum is well fit
using enhanced nitrogen abundances, as determined by
\citet{1992A&A...259..629E} for an absorbing column density of neutral
hydrogen of $N_\mathrm{H} = 1.1 \times 10^{21}$~cm$^{-2}$ (see
Table~\ref{tab:nebulae}). From Figure~\ref{fig:lum2D3} we see that the
STARS $40\,M_{\odot}$ model completely fails to reproduce the observed
X-ray luminosity, while the MM2003 model without rotation achieves
this luminosity when thermal conduction is not included, and the
MM2003 model with rotation both with and without thermal conduction
produces sufficient X-ray luminosity in the soft energy band. We
remark that the results in Figure~\ref{fig:lum2D3} were calculated for
Solar abundances and that if the enhanced abundances are used, the
luminosities are slightly different, depending on the dominant
temperature. For example, the soft-band X-ray luminosity corresponding
to the $40\,M_{\odot}$ MM2003 model with rotation and thermal
conduction, whose differential emission measure is shown in the lower
panel of Figure~\ref{fig:dem}, is $8.2\times 10^{33}$~erg~s$^{-1}$ for
Solar abundances, but $4.8\times 10^{33}$~erg~s$^{-1}$ when we use the
same abundances as \citet{2003ApJ...599.1189C}.

We also calculate the predicted X-ray spectra of our models. We note
that none of our models have been particularly tailored to S\,308 (for
example, we have not tried to match the stellar wind velocity to that
of the central star WR\,6). The main difference between our models and
the reported observations is that the simulations contain a range of
temperatures that contribute more or less equally to the total
spectrum, whereas \citet{2003ApJ...599.1189C} find that the observed
spectrum can be well fit by a single temperature component at $1.1
\times 10^6$~K. In Figure~\ref{fig:spec} we show the simulated spectra
for the $40\,M_{\odot}$ MM2003 model without stellar rotation and with
thermal conduction, where we have considered Solar abundances and also
the modified, nitrogen-rich abundances used by
\citet{2003ApJ...599.1189C}.  The assumed distance is 1.5~kpc and the
absorption column density is $N_\mathrm{H} = 1.1 \times
10^{21}$~cm$^{-2}$, and we also convolved our model spectra with the
instrumental response files appropriate for EPIC/pn on
\textit{XMM-Newton}. From this figure we see that the abundances make
a large difference to the appearance of the spectrum. This is
principally due to the reduced abundance of iron, which is responsible
for the lines at around 0.72~keV. The model spectra are very similar
to a single temperature model with $T \sim 3.6\times 10^6$~K, which is
hotter than that derived for S\,308.  It is possible that
non-equilibrium ionization effects are important in this nebula. The
DEM for this simulation is shown in Figure~\ref{fig:dem} and we see
that most of the gas actually has temperatures below $10^6$~K but the
emission from this gas is completely absorbed by neutral hydrogen along
the line of sight.

\subsubsection{NGC\,6888}
\label{sssec:compngc6888}
NGC\,6888 is the wind-blown bubble around the WN6 star WR\,136,
located at a distance of $\sim 1.8$~kpc
\citep{1988A&A...199..217V}. It is elliptical in shape and was
classified by \citet{2000AJ....120.2670G} as having Type II morphology
along the major axis but Type IV (no H$\alpha$ emission) along the
minor axis. The radius, determined from the [\ion{O}{3}] image, is $\sim
4.5$~pc. Morphologically, this bubble is more like one of the MM2003
models than the STARS models, although the elliptical shape suggests
that the mass loss is non-spherical.

The central star of NGC\,6888 sits well below any of our evolutionary
tracks in the HR diagrams of Figure~\ref{fig:track}. However, the
minimum mass for a star to become a Wolf-Rayet star is $37\,M_{\odot}$
for MM2003 models without rotation but $22\,M_{\odot}$ for MM2003
models with rotation \citep{2003A&A...404..975M}. The
nitrogen-to-oxygen abundance for this nebula 
shows a high level of nitrogen enrichment, which is also consistent
with the stellar evolution models with rotation. 

NGC\,6888 was detected in X-rays by \textit{ROSAT}
\citep{1994A&A...286..219W}, \textit{ASCA}
\citep{2005ApJ...633..248W}, and \textit{SUZAKU}
\citep{2011ApJ...728..135Z}. Like S\,308, it has a limb-brightened
X-ray surface brightness profile and a soft X-ray spectrum, which can
be fit with two-temperature emission models where the principal
component has $T = 1.3\times 10^6$~K and the second component has $T =
8.8\times 10^6$~K for an absorption column density $N_\mathrm{H} = 3.1
\times 10^{21}$~cm$^{-2}$. The total X-ray luminosity in the
0.4--2.4~keV band is $\sim 3.0 \times 10^{34}$~erg~s$^{-1}$, where
Solar metallicities were assumed \citep{2005ApJ...633..248W}. Although
none of our simulations is consistent with the stellar parameters of
this object, we remark that our models produce a range of temperatures
in the X-ray emitting gas (see Fig.~\ref{fig:dem}), which could easily
be interpreted as two main components.

\subsubsection{RCW\,58}
\label{sssec:comprcw58}
RCW\,58 is the wind-blown bubble around the WN8 star WR\,40. The
optical emission is composed primarily of radial H$\alpha$ filaments
and clumps and the much fainter [\ion{O}{3}] emission is smoother and
more extended. \citet{2000AJ....120.2670G} classify this object as
having Type~III morphology. The H$\alpha$ filaments form a roughly
elliptical structure of dimension $6 \times 8$~pc, assuming a distance
of 3~kpc \citep{1982ApJ...254..578C,2001ApJ...548..932H}, which is
roughly half the diameter of S\,308. The morphology is similar to that
given by the STARS models in Figures~\ref{fig:2D40_snapshots1} and
\ref{fig:2D60_snapshots1} when the RSG shell has been totally
disrrupted.

The stellar evolution tracks shown in Figure~\ref{fig:track} suggest
that the central star would be fit by either of the $60\,M_{\odot}$
MM2003 models but the $60\,M_{\odot}$  STARS is underluminous by an
order of magnitude and a higher initial mass model is suggested. The
nitrogen-to-oxygen abundance determined for RCW\,58 indicates material
expelled from the star early on in the post-main-sequence evolution,
since the nitrogen enhancement is only moderate. This could be due to
the measurements being predominantly of clump material.

RCW\,58 was observed in X-rays by \textit{XMM-Newton} but not detected
\citep{2005A&A...429..685G}. Our Figures~\ref{fig:dem} and
\ref{fig:lum2D3} suggest that this could be because morphologies such
as those produced by our STARS simulations do not produce much gas
around $T \sim 10^6$~K. This seems to be because most of the mass of
the RSG stage is bound up in the clumps. Clump material is
photoevaporated by the WR ionizing flux and this material interacts
with the fast wind. The resulting amount of material at $\sim 10^6$~K
is rather small. Another contributing factor to the non-detection of
RCW\,58 is the fact that it is about twice as distant as S\,308 and
hence the absorption column density, $N_\mathrm{H}$, to this object is
higher, which will have a large effect on the detection of soft X-ray
emission.

\section{Summary and conclusions}
\label{sec:conclude}
We have carried out numerical simulations in one and two dimensions of
the evolution of the interstellar and circumstellar bubbles that form
around massive stars during their lives. We find that the structures
that form around the stars in the final stages of their lives are
highly dependent on the details of the mass loss in the
post-main-sequence stages and that this can be used to discriminate
between different stellar evolution models. We have compared results
obtained with the publicly available STARS \citep{2004MNRAS.353...87E}
and MM2003 \citep{2003A&A...404..975M} stellar evolution models with
observations of wind-blown bubbles around three Galactic Wolf-Rayet
stars: S\,308, NGC\,6888 and RCW\,58. In particular we find that :
\begin{enumerate}
\item We confirm the findings of previous authors that the interaction
  of a fast Wolf-Rayet wind with very slow moving RSG material results
  in the break up of the swept-up shell of material into numerous
  clumps and filaments, while the interaction of the fast wind with
  faster moving YSG or LBV material leads to corrugated shells of
  swept-up material several parsecs from the star.
\item When thermal conduction is included in the simulations, the main
  differences occur in the main-sequence stage, since the timescale
  here is sufficient for the conduction process to occur. For the
  saturated conduction model that we use in this paper, conduction is
  most important at the edge of the bubble and the increased cooling
  rate that results lowers the pressure in the hot bubble. Thus
  bubbles with thermal conduction radiate more energy and are slightly
  smaller than their counterparts without conduction. In the later
  stages of stellar evolution, conduction does not affect
  the structures formed when the Wolf-Rayet wind interacts with the
  slow wind.
\item We used stellar wind parameters from three sets of publicly
  available stellar evolution models for a range of initial stellar
  masses and find that the full range of morphologies are
  reproduced. However, other tests of the models are comparisons with
  optical spectroscopy (e.g., the chemical abundances in the nebulae)
  and X-ray observations, both luminosities and spectra. 
\item We find that the STARS stellar evolution models, both 40 and
  $60\,M_{\odot}$,  always result in the complete break up of the
  swept-up  RSG material into clumps and filaments early on in
  Wolf-Rayet phase, since the slow wind velocities become so low
  ($<15$~km~s$^{-1}$) that a large amount of dense material remains
  close to the star before the onset of the fast wind. 
\item None of the MM2003 evolution models, both with and without
  stellar rotation, result in clumpy wind-blown bubbles (for wind-wind
  interaction radius $R\leqslant10$~pc) because the long duration
  of the YSG phase and faster wind velocity ($\sim 30 $~km~s$^{-1}$)
  result in a more extended distribution of dense material, which only
  becomes swept up into a dense shell at several parsecs from the
  central star.
\item The X-ray luminosities of bubbles with clumps and filaments are
  two orders of magnitude less than those with swept-up shells
  ($10^{32}$~erg~s$^{-1}$ as opposed to $10^{34}$~erg~s$^{-1}$) and
  the gas is low temperature, hence the X-rays will be very soft. This
  could be the reason that RCW\,58 has not been detected in X-rays,
  whereas S\,308 and NGC\,6888 ($L_\mathrm{X} \gtrsim
  10^{34}$~erg~s$^{-1}$) have been detected. Models with thermal
  conduction have slightly different soft-band luminosities to their
  counterparts without conduction but there is no systematic trend.
\item Observationally determined nitrogen-to-oxygen ratios reveal that
  NGC\,6888 and S\,308 have very enhanced nitrogen abundances compared
  to Solar but RCW\,58 is only moderately enriched. Of our stellar evolution
  models, the STARS models give enhanced abundances from the onset of
  the Wolf-Rayet phase, the MM2003 models without rotation give 
  enhanced abundances towards the end of the Wolf-Rayet phase, and the
  MM2003 models with rotation give enhanced abundances even during the
  main-sequence stage due to mixing of processed material from the
  stellar interior to the surface. The models with rotation are unable
  to explain moderate abundances such as those determined for RCW\,58.
\item The observed X-ray spectra of S\,308 and NGC\,6888 are very soft
  and as such must be strongly affected by interstellar
  absorption. These spectra can be well fit by one, or at most two,
  components, with the main component having temperature $T \sim
  1.1$--$1.3\times 10^6$~K. Our simulations, on the other hand,
  produce gas with a wide range of temperatures, which all contribute
  to the observed spectrum.
\end{enumerate}

Finally, we remark that none of the stellar evolution models we have
tested can simultaneously reproduce all of the features of Wolf-Rayet
wind-blown bubbles, that is morphology, abundances, X-ray luminosity
and spectral characteristics.

\acknowledgements SJA and JAT acknowledge financial support from
DGAPA, UNAM through project PAPIIT IN100309. JAT thanks CONACyT,
M\'exico for a scholarship. We thank Will Henney for a critical
reading of the manuscript. We would like to thank the anonymous
referee for constructive comments, which improved the presentation of
this paper.

\appendix
\section{Details of the numerical method}
\label{app:A}
In this appendix we include more details of the numerical method for
the interested reader.
\subsection{Equations}
In spherical symmetry, the hydrodynamic equations in conservation form
are
\begin{equation}
  \frac{\partial \rho}{\partial t} + \frac{1}{r^2}\frac{\partial r^2
  \rho u}{\partial r} = q_\mathrm{m},
\end{equation}
\begin{equation}
  \frac{\partial \rho u}{\partial t} + \frac{1}{r^2}\frac{\partial r^2
  (p + \rho u^2)}{\partial r} = \frac{2p}{r},
\end{equation}
\begin{equation}
  \frac{\partial e}{\partial t} + \frac{1}{r^2}\frac{\partial r^2
  u(e+p)}{\partial r} = G - L + q_\mathrm{e},
\end{equation}
where $\rho$, $u$, $p$ are the mass density, radial velocity and
pressure respectively, $e$ is the total energy, defined by
\begin{equation}
  e = \frac{p}{\gamma - 1} + \frac{1}{2}\rho u^2,
\end{equation}
$G$ and $L$ are the heating and cooling rates, and $\gamma$ is the
ratio of specific heats. The stellar wind is injected as a thermal
wind, with mass-loss rate $q_\mathrm{m} = 3\dot{M}_\mathrm{w}/4\pi R_\mathrm{w}^3$,
and thermal energy $q_\mathrm{e} = 3 L_\mathrm{w}/4\pi
R_\mathrm{w}^3$, where $\dot{M}_\mathrm{w}$ is the stellar wind mass-loss rate,
$L_\mathrm{w} = \dot{M}_\mathrm{w}{V_\mathrm{w}}^2/2$ is the stellar wind
  mechanical luminosity, $V_\mathrm{w}$ is the stellar wind velocity,
  and $R_\mathrm{w}$ is the radius of the stellar wind injection
  volume \citep{1985Natur.317...44C}.
Outside of the stellar wind injection region, i.e, $r > R_\mathrm{w}$, we
have $q_\mathrm{m} = q_\mathrm{e} = 0$.

In our spherically symmetric simulations, we choose $R_\mathrm{w} = 30
dr$, where $dr$ is the grid cell size. In all our 1D
simulations, the cell size is set to be $dr = 0.002$~pc. The stellar
wind injection region is small enough such that the stellar wind
reaches its terminal velocity well before it shocks. This method of
injecting the stellar wind is extremely robust to variations in the
stellar wind parameters and does not result in spurious shock waves
propagating back towards the center of symmetry.

In addition to the usual hydrodynamic equations, we use advection
equations for the neutral and ionized density, which couple the
hydrodynamics to the radiative transfer, for example
\begin{equation}
  \frac{\partial \rho n_\mathrm{X}}{\partial t} +
  \frac{1}{r^2}\frac{\partial}{\partial r}(r^2 \rho u n_\mathrm{X}) = 0 \ ,
\end{equation}
where $n_\mathrm{X}$ is the ionized or neutral number density. The
ionized hydrogen fraction is updated using the current photoionization,
collisional ionization and recombination rates in the computational
cell.
\begin{equation}
\frac{\partial y_\mathrm{i}}{\partial t} = n_e C_{\mathrm{i-1}} y_\mathrm{n} +
P_\mathrm{i-1} y_\mathrm{n} - n_e R_\mathrm{i} y_\mathrm{i} \ ,
\end{equation}
where $y_\mathrm{n}$ and $y_\mathrm{i}$ are the neutral and ionized
hydrogen fractions, $n_e$ is the electron density, $C_{\mathrm{i-1}}$
is the collisional ionization rate of neutral hydrogen and
$R_\mathrm{i}$ is the radiative recombination rate of ionized
hydrogen. Both $C_{\mathrm{i-1}}$ and $R_{\mathrm{i}}$ are functions
of temperature only and we use analytic fits for both rates. The
photoionization rate in each cell, $P_{\mathrm{i-1}}$, is obtained
from the radiative transfer procedure.

In cylindrical symmetry, the conservation equations in the $r$-$z$
plane are
\begin{equation}
  \frac{\partial \rho}{\partial t} + \frac{1}{r}\frac{\partial r
  \rho u_\mathrm{r}}{\partial r} + \frac{\partial 
  \rho u_\mathrm{z}}{\partial z}= q_\mathrm{m},
\end{equation}
\begin{equation}
  \frac{\partial \rho u_\mathrm{r}}{\partial t} + \frac{1}{r}\frac{\partial r
  (p + \rho u_\mathrm{r}^2)}{\partial
  r} + \frac{\partial \rho u_\mathrm{r} u_\mathrm{z}}{\partial z} = \frac{p}{r},
\end{equation}
\begin{equation}
  \frac{\partial \rho u_{\mathrm{z}}}{\partial t} + \frac{1}{r}\frac{\partial r
  \rho u_\mathrm{r} u_\mathrm{z}}{\partial r} + \frac{\partial
  (p + \rho u_\mathrm{z}^2)}{\partial z} = 0,
\end{equation}
\begin{equation}
  \frac{\partial e}{\partial t} + \frac{1}{r}\frac{\partial r
    u_\mathrm{r}(e+p)}{\partial r} +\frac{\partial
    u_\mathrm{z}(e+p)}{\partial z} = G - L + q_\mathrm{e}
\end{equation}
where $u_\mathrm{r}$, $u_\mathrm{z}$ are the radial and azimuthal
velocities, respectively, $p$ is the pressure, and $e$ is the total
energy, defined by
\begin{equation}
  e = \frac{p}{\gamma - 1} + \frac{1}{2}\rho (u_\mathrm{r}^2 +
  u_\mathrm{z}^2) \ .
\end{equation}
The $r$ and $z$ directions are dealt with using operator splitting.

\subsection{Heating and cooling}
\begin{figure}
\includegraphics[width=\linewidth]{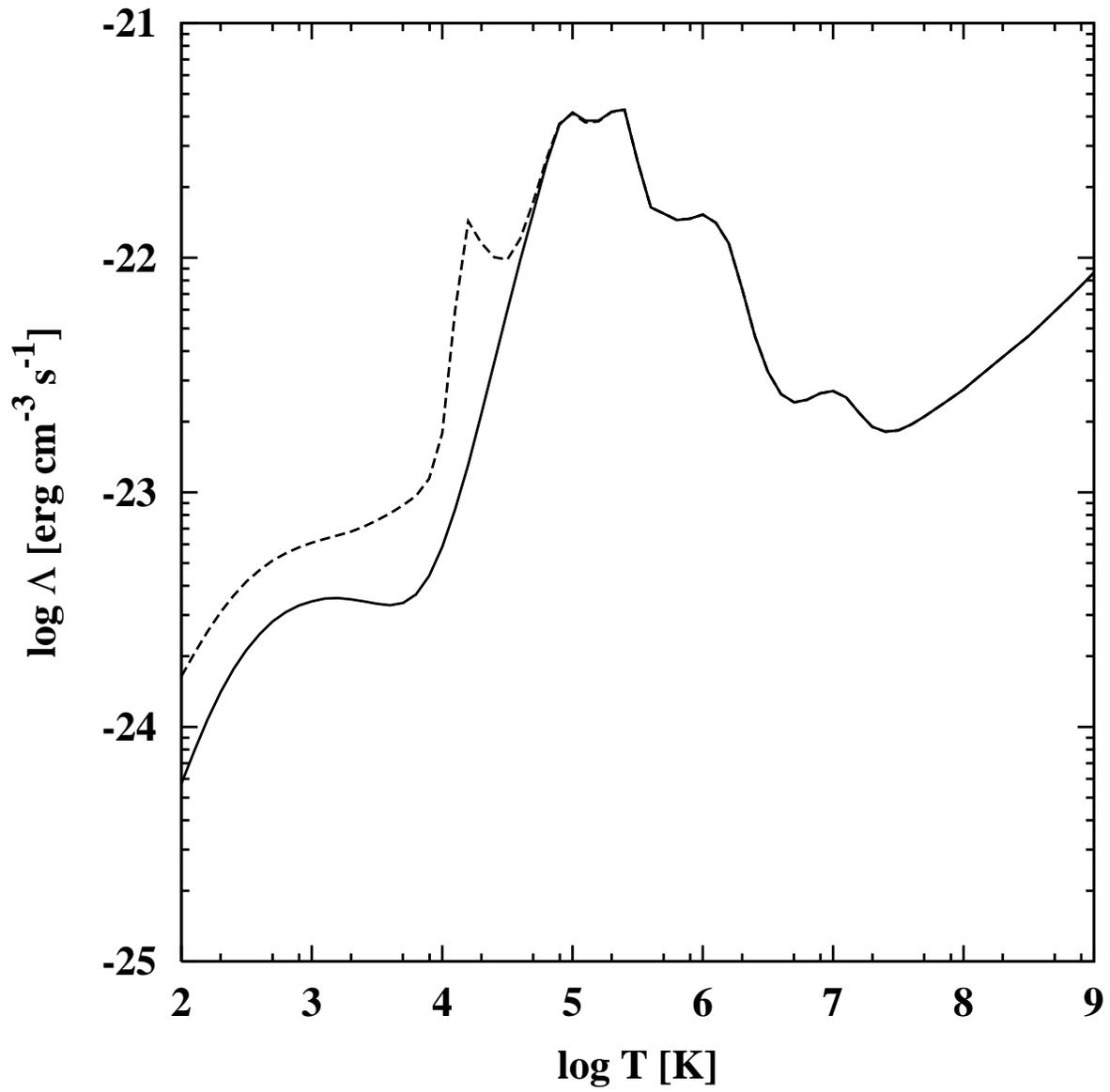}
\caption{Cooling rates for photoionized gas (solid line) and collisionally ionized gas (dotted line).}
\label{fig:coolrate}
\end{figure}
The energy equation includes radiative cooling and heating due to the
absorption of stellar radiation. In photoionized gas, the radiative
cooling term should take into account the fact that hydrogen is fully
ionized even when the gas temperature is only $10^4$~K, and so
collisional ionization of hydrogen is not an important cooling process
in an \ion{H}{2} region (see Fig.~\ref{fig:coolrate}). We generate a
cooling curve using the Cloudy photoionization code
\citep{1998PASP..110..761F} tailored to the ionizing source, which we
tabulate and use as a look-up table during the calculation. Cloudy
uses the most up-to-date and complete set of atomic data in the
astrophysical literature and is constantly updated.

The heating term at a given point in the numerical grid depends on the
photoionization rate at that point and the stellar effective
temperature, where the photoionization rate is calculated using the
radiative transfer procedure.

The heating and cooling are source terms in the energy equation, which
is solved together with the hydrodynamic equations. If thermal
conduction is included in the simulations, this is treated as an
update to the energy equation using
operator splitting after the hydrodynamic equations have been solved.

\subsection{Radiative transfer}
In this work, we use the method of short characteristics, described in
detail by \citet{1999RMxAA..35..123R}. This method consists of finding
the column density from a source point to any point on the grid by
first calculating the column density at all intermediate points, beginning with
those closest to the source. The column density at point $P$ is
therefore given in terms of the column density at a neighbouring point
$C$ by the equation
\begin{equation}
N_\mathrm{P} = N_\mathrm{C} + l n_\mathrm{0,C} \ ,
\end{equation}
where $l$ is the path length from $C$ to $P$ and $n_\mathrm{0,C}$ is
the neutral hydrogen density at point $C$.  This method can be used
both for the direct radiation from point sources and also for the diffuse
radiation from a distributed source. In the spherically symmetric
case, for direct radiation from a point source, the procedure is
trivial. For cylindrical symmetry, the calculation of the path lengths
$l$ is slightly more involved, since it is necessary to find the points where
the characteristics cross the cell boundaries by linear interpolation.

The photoionization rates are needed to calculate the ionization
state and heating rates of the gas. The column densities are therefore
calculated at the beginning of every timestep and these are used to
find the optical depths to each grid cell center and hence the
photoionization rates at those positions. The update to the ionized
(and neutral) hydrogen fractions is done by operating splitting after
the solution of the hydrodynamic equations.

\end{document}